\documentclass[twocolumn]{aastex631}

\usepackage{graphicx}
\usepackage{xcolor}
\usepackage{txfonts}
\usepackage{tablefootnote}
\usepackage{enumitem}
\usepackage{multirow}

\newcommand{\pc}	{\ifmmode {\rm pc} \else pc\fi}
\newcommand{\kpc}	{\ifmmode {\rm kpc} \else kpc\fi}
\newcommand{\ld}	{\ifmmode {\rm l.d.} \else l.d.\fi}
\newcommand{\kms}	{\ifmmode {\rm km\,s}^{-1} \else km\,s$^{-1}$\fi}
\newcommand{\cc}	{\ifmmode {\rm cm}^{-3}    \else cm$^{-3}$\fi}
\newcommand{\cmii}	{\ifmmode {\rm cm}^{-2}    \else cm$^{-2}$\fi}
\newcommand{\ergs}	{\ifmmode {\rm erg\,s}^{-1} \else erg s$^{-1}$\fi}
\newcommand{\ergcms}	{\ifmmode {\rm erg\,cm}^{-2}\,{\rm s}^{-1} \else erg\,cm$^{-2}$\,s$^{-1}$\fi}
\newcommand{\ergcmsA}	{\ifmmode {\rm erg\,cm}^{-2}\,{\rm s}^{-1}\,{\rm\AA}^{-1}
\else erg\,cm$^{-2}$\,s$^{-1}$\,\AA$^{-1}$\fi}
\newcommand{  \ergcmsHz  }{\ifmmode{\rm erg\,cm}^{-2}\,{\rm s}^{-1}\,{\rm Hz}^{-1}
                       \else ergs\,cm$^{-2}$\,s$^{-1}$\,Hz$^{-1}$\fi}
\newcommand{\kev}	{\ifmmode {\rm keV} \else keV\fi}

\newcommand{\mic}	{\ifmmode {\rm \mu m} \else $\mu$m\fi}
\newcommand{\vFWHM}	{\ifmmode v_{\mbox{\tiny FWHM}} \else $v_{\mbox{\tiny FWHM}}$\fi}
\newcommand{\vBLR}	{\ifmmode v_{\mbox{\tiny BLR}} \else $v_{\mbox{\tiny BLR}}$\fi}
\newcommand{\sigBLR}	{\ifmmode \sigma_{\mbox{\tiny BLR}} \else $\sigma_{\mbox{\tiny BLR}}$\fi}
\newcommand{\vNLR}	{\ifmmode v_{\mbox{\tiny NLR}} \else $v_{\mbox{\tiny NLR}}$\fi}
\newcommand{\tauBLR}	{\ifmmode \tau_{\mbox{\tiny BLR}} \else $\tau_{\mbox{\tiny BLR}}$\fi}

\newcommand{\Hubble}	{\ifmmode {\rm km\,s}^{-1}\,{\rm Mpc}^{-1} \else km\,s$^{-1}$\,Mpc$^{-1}$\fi}
\newcommand{\NDunit}	{\ifmmode {\rm Mpc}^{-3} \else Mpc$^{-3}$\fi}
\newcommand{\LFunit}	{\ifmmode {\rm Mpc}^{-3}\,{\rm mag}^{-1} \else Mpc$^{-3}$\,mag$^{-1}$\fi}
\newcommand{\MFunit}	{\ifmmode {\rm Mpc}^{-3}\,{\rm dex}^{-1} \else Mpc$^{-3}$\,dex$^{-1}$\fi}

\newcommand{\Msun}{\ifmmode M_{\odot} \else $M_{\odot}$\fi}
\newcommand{\Lsun}{\ifmmode L_{\odot} \else $L_{\odot}$\fi}
\newcommand{\Zsun}{\ifmmode Z_{\odot} \else $Z_{\odot}$\fi}
\newcommand{\mpyr}{\ifmmode \Msun\,{\rm yr}^{-1} \else $\Msun\,{\rm yr}^{-1}$\fi}

\newcommand{\Msol}{\Msun}

\newcommand{\qnote}{\ifmmode q_{0} \else $q_{0}$\fi}
\newcommand{\Hnote}{\ifmmode H_{0} \else $H_{0}$\fi}
\newcommand{\hnote}{\ifmmode h_{0} \else $h_{0}$\fi}
\newcommand{\anote}{\ifmmode a_{0} \else $a_{0}$\fi}
\newcommand{\tnote}{\ifmmode t_{0} \else $t_{0}$\fi}
\newcommand{\OmegaLambda}{\ifmmode \Omega_{\Lambda} \else $\Omega_{\Lambda}$\fi}
\newcommand{\OmegaM}{\ifmmode \Omega_{M} \else $\Omega_{M}$\fi}
\def\gsim{\;\rlap{\lower 2.5pt \hbox{$\sim$}}\raise 1.5pt\hbox{$>$}\;}
\def\lsim{\;\rlap{\lower 2.5pt \hbox{$\sim$}}\raise 1.5pt\hbox{$<$}\;}

\newcommand{  \Halpha   }{\ifmmode {\rm H}\alpha \else H$\alpha$\fi}

\newcommand{  \ha       }{\Halpha}
\newcommand{  \Hbeta    }{\ifmmode {\rm H}\beta \else H$\beta$\fi}

\newcommand{  \hb       }{\Hbeta}
\newcommand{  \Hgamma   }{\ifmmode {\rm H}\gamma \else H$\gamma$\fi}
\newcommand{  \Hdelta   }{\ifmmode {\rm H}\delta \else H$\delta$\fi}
\newcommand{  \Lya      }{\ifmmode {\rm Ly}\alpha \else Ly$\alpha$\fi}
\newcommand{  \Lyb      }{\ifmmode {\rm Ly}\beta \else Ly$\beta$\fi}
\newcommand{  \Pa       }{\ifmmode {\rm P}\alpha \else P$\alpha$\fi}
\newcommand{  \Pb       }{\ifmmode {\rm P}\beta \else P$\beta$\fi}
\newcommand{  \Bra      }{\ifmmode {\rm Br}\alpha \else Br$\alpha$\fi}
\newcommand{  \Brg      }{\ifmmode {\rm Br}\gamma \else Br$\gamma$\fi}
\newcommand{  \hii      }{\ifmmode {\rm H}\,\textsc{ii} \else H\,\textsc{ii}\fi}

\newcommand{  \hei      }{\ifmmode {\rm He}\,\textsc{i} \else He\,\textsc{i}\fi}
\newcommand{  \heii     }{\ifmmode {\rm He}\,\textsc{ii} \else He\,\textsc{ii}\fi}
\newcommand{  \HeIIuv   }{\ifmmode {\rm He}\,\textsc{ii}\,\lambda1640 \else He\,\textsc{ii}\,$\lambda1640$\fi}
\newcommand{  \HeIIop   }{\ifmmode {\rm He}\,\textsc{ii}\,\lambda4686 \else He\,\textsc{ii}\,$\lambda4686$\fi}
\newcommand{  \CII	}{\ifmmode \left[{\rm C}\,\textsc{ii}\right]\,\lambda157.74\,\mu{\rm m} \else [C\,{\sc ii}]\ $\lambda157.74\,\mu{\rm m}$\fi}
\newcommand{  \cii	}{\ifmmode \left[{\rm C}\,\textsc{ii}\right] \else [C\,{\sc ii}]\fi}

\newcommand{  \ciii     }{\ifmmode {\rm C}\,\textsc{iii}\right] \else C\,\textsc{iii}]\fi}
\newcommand{  \CIII     }{\ifmmode {\rm C}\,\textsc{iii}\right]\,\lambda1909 \else C\,\textsc{iii}]\,$\lambda1909$\fi}
\newcommand{  \civ      }{\ifmmode {\rm C}\,\textsc{iv}  \else C\,\textsc{iv}\fi}
\newcommand{  \CIV      }{\ifmmode {\rm C}\,\textsc{iv}\,\lambda1549 \else C\,\textsc{iv}\,$\lambda1549$\fi}
\newcommand{  \NIIopt   }{\ifmmode \left[{\rm N}\,\textsc{ii}\right]\,\lambda6584 \else [N\,\textsc{ii}]\,$\lambda6584$\fi}
\newcommand{  \nii      }{\ifmmode \left[{\rm N}\,\textsc{ii}\right]  \else [N\,\textsc{ii}]\fi}
\newcommand{  \niii     }{\ifmmode {\rm N}\,\textsc{iii} \else N\,\textsc{iii}\fi}
\newcommand{  \NIII     }{\ifmmode {\rm N}\,\textsc{iii}\,\lambda4640 \else N\,\textsc{iii}\,$\lambda4640$\fi}
\newcommand{  \niv      }{\ifmmode {\rm N}\,\textsc{iv}  \else N\,\textsc{iv}\fi}
\newcommand{  \NIVuv    }{\ifmmode {\rm N}\,\textsc{iv}\,\lambda1486 \else N\,\textsc{iv}\,$\lambda1486$\fi}
\newcommand{  \nv       }{\ifmmode {\rm N}\,\textsc{v}   \else N\,\textsc{v}\fi}
\newcommand{\oi}{\ifmmode \left[{\rm O}\,\textsc{i}\right] \else [O\,{\sc i}]\fi}
\newcommand{\OI}{\ifmmode \left[{\rm O}\,\textsc{i}\right]\,\lambda6300 \else [O\,{\sc i}]$\,\lambda6300$\fi}
\newcommand{\oii}{\ifmmode \left[{\rm O}\,\textsc{ii}\right] \else [O\,{\sc ii}]\fi}
\newcommand{\OII}{\ifmmode \left[{\rm O}\,\textsc{ii}\right]\,\lambda3727 \else [O\,{\sc ii}]\,$\lambda3727$\fi}
\newcommand{\oiii}{\ifmmode \left[{\rm O}\,\textsc{iii}\right] \else [O\,{\sc iii}]\fi}
\newcommand{\OIII}{\ifmmode \left[{\rm O}\,\textsc{iii}\right]\,\lambda5007 \else [O\,{\sc iii}]\,$\lambda5007$\fi}
\newcommand{  \OIIIbf   }{\ifmmode {\rm O}\,\textsc{iii}\,\lambda3133 \else O\,\textsc{iii}\,$\lambda3133$\fi}
\newcommand{  \OIIIuv   }{\ifmmode {\rm O}\,\textsc{iii}\,\lambda1663 \else O\,\textsc{iii}\,$\lambda1663$\fi}
\newcommand{  \oiv      }{\ifmmode {\rm O}\,\textsc{iv}  \else O\,\textsc{iv}\fi}
\newcommand{  \OIVuv    }{\ifmmode {\rm O}\,\textsc{iv}\,\lambda1402  \else O\,\textsc{iv}\,$\lambda1402$\fi}
\newcommand{  \OIVIR    }{\ifmmode {\rm O}\,\textsc{iv}\,25.9\,\mu {\rm m} \else O\,\textsc{iv}\,$25.9\,\mu$m\fi}
\newcommand{  \ovi      }{\ifmmode {\rm O}\,\textsc{vi}   \else O\,\textsc{vi}\fi}
\newcommand{  \Ovi      }{\ifmmode {\rm O}\,\textsc{vi}\,\lambda1035 \else O\,\textsc{vi}\,$\lambda1035$\fi}
\newcommand{  \nei      }{\ifmmode {\rm Ne}\,\textsc{i}   \else Ne\,\textsc{i}\fi}
\newcommand{  \neii     }{\ifmmode {\rm Ne}\,\textsc{ii}  \else Ne\,\textsc{ii}\fi}
\newcommand{  \NeiiIR   }{\ifmmode {\rm Ne}\,\textsc{ii}\,12.8\,\mu {\rm m} \else Ne\,\textsc{ii}\,$12.8\,\mu$m\fi}
\newcommand{  \neiii    }{\ifmmode {\rm Ne}\,\textsc{iii} \else Ne\,\textsc{iii}\fi}
\newcommand{  \neiv     }{\ifmmode {\rm Ne}\,\textsc{iv}  \else Ne\,\textsc{iv}\fi}
\newcommand{  \nev      }{\ifmmode {\rm Ne}\,\textsc{v}   \else Ne\,\textsc{v}\fi}
\newcommand{  \NevIR    }{\ifmmode {\rm Ne}\,\textsc{v}\,24.3\,\mu {\rm m} \else Ne\,\textsc{v}\,$24.3\,\mu$m\fi}
\newcommand{  \nevi     }{\ifmmode {\rm Ne}\,\textsc{vi}  \else Ne\,\textsc{vi}\fi}
\newcommand{  \mgi      }{\ifmmode {\rm Mg}\,\textsc{i} \else Mg\,\textsc{i}\fi}
\newcommand{  \mgii     }{\ifmmode {\rm Mg}\,\textsc{ii} \else Mg\,\textsc{ii}\fi}
\newcommand{  \MgII     }{\ifmmode {\rm Mg}\,\textsc{ii}\,\lambda2798 \else Mg\,\textsc{ii}\,$\lambda2798$\fi}
\newcommand{  \sii      }{\ifmmode {\rm S}\,\textsc{ii} \else S\,\textsc{ii}\fi}
\newcommand{  \siii     }{\ifmmode {\rm S}\,\textsc{iii} \else S\,\textsc{iii}\fi}
\newcommand{  \siv      }{\ifmmode {\rm S}\,\textsc{iv} \else S\,\textsc{iv}\fi}
\newcommand{  \sili     }{\ifmmode {\rm Si}\,\textsc{i}   \else Si\,\textsc{i}\fi}
\newcommand{  \silii    }{\ifmmode {\rm Si}\,\textsc{ii}  \else Si\,\textsc{ii}\fi}
\newcommand{  \Siliv    }{\ifmmode {\rm Si}\,\textsc{iv}  \else Si\,\textsc{iv}\fi}
\newcommand{  \SilIVuv  }{\ifmmode {\rm Si}\,\textsc{iv}\,\lambda1400  \else Si\,\textsc{iv}\,$\lambda1400$\fi}
\newcommand{  \AlIII   }{\ifmmode {\rm Al}\,\textsc{iii}\,\lambda1857 \else Al\,\textsc{iii}\,$\lambda1857$\fi}
\newcommand{  \Aliii   }{\ifmmode {\rm Al}\,\textsc{iii} \else Al\,\textsc{iii}\fi}
\newcommand{  \caii     }{\ifmmode {\rm Ca}\,\textsc{ii} \else Ca\,\textsc{ii}\fi}
\newcommand{  \feii     }{\ifmmode {\rm Fe}\,\textsc{ii} \else Fe\,\textsc{ii}\fi}
\newcommand{  \feiii    }{\ifmmode {\rm Fe}\,\textsc{iii} \else Fe\,\textsc{iii}\fi}
\newcommand{  \Kalpha   }{\ifmmode {\rm K}\alpha \else K$\alpha$\fi}

\newcommand{ \Lhb   }{\ifmmode L_{\hb} \else $L_{\hb}$\fi}
\newcommand{ \Lha   }{\ifmmode L_{\ha} \else $L_{\ha}$\fi}
\newcommand{ \fwhb  }{\ifmmode {\rm FWHM}\left(\hb\right) \else FWHM(\hb)\fi}
\newcommand{\sighb  }{\ifmmode \sigma\left(\hb\right) \else $\sigma\left(\hb\right)$\fi}
\newcommand{ \ewhb  }{\ifmmode {\rm EW}\left(\hb\right) \else EW(\hb)\fi}
\newcommand{ \fwha  }{\ifmmode {\rm FWHM}\left(\ha\right) \else FWHM(\ha)\fi}
\newcommand{ \ewha  }{\ifmmode {\rm EW}\left(\ha\right) \else EW(\ha)\fi}
\newcommand{ \Lmg   }{\ifmmode L\left(\mgii\right) \else $L\left(\mgii\right)$\fi}
\newcommand{ \fwmg  }{\ifmmode {\rm FWHM}\left(\mgii\right) \else FWHM(\mgii)\fi}
\newcommand{ \Lciv  }{\ifmmode L\left(\civ\right) \else $L\left(\civ\right)$\fi}
\newcommand{ \fwciv }{\ifmmode {\rm FWHM}\left(\civ\right) \else FWHM(\civ)\fi}
\newcommand{ \fwhm  }{\ifmmode {\rm FWHM} \else FWHM\fi} 
\newcommand{ \voff  }{\ifmmode v_{\rm off} \else $v_{\rm off}$\fi} 
\newcommand{ \vmax  }{\ifmmode v_{\rm max} \else $v_{\rm max}$\fi} 

\newcommand{ \mumg  }{\ifmmode \mu\left(\mgii\right) \else $\mu\left(\mgii\right)$\fi}
\newcommand{ \fmg   }{\ifmmode f\left(\mgii\right) \else $f\left(\mgii\right)$\fi}
\newcommand{ \muciv }{\ifmmode \mu\left(\civ\right) \else $\mu\left(\civ\right)$\fi}
\newcommand{ \fciv  }{\ifmmode f\left(\civ\right) \else $f\left(\civ\right)$\fi}
\newcommand{  \auvo     }{\ifmmode \alpha_{\nu,{\rm UVO}} \else $\alpha_{\nu,{\rm UVO}}$\fi}
\newcommand{  \Ledd     }{\ifmmode L_{\rm Edd} \else $L_{\rm Edd}$\fi}
\newcommand{  \lamLlam  }{\ifmmode \lambda L_{\lambda} \else $\lambda L_{\lambda}$\fi}
\newcommand{  \lLl      }{\ifmmode \lambda L_{\lambda} \else $\lambda L_{\lambda}$\fi}
\newcommand{  \nuLnu    }{\ifmmode \nu L_{\nu} \else $\nu L_{\nu}$\fi}
\newcommand{  \nLn      }{\ifmmode \nu L_{\nu} \else $\nu L_{\nu}$\fi}
\newcommand{  \Luv      }{\ifmmode L_{1350} \else $L_{1350}$\fi}
\newcommand{  \Lop      }{\ifmmode L_{5100} \else $L_{5100}$\fi}
\newcommand{  \lLop     }{\ifmmode \log\left(\Lop/\ergs\right) \else $\log\left(\Lop/\ergs\right)$\fi}
\newcommand{  \Lthree   }{\ifmmode L_{3000} \else $L_{3000}$\fi}
\newcommand{  \lLthree  }{\ifmmode \log\left(\Lthree/\ergs\right) \else $\log\left(\Lthree/\ergs\right)$\fi}
\newcommand{  \Lsix      }{\ifmmode L_{6200} \else $L_{6200}$\fi}
\newcommand{  \lLisx     }{\ifmmode \log\left(\Lop/\ergs\right) \else $\log\left(\Lop/\ergs\right)$\fi}
\newcommand{  \Lxray    }{\ifmmode L_{\rm X} \else $L_{\rm X}$\fi}

\newcommand{  \Lhard    }{\ifmmode L_{\rm 2-10} \else $L_{\rm 2-10}$\fi}
\newcommand{  \Lsoft    }{\ifmmode L_{\rm 0.5-2} \else $L_{\rm 0.5-2}$\fi}

\newcommand{  \Lbh     }{\ifmmode L_{1350} \else $L_{1350}$\fi}

\newcommand{\Fthree}{\ifmmode F_{3000} \else $F_{3000}$\fi}
\newcommand{\fuv}{\ifmmode f_{\lambda}\left(1450{\rm \AA}\right) \else $f_{\lambda}\left(1450 {\rm \AA}\right)$\fi}
\newcommand{\fthree}{\ifmmode f_{\lambda}\left(3000{\rm \AA}\right) \else $f_{\lambda}\left(3000{\rm \AA}\right)$\fi}
\newcommand{\fH}{\ifmmode f_{\lambda}\left(1.65\micron\right) \else
$f_{\lambda}\left(1.65\micron\right)$\fi}

\newcommand{\fbol}{\ifmmode f_{\rm bol} \else $f_{\rm bol}$\fi}
\newcommand{\fbolwv}{\ifmmode f_{\rm bol}\left(\lambda\right) \else $f_{\rm bol}\left(\lambda\right)$\fi}
\newcommand{\fbolopt}{\ifmmode f_{\rm bol}\left(5100{\rm \AA}\right) \else $f_{\rm bol}\left(5100{\rm \AA}\right)$\fi}
\newcommand{\fbolthree}{\ifmmode f_{\rm bol}\left(3000{\rm \AA}\right) \else $f_{\rm bol}\left(3000{\rm \AA}\right)$\fi}
\newcommand{\fboluv}{\ifmmode f_{\rm bol}\left(1450{\rm \AA}\right) \else $f_{\rm bol}\left(1450{\rm \AA}\right)$\fi}

\newcommand{\fbolbat}{\ifmmode f_{\rm bol}\left(14-150\,\kev\right) \else $f_{\rm bol}\left(14-150\,\kev\right)$\fi}
\newcommand{\fbolhard}{\ifmmode f_{\rm bol}\left(2-10\,\kev\right) \else $f_{\rm bol}\left(2-10\,\kev\right)$\fi}

\newcommand{\fobs}{\ifmmode f_{\rm obs} \else $f_{\rm obs}$\fi}

\newcommand{  \mbh      }{\ifmmode M_{\rm BH} \else $M_{\rm BH}$\fi}
\newcommand{  \lmbh     }{\ifmmode \log\left(\mbh/\Msun\right) \else $\log\left(\mbh/\Msun\right)$\fi} 
\newcommand{  \lledd    }{\ifmmode L/L_{\rm Edd} \else $L/L_{\rm Edd}$\fi}
\newcommand{  \mmedd    }{\ifmmode \dot{m}/\dot{m}_{\rm \,Edd} \else $\dot{m}/\dot{m}_{\rm \,Edd}$\fi}
\newcommand{  \Lbol     }{\ifmmode L_{\rm bol} \else $L_{\rm bol}$\fi}
\newcommand{  \lbol     }{\ifmmode L_{\rm bol} \else $L_{\rm bol}$\fi}
\newcommand{  \lLbol    }{\ifmmode \log\left(\Lbol/\ergs\right) \else $\log\left(\Lbol/\ergs\right)$\fi} 
\newcommand{  \Lagn     }{\ifmmode L_{\rm AGN} \else $L_{\rm AGN}$\fi}
\newcommand{  \lagn     }{\ifmmode L_{\rm AGN} \else $L_{\rm AGN}$\fi}

\newcommand{  \tgrow     }{\ifmmode t_{\rm growth} \else $t_{\rm growth}$\fi}
\newcommand{  \tAD     }{\ifmmode t_{\rm acc} \else $t_{\rm acc}$\fi}
\newcommand{  \tacc    }{\ifmmode t_{\rm acc} \else $t_{\rm acc}$\fi}
\newcommand{  \tUni      }{\ifmmode t_{\rm Universe} \else $t_{\rm Universe}$\fi}

\newcommand{  \Mdotin	}{\ifmmode \dot{M}_{\rm infall} \else $\dot{M}_{\rm infall}$\fi}
\newcommand{  \Mdotbh	}{\ifmmode \dot{M}_{\rm BH} \else $\dot{M}_{\rm BH}$\fi}
\newcommand{  \Mdotad	}{\ifmmode \dot{M}_{\rm AD} \else $\dot{M}_{\rm AD}$\fi}
\newcommand{  \Mdotacc	}{\ifmmode \dot{M}_{\rm acc} \else $\dot{M}_{\rm acc}$\fi}
\newcommand{  \Mdotthin	}{\ifmmode \dot{M}_{\rm thin} \else $\dot{M}_{\rm thin}$\fi}
\newcommand{  \Mdotdisk	}{\ifmmode \dot{M}_{\rm disk} \else $\dot{M}_{\rm disk}$\fi}

\newcommand{  \kbol     }{\ifmmode k_{\rm bol} \else $k_{\rm bol}$\fi}

\def\niii{\ifmmode \textrm{N}\,\textsc{iii} \else N\,\textsc{iii}\fi}
\def\NIII{\ifmmode \textrm{N}\,\textsc{iii}\,\lambda4640 \else N\,\textsc{iii}\,$\lambda4640$\fi}
\def\FeX{\ifmmode \textrm{[Fe}\,\textsc{x}]\,\lambda6375 \else[Fe\,\textsc{x}]\,$\lambda6375$\fi}
\def\fex{\ifmmode \textrm{[Fe}\,\textsc{x}] \else[Fe\,\textsc{x}]\fi}
\def\FeXIV{\ifmmode \textrm{[Fe}\,\textsc{xiv}]\,\lambda5303 \else[Fe\,\textsc{xiv}]\,$\lambda5303$\fi}
\def\fexiv{\ifmmode \textrm{[Fe}\,\textsc{xiv}] \else[Fe\,\textsc{xiv}]\fi}
\def\HeIIop{\ifmmode \textrm{He}\,\textsc{ii}\,\lambda4686 \else He\,\textsc{ii}\,$\lambda4686$\fi}
\def\Hgamma{\ifmmode \textrm{H}\gamma \else H$\gamma$\fi}
\def\Hdelta{\ifmmode \textrm{H}\delta \else H$\delta$\fi}

\def\OIIIbf{\ifmmode \textrm{O}\,\textsc{iii}\,\lambda3133 \else O\,\textsc{iii}\,$\lambda3133$\fi}

\def\OIII{\ifmmode \left[{\rm O}\,\textsc{iii}\right]\,\lambda5007 \else [O\,{\sc iii}]\,$\lambda5007$\fi}
\def\oiii{\ifmmode \left[{\rm O}\,\textsc{iii}\right]\,\lambda5007 \else [O\,{\sc iii}]\,$\lambda5007$\fi}

\newcommand{\Fhb}{\ifmmode F(\hb) \else $F(\hb)$\fi}
\newcommand{\Foiii}{\ifmmode F(\oiii) \else $F(\oiii)$\fi}
\newcommand{\Fbf}{\ifmmode F(\heii+\niii) \else $F(\heii+\heii)$\fi}
\newcommand{\Ffex}{\ifmmode F(\fex) \else $F(\fex)$\fi}

\def\ergs	{\ifmmode \textrm{erg\,s}^{-1} \else erg s$^{-1}$\fi}

\newcommand{\todo}{\ifmmode \text{\color{red}\Huge{\(\bullet\)}} \else {\color{red}{\Huge$\bullet$}}\fi}
\newcommand{\tido}{\ifmmode {{\color{red}\bullet}} \else {\color{red}$\bullet$}\fi}

\def\nuLnu{\nuL_{\nu} \else $\nu L_{\nu}$\fi}

\newcommand{\objname}{AT\,2019aalc}
\newcommand{\hostname}{SDSS\,J152416.66+045119.0}
\newcommand{\hname}{J1524+0451}

\begin{document}

%\title{AT2019aalc: double flaring event Bowen Fluorescence features in Active Galactic Nuclei with}

\title{\objname: a Bowen Fluorescence Flare With a Precursor Flare in an Active Galactic Nucleus}

\author[0000-0003-2656-6726]{Marzena \'Sniegowska}
\affiliation{School of Physics and Astronomy, Tel Aviv University, Tel Aviv 69978, Israel}

\author[0000-0002-3683-7297]{Benny Trakhtenbrot}
\affiliation{School of Physics and Astronomy, Tel Aviv University, Tel Aviv 69978, Israel}
\affiliation{Max-Planck-Institut f{\"u}r extraterrestrische Physik, Gie\ss{}enbachstra\ss{}e 1, 85748 Garching, Germany}
\affiliation{Excellence Cluster ORIGINS, Boltzmannsstra\ss{}e 2, 85748 Garching, Germany}

\author[0000-0002-7466-4868]{Lydia Makrygianni}
\affiliation{Department of Physics, Lancaster University, Lancaster LA1 4YB,
UK}

\author[0000-0001-7090-4898]{Iair Arcavi}
\affiliation{School of Physics and Astronomy, Tel Aviv University, Tel Aviv 69978, Israel}
\affiliation{CIFAR Azrieli Global Scholars program, CIFAR, Toronto, Canada}

\author[0000-0001-5231-2645]{Claudio Ricci}
\affiliation{Instituto de Estudios Astrof\'isicos, Facultad de Ingenier\'ia y Ciencias, Universidad Diego Portales, Av. Ej\'ercito Libertador 441, Santiago, Chile} 
\affiliation{Kavli Institute for Astronomy and Astrophysics, Peking University, Beijing 100871, China}

\author[0009-0007-8485-1281]{Sarah Faris}
\affiliation{School of Physics and Astronomy, Tel Aviv University, Tel Aviv 69978, Israel}

\author[0000-0002-4533-3170]{Biswaraj Palit}
\affiliation{Nicolaus Copernicus Astronomical Center, Polish Academy of Sciences, ul.\ Bartycka 18, 00-716 Warsaw, Poland}

%%%% LCO folks:

\author[0000-0003-4253-656X]{D. Andrew Howell}
\affiliation{Las Cumbres Observatory, 6740 Cortona Drive, Suite 102, Goleta, CA 93117-5575, USA}
\affiliation{Department of Physics, University of California, Santa Barbara, CA 93106-9530, USA}

\author[0000-0001-9570-0584]{Megan Newsome}
\affiliation{Las Cumbres Observatory, 6740 Cortona Drive, Suite 102, Goleta, CA 93117-5575, USA}
\affiliation{Department of Physics, University of California, Santa Barbara, CA 93106-9530, USA}

\author[0000-0003-4914-5625]{Joseph Farah}
\affiliation{Las Cumbres Observatory, 6740 Cortona Drive, Suite 102, Goleta, CA 93117-5575, USA}
\affiliation{Department of Physics, University of California, Santa Barbara, CA 93106-9530, USA}

\author[0000-0001-5807-7893]{Curtis McCully}
\affiliation{Las Cumbres Observatory, 6740 Cortona Drive, Suite 102, Goleta, CA 93117-5575, USA}

\author[0000-0003-0209-9246]{Estefania Padilla-Gonzalez}
\affiliation{Las Cumbres Observatory, 6740 Cortona Drive, Suite 102, Goleta, CA 93117-5575, USA}
\affiliation{Department of Physics, University of California, Santa Barbara, CA 93106-9530, USA}

\author[0000-0003-0794-5982]{Giacomo Terreran}
\affiliation{Las Cumbres Observatory, 6740 Cortona Drive, Suite 102, Goleta, CA 93117-5575, USA}

%\author[0000-0000-0000-0000]{co-authors}
%\affiliation{xxx}

\begin{abstract}
\objname\ is a peculiar sequence of highly variable emission events observed towards the nucleus of the broad-line AGN \hostname. 
The system exhibited two distinct UV/optical flares (the first detected in 2019, the second one in 2023). 
Spectra obtained following the detection of the second flare revealed prominent Bowen fluorescence and high-ionization coronal emission lines, which were much weaker, if at all detectable, in a spectrum taken following the first flare. 
We present and analyze a large set of multi-wavelength, multi-epoch data for this source, with particular emphasis on optical spectroscopic monitoring conducted with the Las Cumbres Observatory network. 
During the relatively slow dimming that followed the second optical flare, the UV/optical light-curve shows a sequence of minor rebrightening events, while the Bowen fluorescence and the coronal lines vary (roughly) in tandem with these ``bumps'' in the broad-band light-curve.
Most of the observed behavior of \objname\ links it to the growing class of Bowen fluorescence flares (BFFs) while setting it apart from canonical tidal disruption events. 
However, \objname\ has some outstanding peculiarities, including two short flares seen in its soft X-ray light-curve during the dimming phase of the second optical flare, and which do not seem to be linked to the emission line variations.
We discuss the optical and X-ray properties of the source and possible scenarios of the origin of the flare, in particular radiation pressure instabilities in the (pre-existing) AGN accretion disk.
\end{abstract}

%\keywords{AGN -- nuclear transient}
\section{Introduction}
\label{sec:intro}

Active Galactic Nuclei (AGNs), powered by accreting supermassive black holes (SMBHs), are thought to sustain their active phases over relatively long periods, of at least tens of thousands of years. During these prolonged accretion episodes, their multi-wavelength emission is known to show stochastic variability, typically of order $\approx$ 10$\%$ over a wide range of timescales, specifically months-to-decades in the optical band \cite[e.g.,][and references therein]{VdB04,2012macleod,Caplar17}.

Advances in time-domain astronomy allowed to study AGN variability for increasingly large samples and across increasingly short timescales. Thanks to wide surveys such as those enabled by the Catalina Real-time Transient Survey \citep[CRTS;][]{2009drake}, the Zwicky Transient Facility \cite[ZTF;][]{2019PASP..131a8003M} and
the All-Sky Automated Survey for Supernovae \citep[ASAS-SN;][]{Shappee14}, it became possible to pursue follow-up studies of specific cases or samples of interestingly variable AGNs, using responsive spectroscopic and multi-wavelength facilities.
This helped to reveal several classes of surprisingly fast, dramatic flux changes related to AGNs and SMBH accretion. In those cases, AGN emission may vary by over an order of magnitude within a timescale of a few months, while in some extreme cases, the SMBH accretion activity appears to turn ``on'' or ``off''. 
Moreover, some of these types of extreme events of SMBH-related flux changes could present recurring phenomena, particularly recurring flares.

One of those classes is known as changing-look AGN \cite[CL-AGN; see review by][and references therein]{2023riccitrakhtenbrot}. In the UV-optical regime, CL-AGNs show the (dis-)-appearance of broad emission lines and/or the blue, accretion-powered continuum emission, with corresponding flux changes of a factor of 2 or more, within timescales of a few years and perhaps even a few months \cite[e.g.,][]{Zeltyn2022}. Recent years have seen a significant increase in the number of known CL-AGN, owing to dedicated searches in survey data \citep{macleod2019, graham2020, Green_etal_2022, 2024zeltyn,Guo24}. In parallel, there has been a steady progress in the theoretical study of the physical mechanisms driving such unexpectedly dramatic variability events \citep[e.g.,][]{noda2018,2021scepi,Li24}.
Importantly, there are several reported cases of recurring CL-AGN spectral transitions. This includes both low-luminosity, local Seyfert galaxies (see, e.g., \citealt{2018MNRAS.475.2051K}, \citealt{oknyansky2019}, the compilation in \citealt{Wang25}, and references therein), as well as higher-luminosity systems out to $z\sim0.5$ \citep{Zeltyn2022,Jana24,Wang25}. One possible explanation for recurring CL-AGN, put forward by \cite{Sniegowska20}, is based on radiation pressure instabilities in a narrow unstable region in a standard (thin) AGN accretion disk. This model was later extended by adding a magnetic field \citep{Pan21} and further focusing only on the inner parts of the disk \citep{sniegowska2023}.
Broadly speaking, the timescale for the recurring flares in such models is dictated by the size of the unstable region and the strength of the magnetic field.

Another main class of SMBH-related transients is driven by stellar tidal disruption events (TDEs; see, e.g., \citealt{2012gezari, 2014arcavi}, and the recent review by \citealt{Gezari21_rev}).
Those TDEs identified and monitored in the UV-optical regime are typically characterized by a relatively fast ($\lesssim2$ months) rise to a single peak, followed by a decay that lasts a few months, as well as the presence of an extremely broad ($\gtrsim$ 10000\,\kms) \HeIIop\ and/or \ha\ emission lines. 
Some TDEs were also observed to have so-called ``coronal'' emission lines from highly ionized species \cite[e.g.,][]{2023Short, 2024newsome}. The two prevalent explanations for the origin of the transient UV-optical emission in such TDEs link it either to the reprocessing of X-ray radiation originating from the newly-formed, transient accretion disk \citep[e.g.,][]{Guillochon2014, Roth2016,2018Dai}, or to shocks in the stellar debris streams \cite[e.g.,][]{2015piran}. Some simulations \cite[e.g.,][]{Steinberg24} suggest that both mechanisms may coincide and dominate the emission in different stages during the TDE.
The tidal disruption of stars by SMBHs may also produce recurring transients. Specifically, a partial stellar disruption may give rise to repeating TDE-like flares--a scenario pursued both theoretically \cite[e.g.,][]{Coughlin2019} and observationally \citep[e.g.,][]{2023somalwar}.
% Also, the repeating pTDE driven by partial striping of stars, like AT 2020vdq \citep{}, were observed. This type of source shows more than one flare, also accompanied by the appearance of extremely broad \heiiopt.
Models suggest that such partial TDEs (pTDEs) should broadly evolve faster than the canonical, total disruption counterparts \citep{Coughlin2019}.
The period between recurring flares in pTDEs is thought to be dictated by the mass distribution of the stellar debris and the orbital parameters of the system.

Recently, a new class of flaring events related to AGNs was defined, which are often referred to as ``Bowen Fluorescence Flares'' \cite[BFFs;][]{2019Trakhtenbrot}.
Their key defining feature is the presence of exceptionally strong, and broad ($\approx2000$\,\kms) Bowen emission lines, such as \NIII\ and \OIIIbf, in addition to the \HeIIop\ line that is more directly linked to the Bowen fluorescence mechanism \citep{Bowen1928}.
Such Bowen fluorescence (BF) lines are rarely seen in persistent, non-flaring AGNs \cite[e.g.,][]{VandenBerk2001}, despite long-lasting predictions that they should be observable \citep{Netzer1985}.
The BF lines are thought to originate from high density ($n_{\rm H}>10^{9.5}\,{\rm cm}^{-3}$) and high metalicity gas, exposed to intense extreme UV radiation \cite[$>$50 eV;][]{ Netzer1985}.
Although BF lines are also detected in a few optically-identified TDEs \cite[e.g.,][]{Holoien2016,Leloudas2019}, BFFs appear to be a distinct (observational) class, given their much slower declining light-curves, the absence of exceptionally broad \HeIIop\ lines (with $\gtrsim$10000\,\kms), and the apparent preference to occur in already-active SMBHs (i.e., in AGNs; \citealt{Makrygianni2023}).
Essentially all the BFFs reported to date showed a significant rebrightening during their post-peak decline, and in some cases, there was indeed a secondary, albeit weaker flare, occurring roughly a year after the main UV-optical flare (see, e.g., Fig.~11 in \citealt{Makrygianni2023}). 

Perhaps the most elusive class of (recurring) SMBH-related transients is that of X-ray quasi-periodic eruptions \citep[QPEs; e.g.,][]{Miniutti19,Giustini20,Arcodia21}. These systems exhibit high-magnitude, fast X-ray flares, recurring over timescales of hours, followed by prolonged periods of quiescence, and seem to occur in both previously active and inactive SMBHs. To date, no corresponding phenomena have been observed in the UV-optical regime. 
Various models that try to explain QPEs have been put forward, focusing either on instabilities within an existing accretion flow, or on orbital phenomena. 
The former class of models would require a rather low-luminosity accretion disk, given the lack of evidence for prior AGN-like emission in most QPEs \citep{2021rajnixon,2022metzger}, but could still be linked to the kind of instabilities suggested to explain recurring CL-AGN \citep{sniegowska2023}.
The second class of models focuses on extreme mass ratio in-spirals (EMRIs), involving at least one stellar-mass object (which may be a BH) orbiting the SMBH \citep{2023LuQuataert}. In these models, the quasi-periodicity is closely related to the orbital timescale.

All these types of potentially recurring SMBH-related phenomena may be inter-connected, either physically or even just observationally. For example, there have been reports of TDE candidates occurring in previously-known AGNs (e.g., PS16dtm; \citealt{blanchard2017, petrushevska2023}). Some ``turn on'' CL-AGN were suggested to be driven by a TDE (e.g., 1ES\,1927+654 \citealt{Trakhtenbrot2019_1ES,Ricci2020_1ES};  or \citealt{2015Merloni}). Likewise, it has been claimed that at least some of the transients associated with the BFF class are in fact driven by a TDE (e.g., F01007–2237; \citealt{2017tadhunter, 2021tadhunter}, in addition to the few optically-detected TDEs that exhibit both a broad \HeIIop\ line and BF features \cite[e.g., AT\,2019dyb,][]{Leloudas2019}. 
Moreover, there is evidence that at least some QPEs share some observational characteristics with TDEs \citep{Arcodia24,Wevers24}. Finally, from a purely observational point of view, many of these SMBH-related transients share basic light-curve and/or spectral features (e.g., the aforementioned \HeIIop\ emission in TDEs and BFFs), which makes it hard to tell them apart and throws some doubt on our ability to classify them into distinct types of phenomena. 
Recurring rebrightening events, and indeed recurring flares, can serve as strong discriminants among both the various observational classes and the models that try to explain them.

As the number of time-domain surveys keeps growing, initiated by facilities like Vera Rubin Observatory, \citealt{2019ApJ...873..111I}, BlackGEM,  \citealt{2024PASP..136k5003G}, the La Silla Schmidt Southern Survey \cite[LS4][]{2025ls4} \footnote{https://sites.northwestern.edu/ls4/}, or the upcoming ULTRASAT mission \citep{2024Shvartzvald}, the more important it becomes to be able to identify the various types of SMBH-related transients and to find and understand the properties which they share and which set them apart. This can then pave the way to collecting large samples that, combined with novel models, will allow us to use them to gain insights about SMBH accretion physics and growth.

In this paper, we study the intriguing and recurring brightening events in the nucleus of the galaxy \hostname. The dramatic brightening of this source during 2019, dubbed \objname, was considered to be related to a TDE as early as November 2021,\footnote{See the earliest publically available version of the \cite{vanvelzen2024}}but further observations revealed the appearance of Bowen fluorescence features and a second optical flare (during 2023). 
The recent study by \cite{2024veres} presented several key datasets acquired for this source prior to August 2024, motivated by the coincidence of the first optical flare with a neutrino event (IC\,191119A; \citealt{icecube19}).
The data presented by \cite{2024veres}, which included optical, infrared, and X-ray light-curves as well as several (low cadence) optical spectra and multi-band radio measurements, indicates \objname\ can be classified as a BFF in an AGN, while also showing strong coronal emission lines and a (delayed) brightening of radio emission. They conclude that the events seen in \objname\ are best explained by an enhanced accretion event in an AGN, where the intensified (extreme) UV emission is reprocessed by a pre-existing broad line region and a region of dusty gas, to give rise to the emission lines from highly ionized species, to the (delayed) brightening seen in the IR, and to the enhanced radio emission. 
In Section~\ref{sec:source} we provide a more detailed account of the system hosting this transient, and of the various attempts to monitor and classify it, including observations and analysis reported by \cite{2021TNSTR3680....1V}, \cite{2023TNSAN.194....1V}, and \cite{2023TNSAN.195....1G}. 
Here we present our analysis of the higher cadence optical spectroscopy we obtained after the detection of the second optical flare of \objname, as well as our re-analysis of the most up-to-date X-ray, UV, and X-ray data available for this system.
In Section~\ref{sec:data} we present the archival data, and the new observations used for our analysis, focusing on our own spectroscopic monitoring.
Section~\ref{sec:results} presents the main insights gained from these data, while in Section~\ref{sec:nature} we discuss the possible nature of the flare(s), in the context of other SMBH-related transient phenomena.
We summarize our main findings in Section~\ref{sec:conc}.
Throughout this paper, we assume a cosmological model with  $H_0 = 70\,\Hubble$, $\Omega_\Lambda = 0.7$ and $\Omega_{\rm M}=0.3$.

\section{The source}
\label{sec:source}

\begin{figure}
\centering
\includegraphics[scale=0.4]{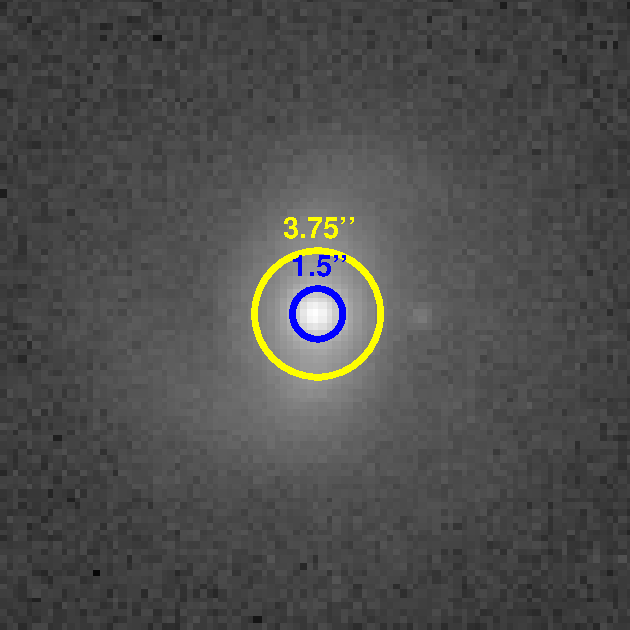}
\caption{An optical image of \hostname, the host galaxy of \objname, based on SDSS $gri$ bands. The image was obtained on 2001 June 16. Circles with radii of 1.5\arcsec and 3.75\arcsec\ represent the apertures used for the archival SDSS (fiber) spectroscopy and for the Swift/UVOT photometry.
}
\label{fig:host_img}
\end{figure}

The extreme variability event \objname, also known as ZTF19aaejtoy, was identified in the center of the $z=0.03557$ Seyfert 1 galaxy \hostname\ (\hname\ hereafter), the optical image of which is shown in Figure \ref{fig:host_img}. 
We present some of the available pre-flare photometry of \hname\ in Table \ref{tab:archphotometry}. 
The source did not show any significant variability, beyond what is expected for an unobscured AGN, prior to a dramatic brightening in the center of that galaxy identified in early 2019, with a formal detection date of 2019 January 22, based on optical photometry obtained with the ZTF (but reported only later, in 2021; \citealt{2021TNSTR3680....1V}). 
The optical flux continued to rise through mid-2019, reaching a clear peak during June 2019, followed by a smooth decline that lasted through 2022. This motivated its early classification as a TDE candidate \citep{vanvelzen2024}.
Continued ZTF photometry then revealed a second significant rebrightening, starting in mid-May 2023 (formal detection date May 11), about four years after the first flare \cite[][]{2023TNSAN.194....1V}. This second flare peaked in early July 2023.
% The second considerable rebrightening of the source was reported in \cite{2023TNSAN.194....1V}, with detection on 2023 May 11 at an $r$-band magnitude of $r=19.3$ AB mag, followed by a fast increase in flux, reaching $r=16.5$ AB mag on 2023 July 2. 

In parallel with the identification of the second flare, \citet{2023TNSAN.195....1G} reexamined the multi-wavelength and variability data available for the source and concluded that it is consistent with an unobscured AGN.
\cite{2023ATel16118....1P} re-analyzed XMM data of the source obtained in 2021, i.e. during optical quiescence, and found the power-law photon index was $\Gamma \approx 2.6$, consistent with later Swift observations (during 2023), and favoring a variable, albeit very soft AGN.

An optical spectrum of the nucleus of \hname\ was obtained back in 2008 April 6, as part of the Sloan Digital Sky Survey (SDSS) legacy spectroscopic observations \citep{2009abazajian}. That spectrum shows clear, strong broad Balmer emission lines, characteristic of unobscured AGNs. Following the first optical flare, a Keck/LRIS \citep{Oke1995} spectrum was obtained on 2021 July 7, i.e. during the long decline from the first flare peak. That spectrum did not show significant changes in emission line shapes and/or (relative) strengths compared to the SDSS one. 

As mentioned in Section~\ref{sec:intro}, \objname\ has been studied by \citealt{2024veres} based on a rich dataset including multi-epoch optical, UV, and IR photometry, optical spectroscopy, (soft) X-ray light-curves and spectral modeling, and multi-band radio observations---all obtained prior to August 2024.
The data and analysis presented by \cite{2024veres} clearly show some of the main features of this source, including: 
two major optical flares, with peaks separated by roughly 4 years;
enhanced and variable mid-IR emission with two peaks that lag the optical flares by several months; 
the presence of strong Bowen fluorescence and coronal emission lines, not present either in the archival SDSS spectrum or in the post-first-flare Keck/LRIS one;
a transient and weak X-ray brightening that (slightly) preceded the peak second optical flare, and a more prominent X-ray flare occurring about 300 days after the second optical peak;
and enhanced and variable radio emission indicative of a newly formed outflow or jet structure.
Importantly, the rich data studied by \cite{2024veres} showed no decisive evidence for the characteristics of typical UV-optically selected TDEs, namely extremely broad ($\gtrsim 10,000\,\kms$) \HeIIop\ line emission and/or a $\sim t^{-5/3}$ declining optical light-curve.
Instead, \objname\ was interpreted as a BFF occurring in a pre-existing AGN, where the accretion-driven flare in the UV (and maybe the X-rays) is driving a dust echo and renewed jet or outflow. While the \cite{2024veres} study was not conclusive about the origin of the accretion-related flare, a standard single TDE interpretation seems highly unlikely.

Interestingly, the first optical flare of \objname\ was claimed to be associated with the high-energy neutrino event IC\,191119A \citep{icecube19}, detected $\sim$150 days after the peak of the first optical peak. 
Indeed, several works have investigated the possible association of accretion-driven flares from SMBHs with neutrino events \cite{Reusch2023, Winter23,vanvelzen2024}. Such associations would imply that flaring SMBH accretion is a significantly more efficient source of high-energy neutrinos compared with more luminous, but persistent, AGNs. This is particularly the case for \objname, given its high implied neutrino fluence \citep{Winter23}. 

Following the 2023 rebrightening of the source, we have obtained an optical spectrum with the Las Cumbres Observatory robotic network of telescopes \citep{2013PASP..125.1031B} on 2023 July 12. Similarly to the spectra presented by \cite{2024veres}, this spectrum showed Bowen fluorescence lines, and other emission lines associated with highly ionized species, not seen prior to the second optical flare. 
% This, in addition to the double-flare nature of the transient, cast doubt on the TDE classification of the transient.

The active discussion of the nature of this source, and its rather exceptional observed properties among SMBH-related flares, motivated us to monitor the source more intensely.  This intensive monitoring effort, mainly with relatively high-cadence optical spectroscopy and UV-optical imaging, is the main focus of the present paper. While some of the data we present throughout the rest of the present paper is similar (or indeed identical) to the data used in the \cite{2024veres} study, we stress that (1) our analysis is completely independent, (2) our higher-cadence optical spectroscopy provides new and complementary insight into the nature of the system, and (3) we present more up-to-date data in terms of optical, UV, and IR photometry, and X-ray observations. On the other hand, our analysis leaves aside some spectral regimes and does not revisit all the data presented in the aforementioned studies.

\section{Observations, Data, and Analysis}
\label{sec:data}

After the detection of the second optical flare in the nucleus of \hname, we initiated a campaign of optical photometric and spectroscopic follow-up observations using the Las Cumbres Observatory network of telescopes, starting 2023 July 28, as well as lower-cadence space-based, optical-UV and X-ray monitoring with Swift.  
We also collected publicly available photometric measurements and spectra from various optical, IR, and UV surveys.
Below we describe all the observations and data used for our analysis, as well as our basic analysis of these data and the key features seen in each dataset.

\subsection{Optical photometry}
\label{sec:data_opt_phot}

\begin{figure*}
\centering
\includegraphics[scale=0.5]{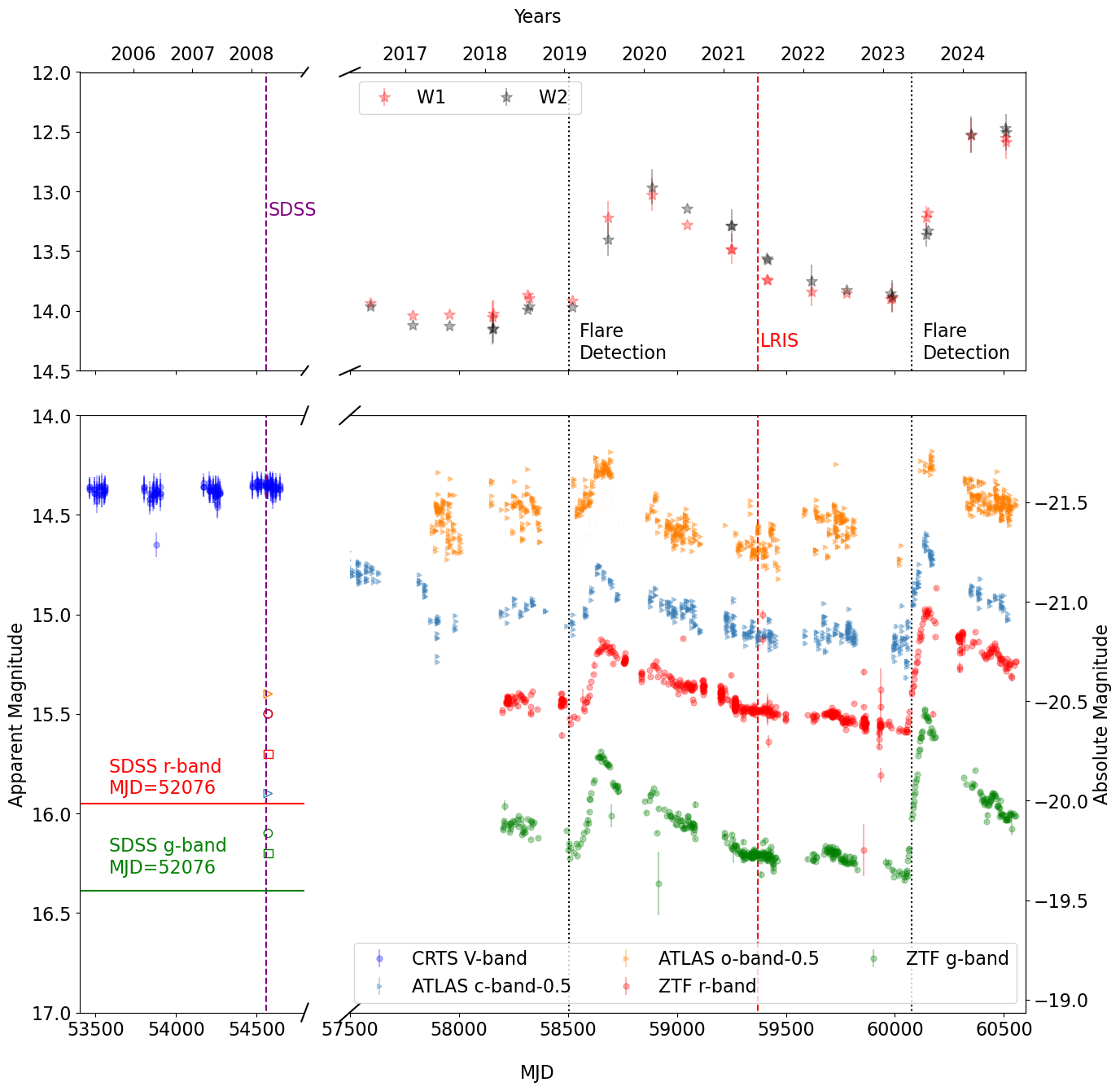}
\caption{Long-term light-curves of \objname. \textit{Top:} MIR NEOWISE-R photometry in the W1 and W2 bands (converted to AB magnitudes). \textit{Bottom:} optical light-curves based on photometric measurements obtained from the CRTS ($V$-band), ZTF ($g$ and $r$ bands), and ATLAS ($c$ and $o$ bands) surveys. Red and green horizontal lines mark the brightness corresponding to the archival optical SDSS (PSF-fit) photometry. The vertical dotted black lines indicate the dates of the two main optical flare detections (2019 January 22 and 2023 May 11). The dashed vertical purple line indicates the date of obtaining the archival SDSS spectrum (2008 April 5) and the vertical dashed red line represents the Keck spectrum (2021 June 7).  
Empty markers on top of the dashed vertical purple line indicate the synthetic photometry calculated from the archival SDSS spectrum, with empty squares, circles, and triangles marking SDSS, ZTF, and ATLAS filter pass-bands (respectively).
The two main flares, with the second being brighter than the first, are clearly seen in all optical and MIR bands, with the latter seemingly lagging behind the former.}
\label{fig:lc_long}
\end{figure*}

To assess the past variability of \hname, we collected photometric optical measurements in the $V$-band available from the CRTS and spanning MJD = 53464--56449 (2005 April 4--2013 June 6). These measurements, presented in the bottom panel of  Figure~\ref{fig:lc_long}, show no signs of significant variability in the years preceding the two major flares of 2019 and 2023. The median $V$-band magnitude over that period was $V=14.37$ mag (in the Vega system).

We next retrieved the Sloan Digital Sky Survey \citep{2000york} photometric measurements for \hname, based on imaging observations conducted on 2001 June 16. Specifically, we use the PSF magnitudes in the $g$ and $r$ bands, which are comparable with the ZTF photometry that covers the two flares (i.e., the brightening events known as \objname). The brightness measured in June 2001 was $g=16.40\pm0.01$ and $r=15.95\pm0.02$ AB mags.  
%15.7 for r-band and 16.2 for g-band.
% opis mag https://alerce.science/alerce-pipeline/
We also computed synthetic photometry from the archival SDSS spectrum, obtained on 2008 April 6, using the \texttt{stsynphot} procedure \citep{stsynphot}, which yielded $g_{\rm syn}=16.20\pm0.35$ and $r_{\rm syn}=15.70\pm0.28$ AB mag.
With a distance modulus of 35.97, these latter apparent magnitudes translated to absolute magnitudes of $M_g = -19.7$ and $M_r = -20.2$.

We rely on publicly available measurements from the ZTF as the main source of data for the optical light-curve of \objname\ during its recurring brightening activity.
We specifically used the PSF-fitted photometry from the ZTF forced photometry service \citep{2019PASP..131a8003M} spanning MJD = 58198--60556 (2018 March 21 to 2024 September 9). 
To be able to assess the luminosities associated with the main flares and brightening episodes of \objname, we used the first ZTF detection (MJD = 58203) as a reference and subtracted the fluxes of that epoch from the rest of the corresponding ZTF light-curves, in both the $g$ and $r$ bands.
To verify the key features of the optical light-curve during the two flares, we also used measurements from the Asteroid Terrestrial-impact Last Alert System \cite[ATLAS]{2018torny} forced photometry service, obtaining measurements for MJD = 57229--60574 (2015 July 26 to 2024 September 1) in the $c$ and $o$ bands.
All photometric measurements were corrected for foreground Galactic dust attenuation, using the standard \cite{Cardelli1989} extinction law by assuming $R_V=3.1$ and $E(B-V)=0.042$, based on the dust maps of \cite{Schlafly2011}. 

The four main, dense optical light-curves (i.e., ZTF and ATLAS, two bands each) are shown in the bottom panel of Fig.~\ref{fig:lc_long}, and exhibit two main flaring events, with the first peaking MJD = 58652, and the second peaking MJD = 60138.
Moreover, after each of these two main flares, the light-curves exhibit additional, more modest rebrightening periods, which we refer to as ``bumps'' hereafter. The first of these bumps followed the first major flare, with a $g$-band peak near MJD = 58891; and two additional bumps followed the second major flare, with $g$-band peaks occurring at MJD = 60369 and 60450.

Moreover, we obtained multi-epoch $g-$, $r-$, and $i-$band images of \objname\ using the Sinistro cameras installed on some of the Las Cumbres 1.0m telescopes, with a typical cadence of 12 days. We extracted PSF photometry in these bands using the \texttt{lcogtsnpipe} pipeline\footnote{\url{https://github.com/LCOGT/lcogtsnpipe}} \citep{Valenti2016}. The zeropoints were determined using field stars for which magnitudes were available from the AAVSO Photometry All-Sky Survey \citep{2009AAS...21440702H}. These measurements were meant to complement the ZTF monitoring or replace it, in case the latter would not provide sufficiently well-sampled photometry for \objname\ during the second optical flare (and thereafter). However, since the ZTF did eventually monitor the source with a cadence and reliability that surpasses our Las Cumbres photometric data, in what follows we focus on the ZTF data for our analysis of the optical light-curves.

\begin{deluxetable}{lllcch}
\label{tab:archphotometry}
\tablecaption{Pre-flare photometry of \hname}
\tablewidth{\textwidth}
\tablehead{
\colhead{MJD} & \colhead{filter} & \colhead{magnitude} & \colhead{error} & \colhead{source} & \nocolhead{system}
}
\startdata
%54247 & FUV    & 18.24     & 0.07  & GALEX  & AB     \\
%54247 & NUV    & 17.95     & 0.04  & GALEX  & AB     \\
54247 & FUV    & 18.52     & 0.09  & GALEX  & AB     \\
54247 & NUV    & 18.44     & 0.06  & GALEX  & AB     \\
52076 & u      &  16.93    & 0.02  & SDSS   & AB     \\
52076 & g      &  16.40    & 0.01  & SDSS   & AB     \\
52076 & r      &  15.95    & 0.02  & SDSS   & AB     \\
52076 & i      &  15.74    & 0.01  & SDSS   & AB     \\
52076 & z      &  15.55    & 0.01  & SDSS   & AB   \\ 
%52076 & u      & 16.27     & 0.01  & SDSS   & AB     \\
%52076 & g      & 15.14     & 0.01  & SDSS   & AB     \\
%52076 & r      & 14.46     & 0.01  & SDSS   & AB     \\
%52076 & i      & 14.12     & 0.01  & SDSS   & AB     \\
%52076 & z      & 13.84     & 0.01  & SDSS   & AB   \\ 
\hline
 %&       &      &  & \textcolor{red}{WISE?}   &     \\
\enddata
\tablecomments{All magnitudes are in the AB system. For SDSS-based photometry, we report PSF magnitudes.}
\end{deluxetable}

\begin{figure*}
\centering
\includegraphics[scale=0.5]{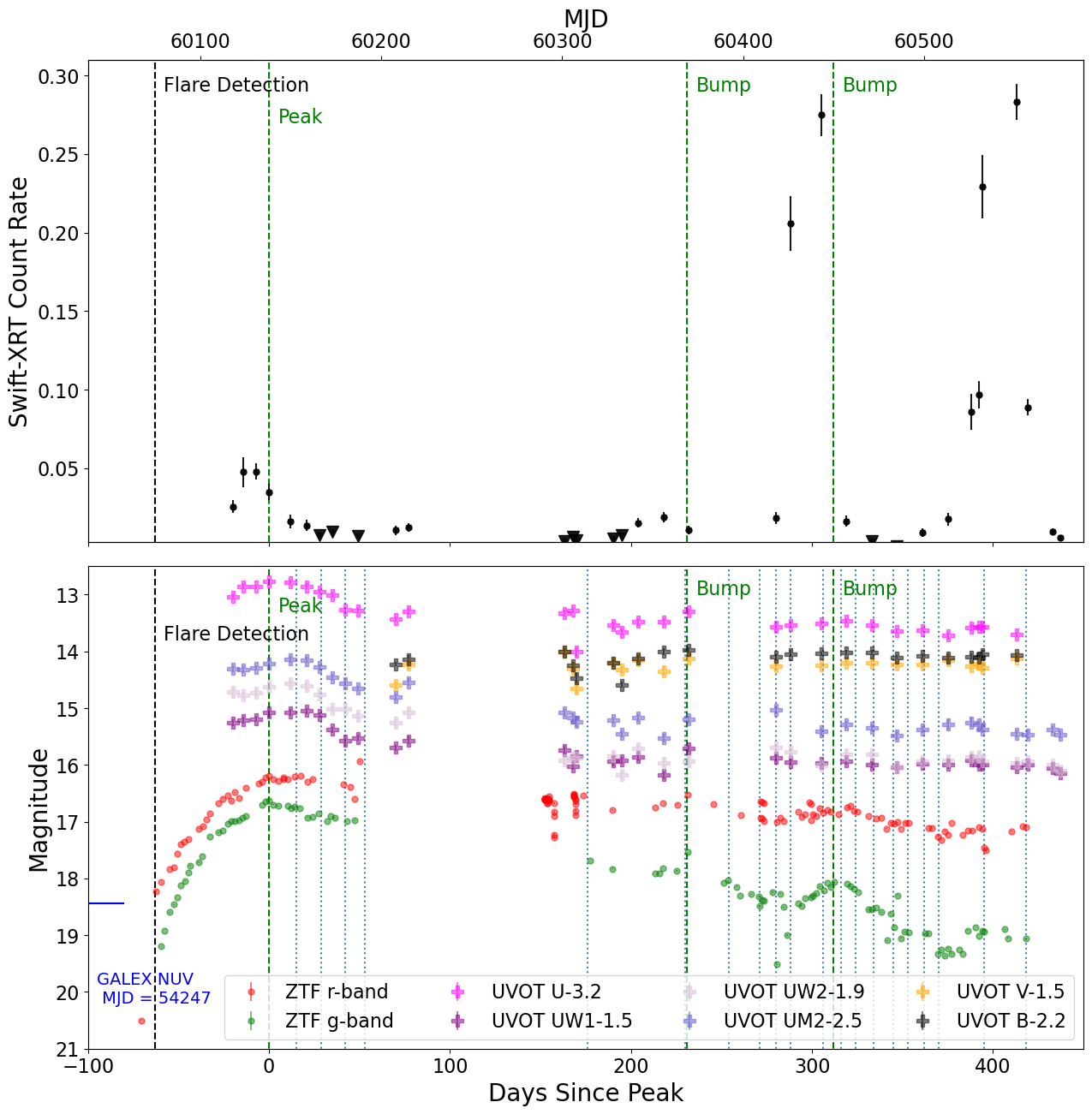}
\caption{The multi-wavelength, broad-band light-curves of \objname\ during the second main optical flare, including optical and NUV measurements (ZTF and Swift/UVOT; bottom panel) and X-ray measurements (Swift/XRT; top panel). 
The peak in optical emission, at MJD = 60138, is defined as $t=0$. 
We mark observations with a $>3\sigma$ detection with points, while triangles mark $3\sigma$ upper limits.
The blue horizontal line near the left edge of the bottom panel indicates the archival GALEX NUV photometry.
The black vertical dotted line indicates the date of the second optical flare detection (2023 May 11). 
The vertical dashed blue lines mark the dates of obtaining Las Cumbres spectroscopy.
The two subtle rebrightening periods (``bumps'') are clearest in the $g-$band light-curve, with peaks near MJD=60369\ and 60450.}
\label{fig:lc_long_zoom}
\end{figure*}

%\begin{figure}
 %   \centering
  %  \includegraphics[scale=0.4]{plots/xray/countrate-plot-16-01-25.png}
   % \caption{X-ray light-curve of \objname, based on Swift/XRT monitoring during the second optical flare. We mark observations with a $>3\sigma$ detection with points, while upper limits are marked as triangles.  }
    %\label{fig:xrt-swift}
%\end{figure}

\begin{figure*}
\centering
\includegraphics[scale=0.9]{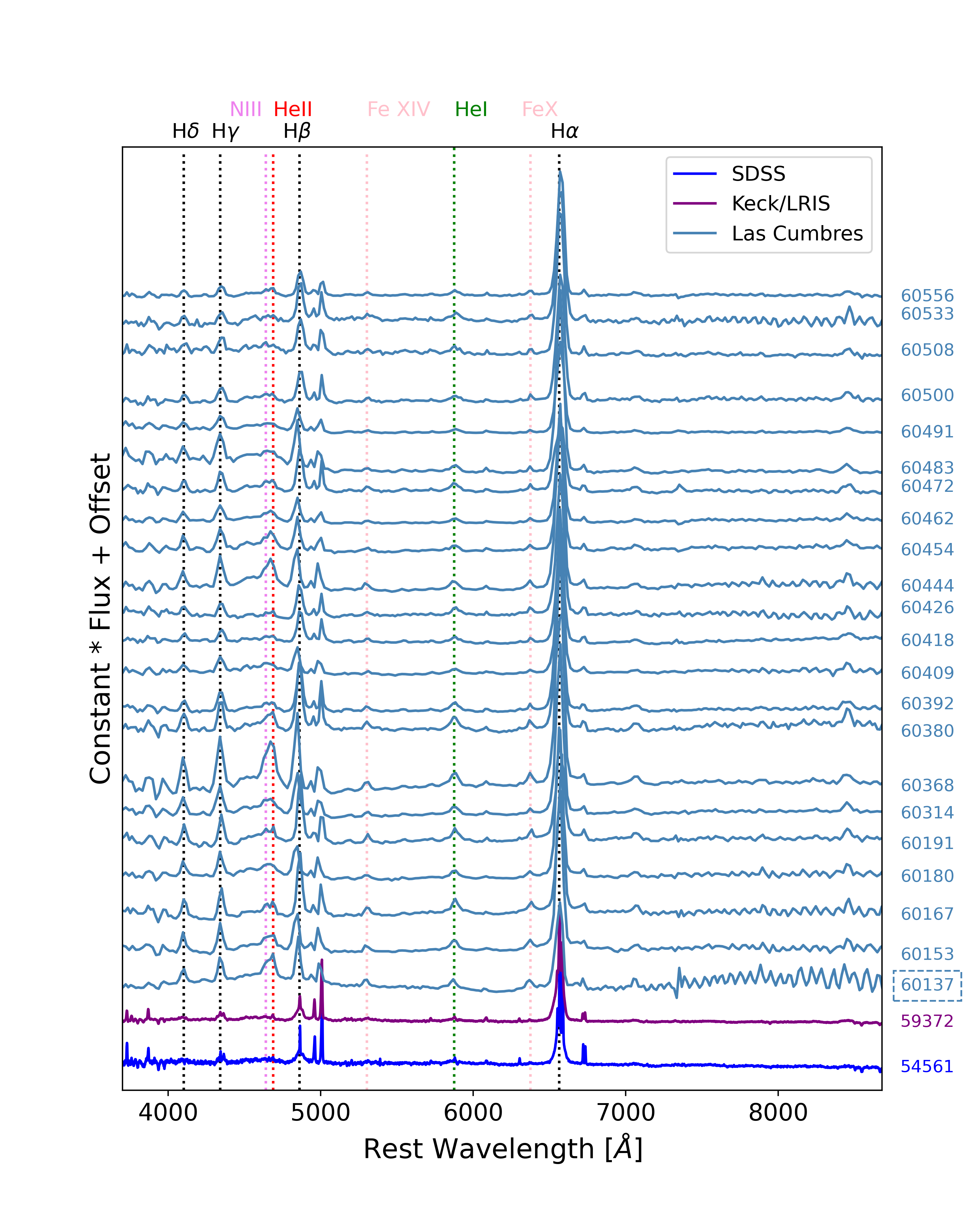}
\caption{The optical spectral sequence of \objname, including the archival (pre-flare) SDSS spectrum (dark blue), the Keck spectrum obtained between the two main optical flares (purple), and our intensive Las Cumbres Observatory spectral monitoring (light blue). Dotted lines indicate the wavelengths of the labeled emission lines. 
All spectra are plotted in the rest frame and are continuum subtracted, normalized, and shifted (for clarity).
The spectrum taken close to the optical peak is highlighted with a dashed box around its MJD.
The enhanced, variable emission from the \HeIIop\ and \NIII\ lines, related to Bowen fluorescence, and the coronal \FeX\ line, are clearly seen in the Las Cumbres spectra.}
\label{fig:all-spectra}
\end{figure*}

\subsection{Additional multi-wavelength observations}

\subsubsection{Swift UVOT}

\objname\ was monitored in the near-UV (NUV) with the Neil Gehrels Swift Observatory \citep{Gehrels2004}, through several programs (PIs: Reusch, Veres, {\'S}niegowska; Target ID 16080) starting on 2023 June 23. In total, 36 observations were obtained.
We reduced the Swift/UVOT photometry using the HEASARC pipeline and the standard analysis task \texttt{uvotsource} (v4.4) with a circular aperture of 3.75\arcsec\ in radius, for all the UV bands of UVOT ($uvm1$, $uvw2$, $uvm2$). This aperture was chosen to match that used for the archival GALEX observation, from MJD = 54247 (2007 May 27; \citealt{2005martin}). For the $U$, $B$, and $V$ bands of UVOT, we used both that aperture for consistency with the UV bands and also a circular aperture with 1.5\arcsec\ radius, to be able to compare with the archival SDSS fiber photometry.
The UVOT photometry was also corrected for foreground dust attenuation, as described above.

In Figure \ref{fig:lc_long_zoom} we present all the optical and UV light-curves for the second flaring event in \objname. The difference between the $g$ and $r$ bands increases as the optical light-curve of \objname\ decays. Moreover, the Swift UV observations exhibit a plateau, with no clear evidence for a decline in the light-curve.

\subsubsection{Swift XRT}
The Swift observations also utilized the X-ray Telescope \citep[XRT;][]{Burrows2005} to collect X-ray data, using the photon-counting mode. 
We processed the XRT data using \texttt{xrtpipeline} (version 0.13.7), following standard procedures.
We used a circular aperture with a radius of 47\arcsec, after verifying there are no neighboring X-ray sources that may contaminate our measurements.  
We also used the XRT data to construct stacked X-ray spectral energy distributions, which we analyze and discuss in Section \ref{sec:x-ray-evolution}. 

We present the X-ray light-curve, in terms of count-rate, in the top panel of Figure \ref{fig:lc_long_zoom}.
The X-ray light-curve of \objname\ shows an early episode of brightening, peaking at $\approx$0.05 cts/s near MJD=60131.
While this brightening generally coincides with the main (second) optical flare, the X-ray emission seems to peak $\approx$7 days prior to the optical one (i.e., MJD=60138 for the optical, $g-$band peak).
Interestingly, we can also clearly see two stronger and faster-evolving X-ray flares, one near MJD = 60443 and another near MJD = 60551, both reaching count rates of $\approx$0.28 cts/s. 
%These X-ray flares did not temporarily coincide with the bumps seen in the optical regime: 
%the first X-ray flare peaked $\approx70$ days after the first bump ($g-$band peaks at MJD = 60369) and the second X-ray flare peaked $\approx$100 days after the second optical bump (MJD = 60450). 
The two X-ray flares did not coincide with the optical bumps. The first X-ray flare peaked approximately 70 days after the first optical bump (which occurred at MJD = 60369, in the $g$-band). Similarly, the second X-ray flare peaked about 100 days after the second optical bump (which occurred at MJD = 60450).
The interpretation of these differences in peak times is complicated by the gaps in our data, as additional optical bumps and X-ray flares might have been missed.

\subsubsection{WISE mid-IR}
We collected archival photometry in the mid-infrared (MIR) regime, obtained with the WISE space observatory as part of the NEOWISE \citep{Mainzer2011} and NEOWISE-R \citep{Mainzer2014} missions. We used the publicly available data in the $W1$ and $W2$ bands (3.4 and 4.6 \mic, respectively), spanning 2014 February 11 to 2022 July 19 (MJD= 56699--59779). The WISE observatory observed the source roughly every 6 months, usually with more than one exposure available per each visit.
We therefore base our analysis on the series of mean fluxes of measurements taken within one day, for each of the two bands.

The WISE-based light-curve is shown in the top panel of Figure~\ref{fig:lc_long}, and clearly presents two major flares that seem to lag behind the main optical ones. We return to these lags in Section~\ref{sec:photometry-all-wv}.

\subsubsection{Radio}

The position of the source was observed at 1.4 GHz as part of the FIRST survey \cite{1995FIRST}. \hname\ is clearly detected with an integrated flux density of $S_{\rm p} = 6.20\pm0.14$ mJy \citep{First_catalog}, which translates to a monochromatic radio luminosity of $\nu L_\nu({\rm 1.4\,GHz}) {\simeq} 2.6 \times 10^{38}\,\ergs$.
To assess the (relative) importance of the radio emission and better understand the source, we calculated the radio loudness parameter following \cite{1989AJ.....98.1195K}, obtaining $R\equiv f_\nu(1.4\,{\rm GHz})/f_\nu(g-{\rm band})=2.4$. 
Thus, historically \hname\ was not a radio-loud AGN. 
Moreover, the radio luminosity could be accounted for with relatively intense star formation in the host galaxy, at a rate of $\approx10\,\mpyr$ (using Eq.~1 in \citealt{2003ApJ...599..971H}).

Regardless of whether the (archival) radio emission is driven by the AGN or by star formation, \citet{2024veres} showed that the $\sim$1.4 GHz radio emission remains rather constant during the second main optical flare, while there is significant brightening in the 18 GHz regime. 
The \cite{2024veres} study presents an elaborate analysis and interpretation of the radio data available for \objname, including VLBI observations. The present work, on the other hand, presents no additional radio data, and we do not go any further in the interpretation of the radio nature of this source.

\subsection{Optical spectroscopy}

\begin{figure*}
\centering
\includegraphics[scale=0.5]{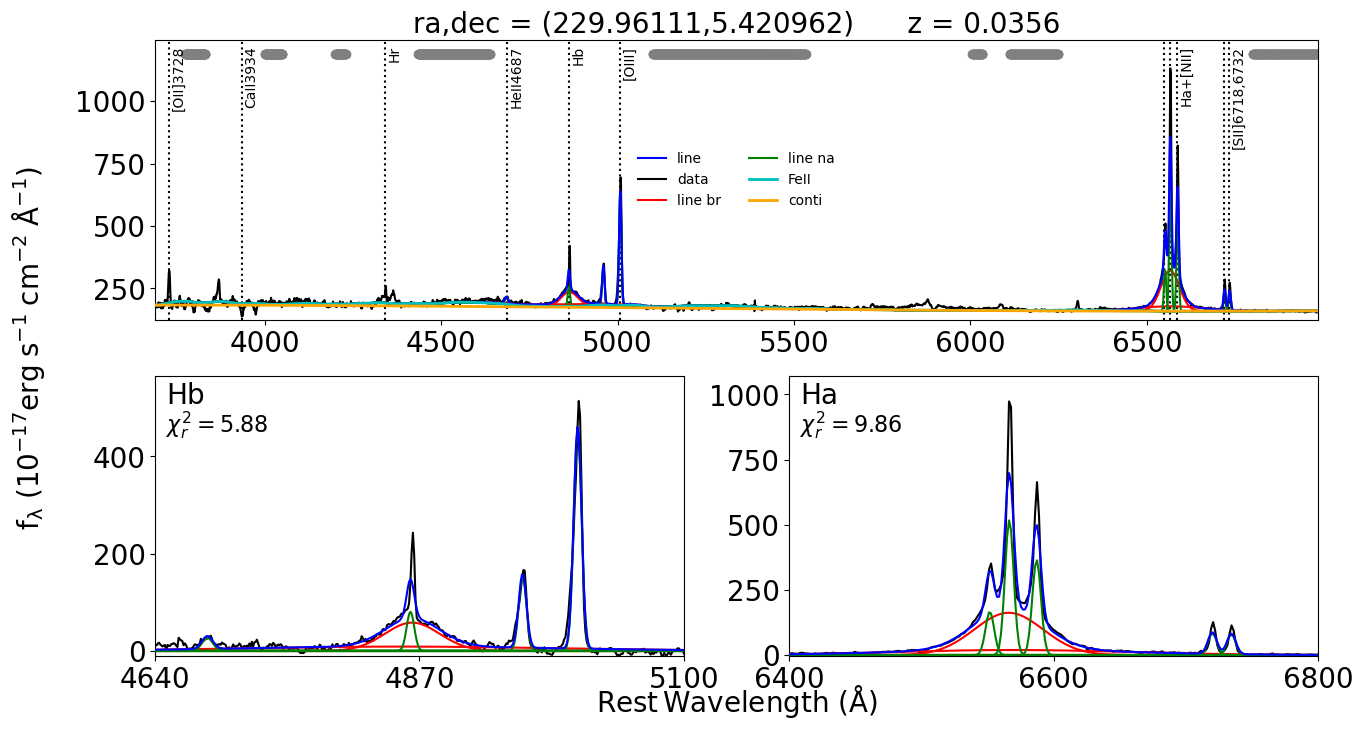}
\caption{Spectral decomposition of the archival pre-flare SDSS spectrum of the AGN hosting \objname, \hname, performed with {\sc PyQSOFit}. 
The observed spectrum is shown in black, and the best-fitting model is shown in blue.
\textit{Top:}
The overall spectrum and best-fit model. The model components include the continuum emission (orange), blended iron pseudo-continuum (cyan), and broad \& narrow components of various emission lines (red \& green, respectively). The thick gray horizontal lines near the top mark the spectral windows used for the continuum estimation.
The black dotted vertical lines represent the rest-frame positions of key emission lines (see labels).  \textit{Bottom:} zoom-in views of the \hb\ and \ha\ spectral regions.
This archival, pre-flare spectrum clearly shows that \objname\ occurred in an already-active galactic nucleus.
}
\label{fig:sdss_spec_model}
\end{figure*}

All the optical spectra used in our analysis are shown in Figure \ref{fig:all-spectra}.
The archival optical spectrum of \hname\ was obtained as part of the legacy SDSS, on 2008 April 6 (MJD =  54562), and covering the wavelength range between 3800-9200\,\AA. 
This SDSS spectrum, shown in Figure \ref{fig:sdss_spec_model}, clearly exhibits spectral features that are characteristic of unobscured AGNs, specifically broad Balmer emission lines.
An additional high-quality optical spectrum of the nucleus of \hname\ was obtained using Keck/LRIS \citep{Oke1995}, on 2021 July 7 (MJD=59372), which is about two years after the first optical peak associated with \objname. The Keck/LRIS spectrum covers 3100-10300\,\AA.

% \subsubsection{Las Cumbres spectroscopic monitoring}
% \label{sec:LCO_spec}

After the second major optical flare event, we initiated a spectroscopic monitoring campaign using the FLOYDS spectrographs mounted on the twin 2 m telescopes of the Las Cumbres Observatory, located at the Haleakal\=a and Siding Spring observatories (Hawaii and Australia, respectively;  \citep{Sand2011}).
In total, we obtained 23 spectra during 14 months. The spectra were obtained through a 2\arcsec\ slit, with exposure times ranging between 1200 and 2700 s, and covering the wavelength range between 3300-10000\,\AA with a modest spectral resolution of $R\approx400$.\footnote{For the spectra taken on MJD = 60380 and 60418 we only used the wavelength range between 3300-9100\,\AA.}
These data were reduced using the \texttt{floydsspec} pipeline \citep{2014Valenti}.\footnote{\url{https://github.com/svalenti/FLOYDS_pipeline/} 
The pipeline allows for the removal of cosmic rays and flux and wavelength calibration of the spectra, although we note that for our analysis we further re-scaled key spectral flux measurements to avoid flux calibration issues (see below).}

\subsubsection{Spectral decomposition}

We used the \texttt{PyQSOFit} package \citep{pyqsofit} to decompose all the optical spectra in our study. 
In this procedure, the spectra are first corrected for foreground Galactic extinction and shifted to the rest-frame of the source.
To improve the continuum fitting near key emission lines, we separately modeled two spectral regions: 4000-5500\,\AA~ for the spectral region surrounding the \hb\ emission line, and 6000-7000\,\AA\ for the spectral regime surrounding \ha. 
In both regions, we model the accretion-powered continuum emission assuming a power-law model that is fit to several bands that are absent of strong emission lines, specifically narrow ($10-20$\,\AA\ wide) bands around 4000, 4200, 4430, and 5100\,\AA\ (for the \hb\ region), and around 6000, 6200 and 6800\,\AA\ (for the \ha\ region).
For modeling the blended iron emission features, we used the empirical \feii\ template \citep{1992borosongreen} available by default in \texttt{PyQSOFit}, but limited it to wavelengths shorter than 5050\,\AA.

All the emission lines, including both narrow and broad components (NCs and BCs, respectively) were modeled using Gaussian profiles. 
The narrow lines include the \oiii$\lambda\lambda$4959,5007, \nii\,$\lambda\lambda$6548,6583 and \sii$\lambda\lambda$6718,6732 doublets, as well as the narrow components of \hb\ and \ha, all of which were modeled with a single (narrow) Gaussian profile. 
The narrow emission components in each spectral region were tied in width and shift (both in velocity space).
The broad components of the \hb\ and \ha\ lines were modeled with two broad  Gaussian profiles (each), while for the \Hgamma\ and \Hdelta\ we used one broad Gaussian profile (each).
Additionally, we used two Gaussians to model the region between 4550 and 4800\,\AA, which includes the blended \NIII\ and \HeIIop\ lines, as well as a single Gaussian to model the coronal \FeX\ line.

To properly fit the \HeIIop, \NIII, and certain coronal emission lines such as \FeX\ and \FeXIV\ ($\heii+\niii$, \fex, and \fexiv\ hereafter) that are present in the spectra we consider here, we had to make some adjustments to our otherwise standard spectral decomposition procedure. 
First, we had to limit the \feii\ template used within \texttt{PyQSOFit} to $\lambda<4800$\,\AA\, as there is very little iron emission present redwards of \hb\ in this source. 
Next, we adopted two approaches to be able to measure these lines.
In the first method, we modeled each of the lines with single Gaussians, by adding these profiles to the \texttt{PyQSOFit} configuration. 
In the second method, we directly integrate the flux density of the continuum- and \feii-subtracted spectra, over the spectral regions that contain the emission lines under study, specifically $\lambda=4540 - 4800$\,\AA\ for $\heii+\niii$, $4800 - 4900$\,\AA\ for \hb, $6330 - 6420$\,\AA\ for \fex, and $5250 - 5350$\,\AA\ for \fexiv. 
We do not use this method for \ha, as the blended \nii\ lines would contaminate any such measurement. 
The flux measurements from both methods are highly consistent with each other. 

After setting up the fitting procedure(s), we first fit the archival SDSS spectrum, assuming that the ratio between the \oiii$\lambda\lambda\,4959,5007$ is 1:3 \citep{2007MNRAS.374.1181D}. 
We used the best-fit narrow line model of this SDSS spectrum as a reference for all other spectra, alleviating some of the potential challenges of flux calibration and limited spectral resolution. 
First, we use the flux of the \OIII\ line from this fit as a reference, so that we could normalize all the other spectra, thus essentially assuming that the \OIII\ line flux remains constant. 
This assumption is justified even for AGN-driven narrow emission lines given the long light travel time to the line-emitting region and the long recombination and/or de-excitation timescales expected for such low-density gas.
% (we keep the ratio between the \oiii\ flux derived from the SDSS spectrum and those derived from other spectra as one), because the resolving power of the SDSS we estimated as 1200, for Las Cumbres 300,
% %i used oiii
% and moreover, we do not expect [OIII]{$\lambda\lambda$4959,5007}  to vary within timescales of our observational campaign. 
Second, we use the ratio between the \OIII\ and narrow \hb\ line fluxes derived from the best-fit model of the SDSS spectrum, $\oiii/\hb_{\rm NC}= 5.35$, and fix it to this value when modeling all the other optical spectra.
Otherwise, the Keck/LRIS and Las Cumbres spectra were decomposed with the same choice of parameters as those used for decomposing the SDSS spectrum.
% In this spectrum (see Figure \ref{fig:sdss_spec_model}), we see Balmer emission lines with broad and narrow components show that before two flares, the source was unobscured AGN.

Given that the synthetic photometry obtained from the SDSS spectrum is consistent with the (asynchronous) SDSS photometry, and that there is no sign of drastic photometric variability over the period that covers the SDSS observations, hereafter we assume that the 2008 SDSS spectrum of the active nucleus of \hname\ represents a relatively stable period of accretion and nuclear activity, and can be used to derive a set of baseline, pre-flare(s) spectral measurements, as well as deduced key properties of the AGN and of the SMBH powering it, including BH mass.
% The Keck/LRIS and Las Cumbres spectra were then fit by fixing \oiii\ flux and \hb\ NC ratio to 5.35, and the rest of the fitting procedure stayed as for the SDSS spectrum.

\section{Results and Discussion}
\label{sec:results}

\subsection{Photometric evolution}
\label{sec:photometry-all-wv}

\subsubsection{UV-optical light-curves}

We first discuss the broad-band photometric measurements of \objname\ taken during our campaign(s) and compare them with relevant archival (i.e., pre-flare) data.

We compare the UV measurements derived from the Swift/UVOT observations to those obtained by GALEX in May 2007. The uvm2 band of UVOT is the most similar to the archival GALEX NUV band.\footnote{The effective wavelengths of the GALEX NUV and the UVOT uvm2 bands are 2245 and 2304\,\AA, respectively.} 
The first UVOT observation in our campaign (MJD = 60118) yielded an apparent AB magnitude of $m_{uvm2}= 16.16\pm0.06$ mag, corresponding to $\nu L_\nu (uvm2) = 5.5\times 10^{43}\,\ergs$.
This is $\approx$10 times brighter than the archival GALEX measurement of $m_{\rm NUV}$ = $17.95\pm0.04$ mag, corresponding to $\nu L_\nu ({\rm NUV}) = 6.5\times 10^{42}\,\ergs$.
During the Swift monitoring campaign, the $uvm2$ flux decreased by a factor of $\approx2.75$ (1.1 mag), over the period 2023 June 23 through 2024 September 18.

The apparent brightness of the source in the optical regime increased from the first ZTF detection, when it had $r = 15.44 \pm 0.10$ and $g = 16.07 \pm 0.16$ mag, to $r = 15.12 \pm 0.02$ and $g = 15.69 \pm 0.02$ mag during the peak of the first flare, and later to yet higher fluxes during the peak of the second flare, reaching $r = 14.97 \pm 0.02$ and $g = 15.48 \pm 0.01$ mag. 
After subtracting the reference optical flux levels (i.e., those measured in the first ZTF visit), we obtain $\nu L_\nu (g) = 1.2\times 10^{43}\,\ergs$ for the peak of the first flare, and  $\nu L_\nu (g) = 2.0\times 10^{43}\, \ergs$ for the peak of the second flare, which is brighter than the first flare by 67\%. Therefore, while the rise during the first flare triggered the initial interest in this extremely variable source, the second flare appears to be even more significant, in terms of its radiative output.
  
The two main optical flares were followed by more subtle rebrightening episodes appearing during the long decline of the light curve, roughly 230 days after each of the main flares. In terms of observed, non reference-subtracted brightness, these secondary bumps reached $r = 15.26 \pm 0.01$ and $g = 15.87 \pm 0.01$ mag for the bump that followed the first main flare, and  $r = 15.09 \pm 0.01$ and $g = 15.78 \pm 0.01$ mag for the second one. 
Moreover, after mid-May 2024 ($\approx$300 days after the second major optical flare), the source experienced another, more minor rebrightening, reaching $r = 15.15 \pm 0.01$ and $g = 15.90 \pm 0.01$ mag.

To quantify the declining phases of the optical light-curves, and compare them with what is expected for various physical scenarios, we fit the flux light curves with a declining power-law model of the form 
\begin{equation}
    F\propto \left(\frac{t-t_0}{t_0}\right)^{-\alpha} \,\, ,
    \label{eq:lc_pw_mod}
\end{equation} 
where $\alpha$ denotes the index of the declining power law ($\alpha>0$) and $t_0$ denotes a reference time, which in principle should be relevant for the observed declining light-curves.
For these fits, we used the reference-subtracted ZTF fluxes in both the $g$ and $r$ bands, after omitting measurements with large errors ($\Delta m > 0.05$ mag in the original ZTF data) to create ``clean'', reference-subtracted light-curves associated with each of the two main optical flares. These four light-curves (two flares, two bands each) were fit using the {\tt Lmfit} package \citep{2016lmfit}. 
%\textcolor{green}{here i had details}
Numerically, the fitting procedure may introduce a degeneracy between $\alpha$ and $t_0$, as it would prefer to fit the data with brighter and earlier flare peaks (small $t_0$), followed by a steeper decline (larger $\alpha$), so that the observed slowly declining light-curves would correspond to late stages in the photometric evolution. 
However, we consider it physically unjustified to allow the disruption times, encoded by $t_0$, to occur months or years before the observed optical flares.
Therefore, when fitting the optical light-curves we constrained $t_0$ so that it does \emph{not} precede the detection of each of the flares (see details in Appendix~\ref{app:lc_fits}).

%\textcolor{red}{The photometric decline that followed both main flares rather obviously appears to be shallower than what is expected for Tidal Disruption Events (TDEs), with a power-law index
%\citep[$\alpha=-5/3$;][]{Rees1988}.
%The second flare is a bit \textbf{shallower} in comparison to the first one but still far from what we expect for TDEs.}

%
The best-fitting power-law models and the data used for these fits are shown in Appendix~\ref{app:lc_fits} (Figure~\ref{fig:lc_pl_fits} and Table \ref{tab:lc_pl_fits}).  
The power-law indices we find are in the range $\alpha\approx\simeq0.2-1$, with the fits to the first flare showing a somewhat faster decline (larger $\alpha$) than those of the second flare, and the fits where $t_0$ is allowed to occur prior to the flare detections yielding larger $\alpha$ than those where $t_0$ is fixed to the flare detection times (as expected; see above).
% \textcolor{red}{The second flare is a bit \textbf{shallower} (in r-band $\alpha$ = 0.68) in comparison to the first one (in r-band $\alpha$ = 0.85) but still far from what we expect for TDEs.}
%
In general, the power-law slopes we derive are significantly shallower than what is expected from the basic theory of TDEs \cite[$\alpha=5/3$;][]{Rees1988}, and are consistent only with the shallowest slopes found among observed samples of UV-optically selected TDE candidates (for which $\alpha\sim[-4]-[-0.7]$; \citealt{2020vanvelzen,vanvelzen2021}). 
Moreover, the slopes derived for \objname\ are inconsistent with what is expected for a partial TDE \citep[pTDE, $\alpha=-9/4$;][]{Coughlin2019}---an astrophysical system that may, in principle, give rise to recurring flares in galaxy nuclei. 
Most importantly, single power-laws do not provide good fits to the observed light-curve(s) of \objname. This is true for the smoothly-declining parts of the light-curves, and is further complicated by the numerous shorter-term bumps occurring during the prolonged declining phases. 
The slow decline which cannot be well represented by a power-law, and the presence of bumps are, in fact, consistent with what is seen in BFFs \cite[e.g.,][]{Makrygianni2023}.

\subsubsection{Optical color evolution}

\begin{figure*}
\centering
\includegraphics[scale=0.5]{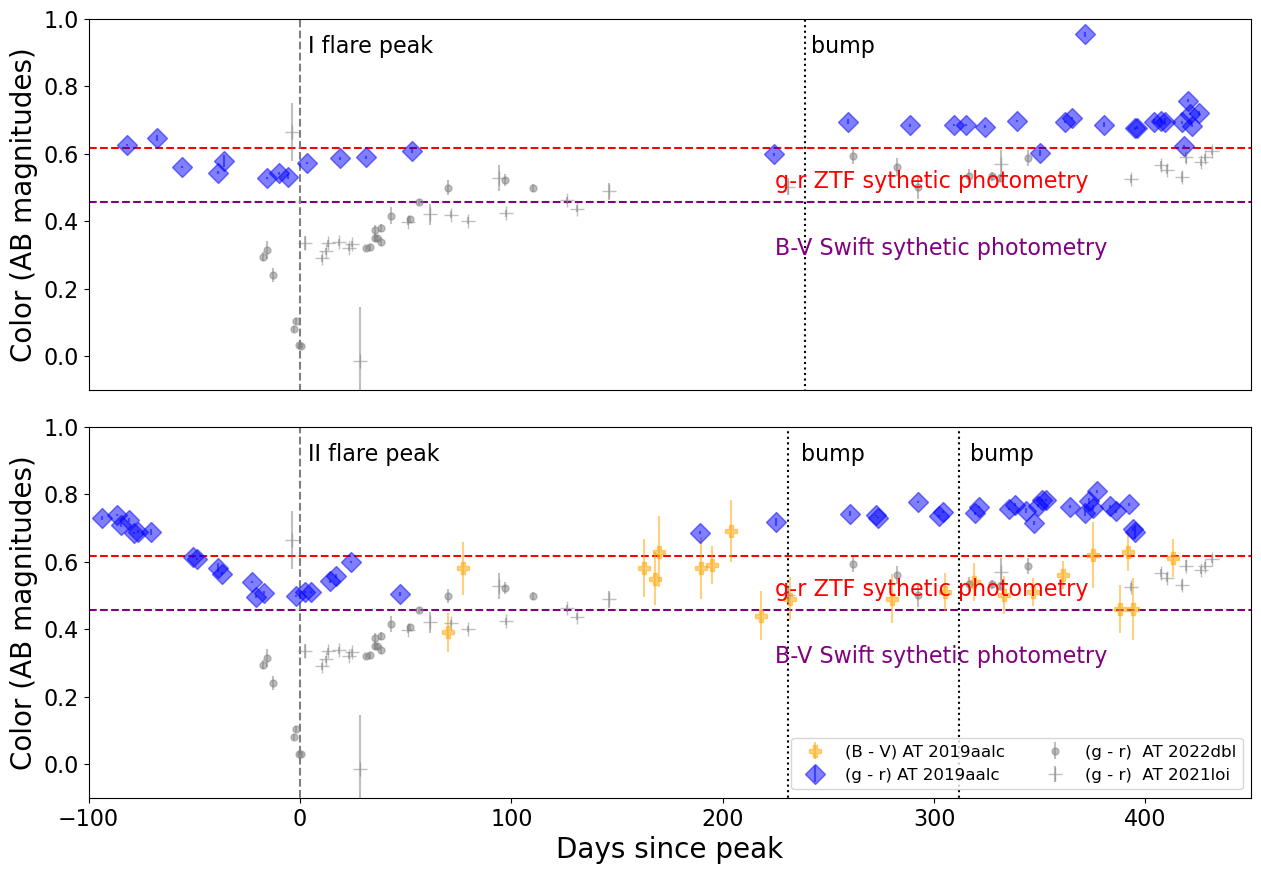}
\caption{The optical color evolution of \objname\ during the two major optical flares. 
The {\it top} and {\it bottom} panels cover the first and second optical flare, respectively
The $g-r$ color is calculated from publicly available ZTF photometry that covers both flares, while the $B-V$ color is calculated from the Swift/UVOT photometry that covers only the second flare. Vertical lines mark the timing of the two main optical flare peaks, as well as the more subtle bumps seen during the dimming phases. For comparison, we also plot the color evolution of one other BFF (AT\,2021loi; \citealt{Makrygianni2023}) and one pTDE candidate (AT\,2022dbl; \citealt{Lin2024_pTDE}). 
Dashed horizontal lines indicate pre-flare synthetic colors measured from the SDSS spectrum, including $g-r$ (using the ZTF bandpasses; red) and $B-V$ (using Swift/UVOT bandpasses; purple).
The photometric light-curve of \objname\ showed a ``bluer when brighter'' behavior, similar to what is typically seen in (unobscured) normal AGN.
}
\label{fig:color_evo}
\end{figure*}

Figure \ref{fig:color_evo} shows the optical color evolution of \objname\ during the two main optical flares.
We also plot the color evolution for two other SMBH-related (transient) brightening events, for comparison: AT\,2021loi \citep{Makrygianni2023}, which is another BFF, and AT\,2022dbl, which is an intriguing nuclear transient with recurring optical flares, interpreted as a pTDE candidate \citep{Lin2024_pTDE}.

To set a reference for the optical color evolution, we again rely on the archival SDSS spectrum for which we obtain synthetic photometry (see Section~\ref{sec:data_opt_phot}) using the ZTF ($g$ and $r$) and Swift/UVOT ($B$ and $V$) transmission profiles. This yields $g-r=0.62$ and $B-V=0.48$ (in AB mag) for this reference, pre-flare epoch, which we mark with dashed horizontal lines in Fig.~\ref{fig:color_evo}.  
% After calculating synthetic photometry from the SDSS spectrum of \objname, and using transmission profiles from ZTF we obtain $g-r = 0.62$ (marked with horizontal dashed line in Figure~\ref{fig:color_evo}).
From this Figure, we notice that during both major optical flares, \objname\ becomes bluer as it rises towards peak optical emission, and then becomes redder during the post-flare decline phases. 
The optical color and bluer-when-brighter behavior of \objname\ are consistent with what is seen in persistent, broad-line AGNs \cite[e.g.,][]{VdB04,2018rumbaugh}.
We do note, however, that even during the periods of peak (blue) optical emission periods, \objname\ is significantly redder than either the BFF AT\,2021loi and/or the pTDE candidate AT\,2022dbl.
% \objname\ color is consistent with typical photometrically selected QSO from SDSS-I \citep[][and references therein]{2006richards}.

\subsubsection{MIR light-curves}
We next turn our attention to the MIR emission of from \hname\, which---given its AGN nature prior to the optical flares---is generally thought to be driven by hot dust.
Indeed, the average pre-flare MIR color of \hname\ was $W1-W2  =  0.56$ mag (in the Vega system), which is consistent with what is seen in AGNs \cite[see Fig.~5 in][]{Stern2012}.

The WISE MIR measurements show an increase of roughly 1 mag following the detection of the first main optical flare (MJD = 58521), reaching a peak MIR flux in both W1 and W2 bands in the measurements taken on MJD = 58885, that is $\approx$230 days after the peak of the first optical flare. The second MIR peak (MJD = 60510) reached a total observed MIR flux that is $\approx$ 1 mag brighter than the first one, and occurring $\approx$210 days after the peak of the second optical flare. 
We stress again that the WISE cadence is very low ($\sim$6 months), and therefore a more precise estimate of the timing of the MIR peak is not available.
In the second MIR flare the flux increase, relative to the pre-brightening base level, was a factor of 1.75 stronger than in the first flare. This is in excellent agreement with the more luminous nature of the second optical flare compared with the first one (see above). 
The $W1-W2$ color during the first optical flare detection is around 0.59 mag, consistent with the typical pre-flare value ($\approx$0.56). After the optical flare detection, $W1-W2$ increased and reached 0.83 mag within $\approx$ 600 days of the first optical peak. The MIR color then decreased, so that during the detection of the second optical flare, it was back at $W1-W2$ = 0.52 mag, again consistent with the pre-flare value. During the second optical flare, it again increased, reaching 0.72 mag during the second MIR peak ($\approx$210 days after the second optical flare peak). For reference, we note that studies of MIR emission in AGNs have found that the higher the $W1-W2$ color, the stronger its association with AGN activity. Specifically, \cite{Stern2012} showed that 85\% (95\%) of the sources with $W1-W2\geq0.7$ ($\geq0.8$) are AGNs.

The MIR light-curve and color evolution thus suggest that the MIR emission responds to the changes in the UV/optical emission, by reprocessing it in a region of dusty circumnuclear gas. An alternative scenario, where the extremely variable UV/optical emission was due to changes in the line-of-sight obscuration by such dusty gas \cite[see discussion in, e.g.,][]{Zeltyn2022,2023riccitrakhtenbrot}, can be ruled out.\footnote{In such a case, the isotropically-emitted MIR radiation is not expected to vary.}
% The With no changes noticed in infrared emission, the flare in the optical could have been caused by unobscuring the inner parts.
To contextualize the timing of the MIR peaks, we first calculate the expected size of a dusty (toroidal) gas structure, following what is known in persistent, nearby AGNs. 
From the reverberation-mapping based correlation between optical luminosity and the scale of the (hot) dust region reported by \cite[][Eq.~3 therein]{2019minezaki}, and the highest optical luminosities of \objname\ (i.e. the two flare peaks), we find that the hottest dust should be located $\approx90-115$ light days from the SMBH. These scales are thought to represent the region where dust is sublimated by the incident UV/optical radiation originating from the central accretion flow.
Given that the MIR peak occurred $\approx$ 210-230 days after the optical flares, and that the MIR emission is indeed expected to be produced further in the dusty (toroidal) gas, we conclude that the enhanced MIR emission could be accounted for by a delayed, reprocessed UV-optical radiation emitted from the (perturbed) AGN accretion flow. This would also explain the persistent AGN-like MIR color of \objname and the fact that the MIR color becomes more AGN-like during the periods of enhanced UV/optical emission.

\begin{figure*}
\centering
\includegraphics[scale=0.5]{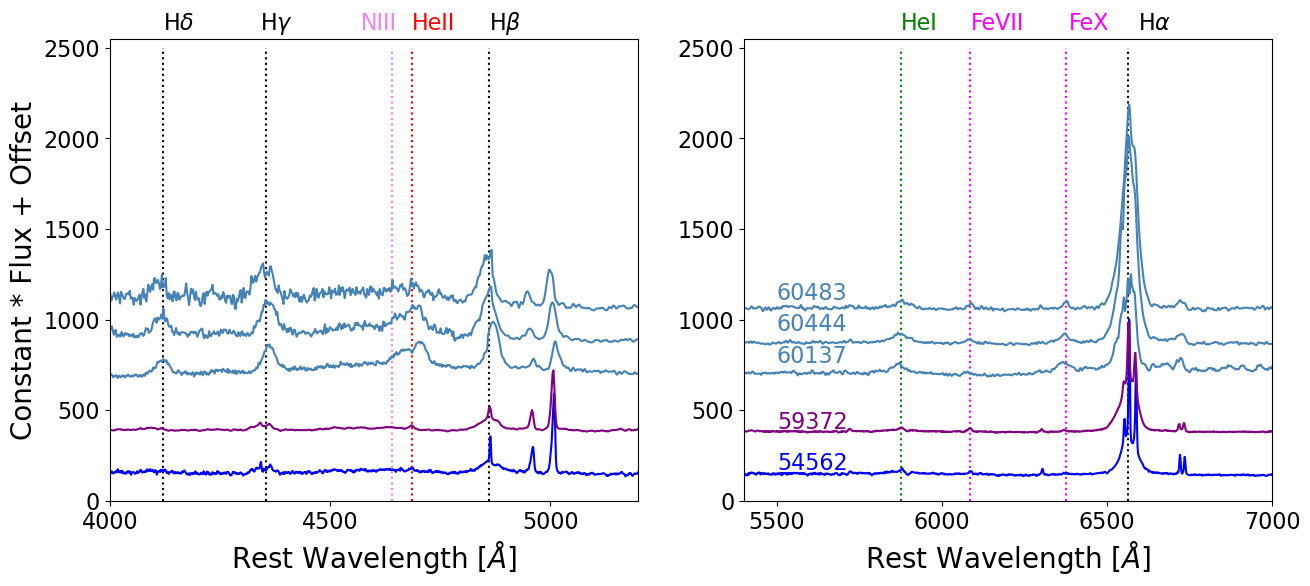}
\caption{Comparison of several key optical spectra of \objname, including the archival pre-flare SDSS spectrum (dark blue), the Keck/LRIS spectrum taken after the first optical flare has decayed (purple), and a few of the Las Cumbres spectra taken after the second flare (light blue). The left and right panels show the spectral regions surrounding the \hb\ and \ha\ emission lines (respectively). 
The spectra taken after the second peak show a clear brightening of the \HeIIop, \NIII, and \FeX\ lines.}
\label{fig:key_spec_comp}
\end{figure*}

\subsection{Spectroscopic properties}

In this Section, we present and discuss all the measurements obtained from our optical spectra, as well as derived quantities. 
The details of all the measurements of the \hb\ spectral range are presented in Table \ref{tab:spectral-measurement}. 
% details about the evolution of Balmer lines are discussed in Section \ref{sec:balmer-lines-evolution}, and the evolution of Bowen lines, are discussed in Section \ref{sec:evolution-bff}. 
% We see changes in intensities of coronal lines (and the appearance of \fexiv, which we discuss in Section \ref{sec:coronal-lines}.

\begin{deluxetable*}{clccccch}
\tablecaption{Measurements of the \hb\ spectral region}
\label{tab:spectral-measurement}
\tablewidth{\textwidth}
\tablehead{
\colhead{MJD} & \colhead{Instrument\tablenotemark{a}} & \colhead{$\log\Lop$} &  \colhead{\fwhb}  & \colhead{\Fhb\tablenotemark{b}}   & \colhead{\Fbf\tablenotemark{c}}    & \colhead{\Ffex\tablenotemark{b}} & \nocolhead{FWHM(\fex)}\\
&  & \colhead{(\ergs)} & \colhead{(\kms)}   & \colhead{$(10^{-15}\,\ergcms)$}  & \colhead{$(10^{-15}\,\ergcms)$}   & \colhead{$(10^{-15}\,\ergcms)$} &  \nocolhead{(\kms)} 
}
\startdata
54562 & SDSS            & 43.42                  & 3100 $\pm$ 100       & 54.23 $\pm$ 0.77   & 21.47 $\pm$ 0.63  & 3.94 $\pm$ 0.01  & 2100 $\pm$ 100                      \\
59372 & Keck/LRIS       & 43.17                  & 3000 $\pm$ 400       & 23.33 $\pm$ 2.63   & 9.37 $\pm$ 2.75   & 4.06 $\pm$ 2.64  & 1600 $\pm$ 300                      \\
60137 & Las Cumbres/COJ & 43.53                  & 2500 $\pm$ 100       & 140.29 $\pm$ 4.01  & 196.48 $\pm$ 4.4  & 38.47 $\pm$ 3.39 & 2800 $\pm$ 300                      \\
60153 & Las Cumbres/COJ & 43.51                  & 2600 $\pm$ 100       & 227.12 $\pm$ 2.69  & 166.55 $\pm$ 2.46 & 35.01 $\pm$ 3.33 & 2700 $\pm$ 300                      \\
60167 & Las Cumbres/COJ & 43.58                  & 2400 $\pm$ 100       & 470.86 $\pm$ 2.37  & 91.71 $\pm$ 2.31  & 30.31 $\pm$ 0.01 & 2100 $\pm$ 100                      \\
60180 & Las Cumbres/COJ & 43.44                  & 2500 $\pm$ 100       & 94.60 $\pm$ 3.31    & 68.22 $\pm$ 1.88  & 21.02 $\pm$ 2.24 & 2700 $\pm$ 200                      \\
60191 & Las Cumbres/OGG & 43.58                  & 2500 $\pm$ 100       & 640.12 $\pm$ 5.82  & 154.36 $\pm$ 2.91 & 25.48 $\pm$ 6.82 & 2400 $\pm$ 700                      \\
60314 & Las Cumbres/OGG & 43.43                  & 2800 $\pm$ 100       & 244.24 $\pm$ 1.24  & 135.05 $\pm$ 2.62 & 8.23 $\pm$ 10.56 & 1400 $\pm$ 1100                     \\
60368 & Las Cumbres/OGG & 43.58                  & 2500 $\pm$ 100       & 892.84 $\pm$ 14.66 & 349.78 $\pm$ 4.47 & 28.18 $\pm$ 2.22 & 2200 $\pm$ 200                      \\
60380 & Las Cumbres/COJ & 43.58                  & 2900 $\pm$ 100       & 729.55 $\pm$ 5.99  & 158.20 $\pm$ 3.25  & 27.62 $\pm$ 0.01 & 2100 $\pm$ 100                      \\
60392 & Las Cumbres/OGG & 43.34                  & 2900 $\pm$ 100       & 297.19 $\pm$ 3.02  & 60.12 $\pm$ 2.11  & 14.87 $\pm$ 2.92 & 2100 $\pm$ 300                      \\
60409 & Las Cumbres/COJ & 43.33                  & 3100 $\pm$ 300       & 79.98 $\pm$ 3.73   & 60.80 $\pm$ 2.42   & 8.76 $\pm$ 2.81  & 1900 $\pm$ 300                      \\
60418 & Las Cumbres/OGG & 43.31                  & 3000 $\pm$ 100       & 128.73 $\pm$ 1.85  & 34.21 $\pm$ 1.42  & 13.67 $\pm$ 3.21 & 2100 $\pm$ 300                      \\
60426 & Las Cumbres/COJ & 43.31                  & 2400 $\pm$ 100       & 92.72 $\pm$ 1.81   & 1.76 $\pm$ 1.77   & 20.92 $\pm$ 0.37 & 2100 $\pm$ 100                      \\
60444 & Las Cumbres/COJ & 43.49                  & 2900 $\pm$ 100       & 374.50 $\pm$ 7.16   & 259.41 $\pm$ 4.02 & 18.99 $\pm$ 1.21 & 2100 $\pm$ 100                      \\
60454 & Las Cumbres/OGG & 43.31                  & 2700 $\pm$ 100       & 103.74 $\pm$ 2.63  & 86.03 $\pm$ 2.27  & 9.14 $\pm$ 3.6   & 1600 $\pm$ 400                      \\
60462 & Las Cumbres/OGG & 43.14                  & 2700 $\pm$ 100       & 42.70 $\pm$ 1.63    & 41.49 $\pm$ 1.33  & 8.66 $\pm$ 1.9   & 2000 $\pm$ 200                      \\
60472 & Las Cumbres/OGG & 43.41                  & 2900 $\pm$ 100       & 284.75 $\pm$ 2.06  & 68.11 $\pm$ 1.94  & 15.95 $\pm$ 0.01 & 2100 $\pm$ 100                      \\
60483 & Las Cumbres/OGG & 43.45                  & 3000 $\pm$ 100       & 321.34 $\pm$ 8.07  & 173.15 $\pm$ 3.06 & 7.99 $\pm$ 0.69  & 1200 $\pm$ 100                      \\
60491 & Las Cumbres/OGG & 43.12                  & 3000 $\pm$ 100       & 49.41 $\pm$ 1.67   & 38.65 $\pm$ 1.56  & 5.21 $\pm$ 8.01  & 1200 $\pm$ 800                      \\
60500 & Las Cumbres/OGG & 43.39                  & 3100 $\pm$ 100       & 175.30 $\pm$ 2.08   & 48.76 $\pm$ 2.21  & 10.25 $\pm$ 1.94 & 1200 $\pm$ 200                      \\
60508 & Las Cumbres/OGG & 43.47                  & 3200 $\pm$ 100       & 250.05 $\pm$ 3.34  & 75.39 $\pm$ 2.25  & 10.40 $\pm$ 0.5   & 1200 $\pm$ 100                      \\
60533 & Las Cumbres/COJ & 43.42                  & 3100 $\pm$ 100       & 273.15 $\pm$ 3.99  & 55.74 $\pm$ 2.05  & 19.58 $\pm$ 0.01 & 2100 $\pm$ 1000                     \\
60556 & Las Cumbres/OGG & 43.19                  & 2600 $\pm$ 100       & 51.37 $\pm$ 1.25   & 28.13 $\pm$ 1.69  & 13.69 $\pm$ 1.12 & 2100 $\pm$ 100 
\enddata
\tablenotetext{a}{For Las Cumbres spectra, ``OGG'' and ``COJ'' denote the 2 m telescopes at the Haleakal\=a and Siding Spring observatories, respectively.}
\tablenotetext{b}{Measured from the (multi-)Gaussian fit approach.}
\tablenotetext{c}{Measured from the residual integration approach.}
% Columns are as follows: MJD, name of the instrument, \Lop, FWHM of broad component of \hb\ in km s$^{-1}$, flux of \hb\ from the gaussian fit, flux of BF measured as the integrated flux density of \HeIIop\, and \NIII\ regions with subtractions mentioned in Section \ref{sec:evolution-bff}, flux of \fex\ from the gaussian fit, and FWHM of \fex.}
\end{deluxetable*}

\subsubsection{Archival spectrum and derived properties prior to the second flare}

Regardless of any spectral decomposition, the broad \hb\ and \ha\ emission lines are clearly visible in the pre-flare archival SDSS spectrum, which confirms the (unobscured) AGN nature of the galaxy prior to the dramatic brightening events. Other, fainter broad features include \Hgamma\ and (rather weak) \Hdelta\ emission, as well as low intensity \feii\ emission, and noticeable emission from the high ionization \HeIIop\ transition.
In terms of narrow lines, in addition to the \oiii\,$\lambda\lambda$4959,5007, $\nii\,\lambda\lambda$6548,6583, and $\sii\,\lambda\lambda$6718,6732 doublets that are commonly seen in AGNs, the SDSS spectrum also shows the  $[\textrm{O\,\textsc {i}}]\,\lambda$6300 and $\textrm{He\,\textsc {i}}\ \lambda$5875 lines.

From the spectral decomposition of the archival SDSS spectrum, we obtain line widths of $\fwhb  = 3000\,\kms$ and $\fwha = 2700\,\kms$ for the broad components of the Balmer lines.
The narrow component of \hb\ has $\fwhm=200\,\kms$, as does \oiii\ (which is tied in width), while the narrow component of \ha\ has $\fwhm=300\,\kms$.
We will use these narrow emission line widths to better understand the coronal lines of highly ionized iron that appeared during the second optical flare.
The decomposition of the Keck/LRIS spectrum, taken after \objname\ has declined from the first flare, yielded  $\fwhb  = 3000\,\kms$ and $\fwha = 2500\,\kms$, which are consistent with the pre-flare SDSS spectrum. While \HeIIop\ is still present, there is no sign of BF emission features (namely \NIII) at this stage. Moreover, \feii\ appears even weaker in this Keck spectrum (relative to, e.g., \hb).

% BT GOT HERE

We derived key AGN and SMBH properties based on the best-fitting spectral decomposition of the archival SDSS spectrum.
The monochromatic luminosity at rest-frame 5100\,\AA\, \lamLlam(5100\,\AA) (or \Lop) measured from that spectrum is $\Lop=2.6 \times 10^{43}\,\ergs$. 
% We use this luminosity to estimate the broad line region (BLR) size, \RBLR, using the relation from \citet{TrakhtenbrotNetzer12},  and obtain $\RBLR=11.3$ light-days.
To estimate the SMBH mass, \mbh, we use two different single-epoch mass prescriptions. Firstly, we combine the \Lop\ and \fwhb\ measurements through the prescription provided in \cite{TrakhtenbrotNetzer12} and obtain $\log(\mbh/\Msol) = 7.3$. 
Second, we combine the luminosity of the broad \ha\ line,  $\Lha=5 \times 10^{41}\,\ergs$, and the \fwha\ measurement through the prescription of \citet{MR22}, and obtain $\log(\mbh/\Msol) =7.1$.
These single-epoch \mbh\ estimates carry significant systematic uncertainties, of order $0.3-0.5$ dex \cite[see, e.g.,][]{Shen2013_rev}. The two \mbh\ estimates are thus in agreement with each other. 

To estimate the (pre-flare) bolometric luminosity, we multiply the SDSS-based \Lop\ measurement by a bolometric correction of $\kbol= 8.8$, based on the approach presented in \cite{marconi2004}. We obtain $\Lbol = 2.3 \times 10^{44}\,\ergs$. Combining this \Lbol\ estimate with the \Lop-based estimate of \mbh, we thus estimate a pre-flare Eddington ratio of about $\lledd\equiv (\Lbol/\ergs)/(1.5\times10^{38}\,[\mbh/\Msun]) \simeq 0.07$. Following the same approach, we obtain \lledd\ estimates from both the Keck spectrum which was taken between the two main optical flares, and the Las Cumbres spectrum taken during the peak of the second optical flare, and find $\lledd\simeq0.07$ and 0.15, respectively.
We conclude that between the two main optical flares, the SMBH was accreting at an Eddington ratio consistent with that of the more stable, pre-flare state, while during the peak of the brighter, second optical flare, its accretion rate was higher by a factor of $\approx2$. 
Despite the dramatic activity seen in \objname, the accretion rates we estimate are consistent with what is known for persistent, low-redshift AGNs, for which \lledd\ is estimated through similar methods \cite[e.g.,][]{Ananna2022}.

\subsubsection{Evolution of Balmer lines}
\label{sec:balmer-lines-evolution}
All the optical spectra in our dataset, specifically those obtained with Las Cumbres, exhibit clear signatures of broad Balmer line emission.
% First, we focus on describing the behavior of \hb\ and \ha.
In terms of changes in the width of the broad Balmer lines, we find that \fwha\ varies between $2100 - 3700\,\kms$ and \fwhb\ varies between $2400-3200\,\kms$. 
These line widths are much narrower than what is seen in TDEs \cite[i.e., $\fwhm>10,000\,\kms$; see, e.g.,][]{2022Charalampopoulos}.
We note that there is no evidence of any noticeable trends between changes in the FWHM of the Balmer lines and changes in the (broad-band) optical flux. In particular, the lines do not seem to become narrower when the continuum emission increases, which may be expected if the broad Balmer lines originate from larger BLR scales (i.e., ``line breathing''; e.g., \citealt{Peterson2000}). This is, however, seen in a non-negligible fraction of persistent broad-line AGNs for which adequate spectroscopic data is available \cite[e.g.,][]{Wang2020}.

The strength of the broad \ha\ and \hb\ lines in the post-first-flare Keck/LRIS spectrum is consistent with what is seen in the pre-flare SDSS spectrum. However, the first Las Cumbres spectrum we obtained after the detection of the second significant optical brightening (MJD = 60137) shows a significant increase in the strengths of these lines, which brightened by factors of $\approx2\times$ and $\approx4\times$, for \ha\ and \hb\ (respectively), compared with the pre-second-flare Keck/LRIS spectrum.
Indeed, the time evolution of all the broad Balmer lines seen in our Las Cumbres spectroscopy (\Hdelta, \Hgamma, \hb, and \ha) is consistent: the fluxes of all these lines increase and decrease in tandem with the broad-band optical flux, as can be seen in Figure \ref{fig:balmer-evolution} (in Appendix \ref{app:evolution-balmer}).
We note that our measurements of the \Hdelta\ line are complicated by the possible blend with the BF $\niii\,\lambda\lambda4097,4103$ lines and the limited spectral resolution of the Las Cumbres spectra \cite[see also][]{Makrygianni2023}. 
% however, due to data resolution, we are not able to comment more on this behavior.
In any case, the latest Las Cumbres spectrum among the dataset we analyze here, taken on 2024 September 3 (MJD = 60556) still shows enhanced Balmer line emission compared with the pre-flare state.

\subsubsection{Evolution of Bowen lines}
\label{sec:evolution-bff}
Here we examine the temporal evolution of the \HeIIop\ and \NIII\ emission lines. These lines are related both physically, as the BF process is initiated by the Ly$\alpha$ transition of the H$^+$ ion; and observationally, as the two lines are blended in our spectra. 
We therefore only discuss the temporal evolution of the combined flux of the two BF-related lines, \Fbf\ hereafter.

As previously noted, the archival SDSS spectrum clearly shows some non-negligible $\heii+\niii$ emission and our measurements indicate $\Fbf/\Fhb \approx 0.29$ or $\approx0.44$ during this pre-flare period (from the fitting or integration methods, respectively). The post-first-flare Keck spectrum exhibits a highly consistent strength, $\Fbf/\Fhb \approx 0.30$ (or 0.28 from the integration method).
To compare the $\heii+\niii$ emission to what is typically seen in luminous, persistent broad-line AGNs, we note that the SDSS-based quasar composite spectrum calculated by \cite{VandenBerk2001} exhibits only $\Fbf/\Fhb < 0.05$, and essentially no \niii\ emission. 

In our post-flare spectroscopy, the first Las Cumbres spectrum is taken around the time of the peak of the second optical flare (MJD = 60137), and we see a significant increase of the \NIII\ BF line.
This is demonstrated in the left panel of Figure \ref{fig:key_spec_comp}, which focuses on the \Hbeta\ spectral region (including the $\heii+\niii$ features) in the spectra taken during several key epochs in the evolution of \objname.
Our measurements indicate $F(\heii + \niii)/F(\hb) \approx1.3$, which is comparable with the value measured for AT\,2021loi ($\approx$1.1; \citealp{Makrygianni2023}). 
Given the temporal gap between the Keck spectrum and this first Las Cumbres spectrum, we cannot determine when did the \niii\ line emerge. 
We thus conclude that the $\heii+\niii$ emission seen in the pre-flare (archival) state of \objname\ was already higher by a factor of $\gtrsim$15$\times$ compared with what is typically seen in persistent, broad line AGNs, and increased by a significant factor during the flaring activity.

\begin{figure*}
\centering
\includegraphics[scale=0.5]{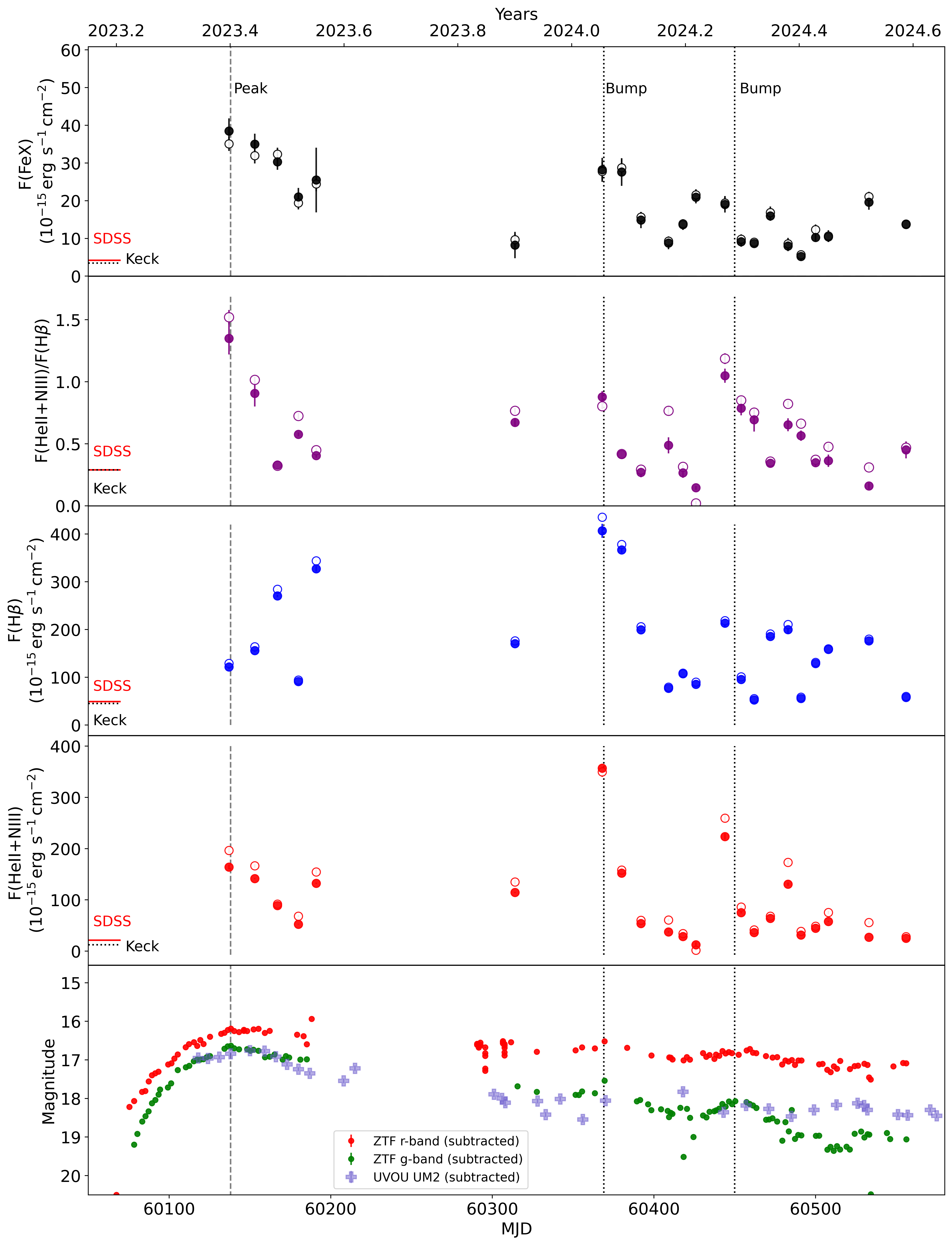}
\caption{Time series of emission line (relative) strengths following the second optical flare of \objname. 
Open circles represent measurements based on direct integration of the line flux density, whereas filled circles represent measurements derived from the spectral modeling.
% values obtained by fitting a broad single Gaussian. 
From top to bottom: \Ffex\ (black points), $\Fbf/\Fhb$ (violet points), the broad component of \hb\ (\Fhb; blue points), and \Fbf\ (red points), and the reference-subtracted photometric light-curves. In the panels showing emission line strengths we also mark the corresponding pre-flare measurements from the SDSS and Keck spectra (near the left edge of each panel; red and black lines, respectively). 
The UVOT measurements shown in the bottom panel were obtained with a circular aperture of 1.5\arcsec\ in radius, comparable with the aperture of the pre-flare SDSS spectrum.
Vertical lines mark the epochs of peak broad-band emission during the entire second optical flare (left) and the bumps seen during the decline (right).
The emission lines, and particularly those related to high ionization species ($\heii+\niii$ and \fex), vary on relatively short timescales and generally follow the broad-band optical light-curve (see text for discussion).}
\label{fig:lines_lc}
\end{figure*}

The three middle panels of Figure~\ref{fig:lines_lc} present the light-curve of $F(\hb)$ (2nd from bottom), $F(\heii+\niii)$ (middle), and of the ratio $F(\heii+\niii)/F(\hb)$ (2nd from top), as derived from our Las Cumbres spectral series.
As can be clearly seen in Fig.~\ref{fig:lines_lc}, the strength of both the broad \hb\ line and the high-ionization, BF-related $\heii+\niii$ emission complex shows rich temporal variations.
First, as noted above, the $\heii+\niii$ emission is significantly enhanced during the peak of the (second) main optical flare. Because the \hb\ is enhanced only after the flare peak, the relative strength is high, and indeed reaches the maximal value observed during our monitoring, $\Fbf/\Fhb\gtrsim1.3$.
Then, in the spectrum taken near the first bump observed in the optical light-curve (MJD=60368), the \hb\ and $\heii+\niii$ features are enhanced again and indeed reach the highest fluxes observed during our spectral monitoring. 
Although the $\Fbf/\Fhb$ ratio has also increased, the fact that the \hb\ line emission has increased by $\approx4\times$  (see Section~\ref{sec:balmer-lines-evolution} above), the line ratio is only mildly enhanced, reaching $\Fbf/\Fhb\approx0.88$.  
We stress again that this peak line emission coincides with---by in fact somewhat predates---the peak optical $g$-band bump (MJD = 60369).  
A qualitatively similar behavior is seen again in the spectrum taken just prior to the second bump in the optical light-curve (on MJD= 60444): both the \hb\ and $\heii+\niii$ spectral features complexes exhibit enhanced emission, reaching their highest fluxes during our entire spectral monitoring campaign. 
The ratio between them reached $\Fbf/\Fhb\approx1$, which is the highest value observed during the decline of the light-curve after the (second) optical flare.
%
% the strength of both the broad \hb\ line and the high-ionization, BF-related $\heii+\niii$ emission complex generally follow the optical continuum emission light-curve, with significantly enhanced line emission concurrent with the optical bumps.
% Perhaps most interestingly, the ratio of these spectral features' strengths, $F(\heii+\niii)/F(\hb)$, also shows rich temporal variations: it reaches a peak of $F(\heii+\niii)/F(\hb)  \simeq 1$ near MJD = 60368, which broadly coincides with---but in fact somewhat predates---the optical $g$-band bump (MJD $\approx$ 60369).  
% We notice enhanced emission of BF near optical bumps increasing to F(\HeIIop+\NIII)/F(\hb) = 0.87, during the first flare (fitted with Gaussians), and F(\HeIIop+\NIII)/F(\hb) = 0.79  during the second flare. 
% Gas producing BF has high hydrogen number density $n_{\mathrm{H}}>10^{9.5}$cm$^{-3}$ and and high abundances of oxygen and nitrogen in the gas \citep{Netzer1985}.

In terms of the width of the main BF-related \HeIIop\ and \NIII\ emission lines, our model fits indicate $\fwhm(\HeIIop)\simeq2000-3000\,\kms$ and $\fwhm(\NIII)\simeq2000- 6000\,\kms$ during our Las Cumbres monitoring campaign. Due to the limited spectral resolution of the FLOYDS spectra, we decided to not report the individual best-fit values. We note however that during the rebrightening bumps seen in the UV/optical lightcurve, when the \heii\ and \niii\ features are stronger (see above) and better defined, their widths have a more limited range of $\fwhm\simeq2500-3000\,\kms$ (for both lines). 
The widths of the BF-related \HeIIop\ and \NIII\ lines are therefore consistent with those of the broad components of the Balmer lines (\ha\ and \hb). Both lines are significantly broader than the narrow emission lines (e.g., \OIII), and the \HeIIop\ line is--again--significantly narrower than what is seen in TDEs \cite[e.g.,][]{2022Charalampopoulos}.

Considering all these measurements from our high-cadence spectral monitoring, may conclude that the BF-related $\heii+\niii$ features, which require EUV radiation reprocessed by a high-density, appear to originate from a region that coincides with the classical BLR of \hname. That BLR has been in place well before the dramatic flaring activity labeled as \objname. 
Since the photon energy required to drive these features is much higher than that required for the Balmer recombination lines ($>$54.4 vs. $>$13.6 eV), we conclude that the SED of \objname\ during the second optical flare peak was much harder than during the first bump, and the SED during the second bump was again harder than during the first one. Moreover, between the two main optical flares, the SED was comparable to that of the pre-flaring state, as suggested by the comparable strength of $\heii+\niii$ in the Keck and SDSS spectra.

\subsubsection{Coronal line emission}
\label{sec:coronal-lines}

%% FROM SLACK 8-APR-2025 -- FeXIV: 
% spectrum closest to the optical flare peak spectrum: 
%%  MJD = 60137, ratio = 0.40, FWHM = 1600
% maximum of that ratio just after the peak, I think MJD~60170 or so - the 3rd point in that sequence is the highest) 
%%  spectrum MJD = 60167, ratio = 0.48, FWHM = 2200
% the one coinciding with the 1st bump spectrum 
%% MJD = 60368, ratio = 0.33, FWHM = 1800
% the one just prior to the 2nd bump  
%% spectrum MJD = 60444, ratio = 0.20, FWHM = 1900

All the optical spectra considered in this work, including the pre-flare (SDSS) and the post-first-flare (Keck) ones, show significant emission from the coronal \FeX\ line (\fex\ hereafter). 
% In this Section, we discuss the evolution of coronal lines. 
This is exemplified in the right panel of Figure \ref{fig:key_spec_comp}, which focuses on the \Halpha\ spectral region (including the \fex\ feature) in the spectra taken during several key epochs in the evolution of \objname.
In both the pre-flare (SDSS) and post-first-flare (Keck) spectra we measure $\Ffex/\Foiii)\approx0.1$.  
In these two spectra, we also detect (but don't model) the fainter [Fe\,\textsc{vii}]\,$\lambda$6087 coronal line, however, we find no evidence for \FeXIV\ emission.
Our Las Cumbres spectral monitoring during the second optical flare of \objname\ reveals enhanced \fex\ emission, as well as significant (and thus enhanced) \fexiv\ emission.
In the first Las Cumbres spectrum, taken around the peak of the second optical peak (MJD = 60137), \fex\ reached $\Ffex/\Foiii \approx 1.34$, then slowly declining over a period of $\sim50$ days.
% The \fex\ line kept brightening even further, reaching $\Ffex/\Foiii)\approx\todo$ about \todo\ days after the peak of the optical flare. 
Then \fex\ showed clear brightening, coincident with the bumps in the broad-band optical light-curve, reaching $\Ffex/\Foiii\simeq1.06$ and 0.74 (for the spectra closest to the peaks of the first and second bumps, respectively).
Interestingly, a close examination of the various panels of Fig.~\ref{fig:lines_lc} shows that the second local maximum of \fex\ emission (MJD=60426) slightly predates the local peak of the $\heii+\niii$ emission (MJD=60444), which itself slightly predates the peak of the second broad-band optical bump (MJD=60450).
%
%
% It then decreased to $\Ffex/\Foiii\approx0.3$ and eventually reached $\approx0.6$ in the last Las Cumbres spectrum we analyzed (MJD  = 60556)
% These late-stage, fainter line strengths are still much higher, by a factor of $\gtrsim$3, than those measured in the pre-flare (SDSS) and between-flares (Keck) spectra. 

The similarities between the light-curves of the \fex\ line and the (blue) continuum emission suggest that the high ionization coronal lines are produced relatively close to the SMBH. To gain further insight into the possible location of the gas emitting the coronal line(s) in \objname\, we examine the line widths, recalling that in most AGNs the coronal lines are thought to be emitted in the inner parts of the narrow line region (NLR), or indeed between the NLR and the BLR (see, e.g.,  \citealt{2010Mazzalay,Landt2015,denBrok2022}, or the recent review by \citealt{Rodriguez2025_CLR_rev}, and references therein).

In the pre-flare (SDSS) and between-flares (Keck) spectra, we measured $\fwhm(\fex)=1200$ and 2000\,\kms, respectively. In the first Las Cumbres spectrum after the detection of the second optical flare, the width of \fex\ was $\approx$2800\,\kms\ and the \fexiv\ line had FWHM of $\approx$2500\,\kms. 
All these coronal line widths are significantly broader than what we measured for the narrow \OIII\ line,  i.e. $\fwhm(\oiii)\approx200-300\,\kms$ in the higher-resolution SDSS and Keck spectra and $\lesssim$900\,\kms\ in the lower-resolution Las Cumbres spectroscopy.\footnote{Note we do not claim the \oiii\ line width has evolved during our spectroscopic monitoring.}
The \cite{2024veres} study of \objname\ reports consistently broad coronal lines, with $\fwhm(\fex)\simeq1200-1800\,\kms$, as well as a detection of the weak [S\,\textsc{xii}]\,$\lambda 7611$ coronal line in a late-time spectrum ($\approx$310 days after the peak of the second optical flare), with $\fwhm\approx\,3400\kms$.

The coronal line widths therefore appear to be consistent with those of the broad Balmer lines ($\sim$3000\,\kms), which suggests that the coronal lines seen in \objname\ are emitted from a region that is located closer to the SMBH than the NLR, and perhaps indeed as close as the BLR. 
Further evidence for this comes from the weeks-timescale variability observed in \fex, which is temporarily coincident both with the broad-band optical variability and with the variability of the BF-related $\heii+\niii$ features, which are also driven by the variable, high-energy radiation of the central source.
%
% Indeed, we note that the critical density of the \FeX\ line, $n_e = 5\times10^{9}\,{\rm cm}^{-3}$ \citep{1988AJ.....95...45A} is consistent with what is usually considered for the BLR.
%
The conclusion that the coronal lines are emitted from BLR-like scales is in agreement with what was found for several other SMBH-related flares exhibiting (transient) coronal line emission, including AT\,2019avd \citep[Fig. 12 in][]{2021Malyali} and AT\, 2022upj \citep[Fig.~11 in][]{2024newsome}.
%, supporting the stratification of BLR.
On the other hand, several other systems associated with extreme and/or transient SMBH accretion showed only narrow coronal lines and were thus interpreted as originating from NLR-like scales  \cite[e.g.,][]{Komossa2008, Wang2012, Yang2013}. 
Moreover, there seems to be no strong evidence for enhanced coronal line emission in persistent, yet highly accreting AGNs \cite[i.e., systems with high \lledd; see, e.g.,][]{2024Bierschenk}.
Strong coronal lines with widths comparable with that of the BLR, such as the \FeX\ and \FeXIV\ lines we see in \objname, thus remain a rather unusual property, which may be unique to BFFs.

If the coronal lines are indeed emitted from the same region as the broad Balmer and BF-related lines, and given the high ionization potentials needed to produce the coronal lines (i.e., $\gtrsim$235 eV), then the temporal evolution of all the lines (Fig.~\ref{fig:lines_lc} supports the following chain of events occurred during the second main flare of \objname. 
The near-peak radiation during the main flare itself had a particularly hard SED, reaching $>$235 eV and driving the significant brightening in the coronal and BF-related lines, while the emission from the lower-ionization H$^+$ (traced by \hb; requiring $>$13.6 eV) was slightly delayed. 
The SED in the first bump may have been similarly hard, although our observations did not cover the rise to the peak of the bump, and so could not disentangle the temporal evolution of the various ionization species.
The SED in the second bump may have been even harder, as we clearly see the emission from the highest-ionization species (Fe$^+9$, traced by \FeX and requiring $>$235 eV) peaking well before the BF-related emission (He$^+$, traced by $\heii+\niii$ and requiring $>$54 eV) and the broad-band optical emission (ZTF $g-$band). In fact, the peak \fex\ emission occurs during a local {\it minimum} in the $\heii+\niii$ emission (in terms of both \Fbf\ and $\Fbf/\Fhb$.
Our high-cadence spectral monitoring of \objname\ thus shows that the SED of the (flaring) accretion flow varies, perhaps in a non-monotonic manner, during the main flare and the post-peak decline.
This further highlights the need for X-ray monitoring of such events, which we discuss in the next Section.

% For \fex\ ionization potential is 262eV, which is high in comparison to \HeIIop, for which the ionization potential is 54.4eV.
% Taking into account the definition of ionization parameter \citep{1983ApJ...264..105F}, high $U$ can be obtained by small r in comparison to NLR distance and typical $n_e$ or by typical distance (inner NLR-like), but extremely low $n_e$.
% The critical density of \fex\ ($= 10^{9.7}\, \mathrm{cm}^{-3}$ \citep{1988AJ.....95...45A} is similar to a typical
% BLR density \citep{1989agna.book.....O} ( $\approx 10^{9} \mathrm{cm}^{-3}$). The presence of relatively broad (significantly broader than narrow lines detected in those sources) \fex\ was measured also in previous SMBH-related transients with BF emission, such as AT 2019avd \citep{2021Malyali} and AT\,2021loi \citep{Makrygianni2023}.

%Coronal lines have been observed not only in BFF events but also in other rare systems associated with extreme accretion onto SMBHs, such as ECLEs \citep{Komossa2008, Wang2012, Yang2013}

% \cite{2024veres} report the presence of [S\,{\sc xii}$] \lambda 7611$ line in the DBSP spectrum (MJD = 60451), however, we are not able to detect this line in our data due to lower resolution and fringing.
% We also see a strong feature around 8450\,\AA\ and we identify it as $\textrm{O\,\textsc{i}}\lambda$8446 emission \citep{2006glikman}.

% BT GOT HERE - BUT HAS TO FINISH CL TEXT!

\subsection{X-ray behavior}
\label{sec:x-ray-evolution}

We checked archival ROSAT data to assess the pre-flare X-ray emission from \hname.
The source was not detected in the ROSAT all sky survey \citep{1993truemper}, which places a $3\sigma$ upper limit on the flux in the $0.1-2.4\,\kev$ band of $F(0.1-2.4\,\kev)\lesssim 10^{-13}\,\ergcms$ \citep{Boller2016}, corresponding to a luminosity upper limit of $L(0.1-2.4\,\rm keV) \lesssim 2\times 10^{42}\,\ergs$. Assuming the X-ray SED follows a power-law with a photon index of $\Gamma=1.8$, which is typical for persistent, low-redshift AGNs \citep{2017ricci}, we can extrapolate this to obtain an upper limit on the luminosity in the $2-10\,\kev$ range, of $L(2-10\,\rm keV) \lesssim 2.5\times 10^{42}\,\ergs$.
These upper limits are lower than what one could have expected to observe in \objname, given the archival, pre-flare NUV emission. Specifically, the GALEX photometry translates to a monochromatic NUV luminosity at 2500\,\AA\ of $L_\nu (2500\,\AA) = 5.0\times10^{27}\,\ergs\,{\rm Hz}^{-1}$.
Using the $L_ \nu (2500\,\AA) - L_\nu(2\,\kev)$ scaling relation from \cite{Lusso2016}, we estimate the expected monochromatic X-ray luminosity at
2 keV would be $\nu L_\nu(2\,\kev)$ = $2.2\times10^{42}\,\ergs$. After the extrapolation to $2-10\,\kev$, we obtain an expected $L(2-10\,\kev) =4.2\times 10^{42}\,\ergs$, which is higher than the ROSAT upper limit by a factor of $>1.6\times$.
%which is comparable with the ROSAT upper limit.
The broad-line AGN residing in the center of \hname\ was thus relatively UV-bright (or X-ray weak) well before the optical flaring activity. This is also consistent with the evidence for \heii\ and the Bowen \niii\ lines in its pre-flare SDSS spectrum.

% we conclude that the source is X-ray weak, even if in UV band flux increased the order of magnitude.

We next performed a spectral analysis of the Swift/XRT data obtained after the detection of the second optical flare. Given the overall low count rate in the individual observations, the time-resolved XRT data were stacked in several ways. 
Specifically, we examined the data obtained:
(1) during the first observation;
(2) close to the optical peak (MJD = 60138);
(3) in all the available XRT observations (up to MJD = 60575); 
(4) before the X-ray flares (prior to MJD = 60400);
and (5) the data obtained near each of the X-ray flares, namely MJD = $60418-60499$ and $60513-60571$ (see Table \ref{tab:xspec_models}).
% We performed time-resolved spectroscopy for Swift observations and 
% model the observation during the optical peak MJD = 60138), all observation stacked together, observations 
% stacked from before MJD 64000, after MJD = 64000. We modeled each flaring event separately (between MJD 60418-60499 and 60513-60571), (see Table \ref{tab:xspec_models}).
We used \texttt{XSPEC} \cite[version 12.12.1,][]{1996ASPC..101...17A} to model each of the X-ray spectra obtained in these stacks with an absorbed power-law model (\textsc{phabs$\times$po}) spanning the entire $0.3-10\,\kev$ band. We fixed the Galactic column density to $N_{\rm H}=3.7 \times 10^{20}\,{\rm cm}^{-2}$, appropriate for the location of the source \citep{2016A&A...594A.116H}. 

\begin{figure}[h]
\center
\includegraphics[scale=0.27]{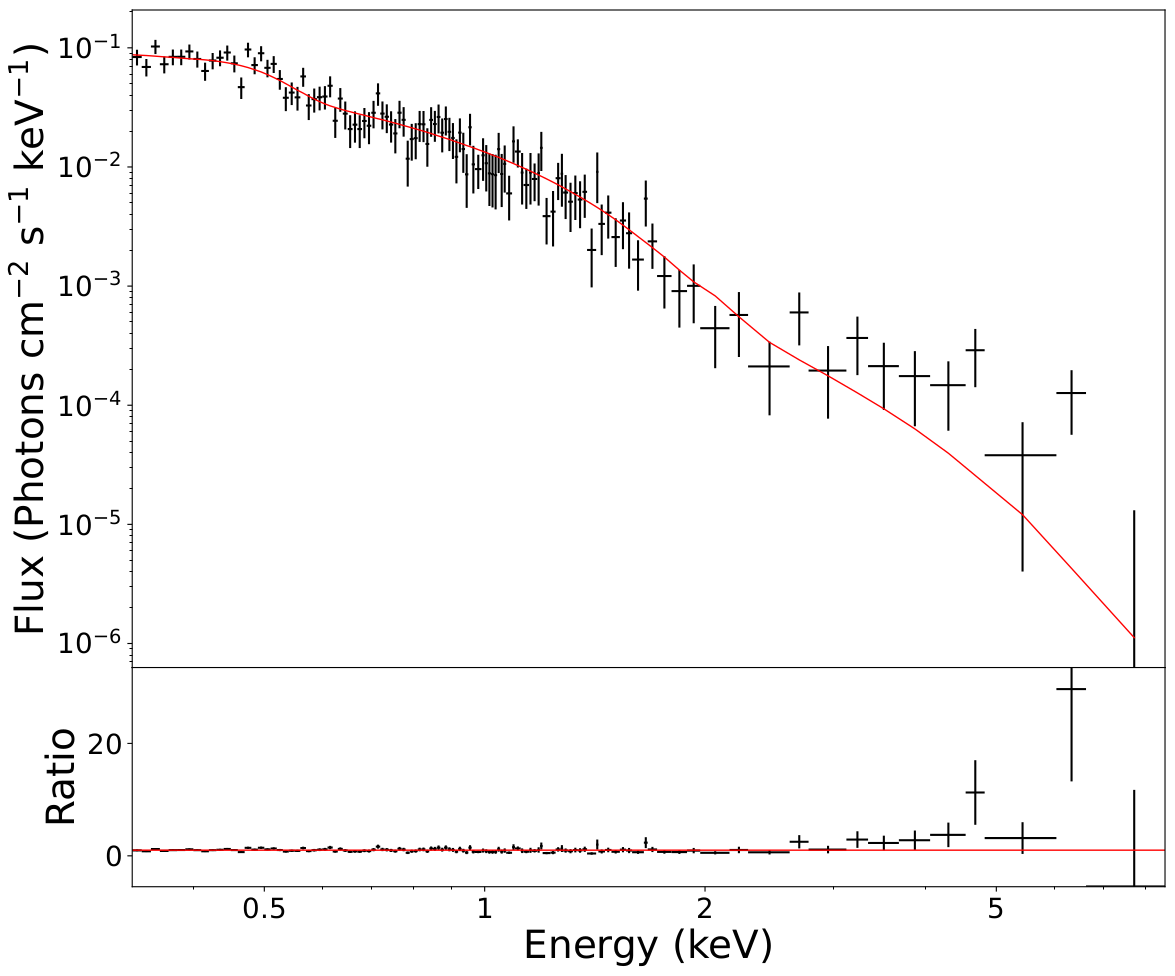}
\caption{The X-ray spectral energy distribution (SED) of \objname, obtained from the stacked Swift/XRT data (covering MJD = $60118-60575$; black data points), and the best-fitting absorbed power-law model (red line). The bottom panel shows the fitting residuals.
The best-fitting model indicates a relatively steep SED, with $\Gamma\simeq 4$. 
The spectral fits of all of the various temporal XRT stacks resulted in $\Gamma\gtrsim2.5$ (see Table~\ref{tab:xspec_models}).}
\label{fig:xspec-model}
\end{figure}

We present the X-ray SED obtained by stacking all the XRT data, and the corresponding best-fit model, in Figure \ref{fig:xspec-model}. There is a hint that another emission component contributes to the observed spectrum above $\approx3$ keV, but the data do not warrant any additional model components. 
All our spectral fits resulted in steep photon indices, with $\Gamma = 2.5$ for the first XRT observation and $\Gamma\approx4$ for all other fits. These are much steeper than what is seen in most low-redshift AGN \cite[typically $\Gamma \lesssim 2$,][]{2017ricci}, but consistent with extreme transients like 1ES 1927+654 \citep{Ricci2020_1ES}.

%Moreover, most of the flux is detected in the 0.3--2\,keV range for all time-resolved datasets. 
Focusing on the first Swift observation (MJD = 60118), carried out $\approx$20 days before the second main optical peak, we obtain $L(2-10\,\kev)=2.9\times10^{41}\,\ergs$.  
While this luminosity is, by itself,  within the range seen for low redshift AGN \cite[e.g.,][]{2015Aird,Ananna2022}, we could have expected a yet higher X-ray luminosity based on the (enhanced) NUV emission of \objname, as traced by UVOT. Combining the UVOT/UWM2 measurement [$L\nu (uvm2)=4.1\times10^{28}\,\ergs\,{\rm Hz}^{-1}$] with the relation of \cite{Lusso2016}\footnote{While the \cite{Lusso2016} relation is defined for $L_\nu (2500\,\AA)$ since the SED of AGN is typically only expected to vary by $\sim$5\% between 2200 and 2500\,\AA\ \citep{VandenBerk2001}, we use the UVOT luminosity as-is and assume $\Gamma=2.46$, as obtained from the \texttt{XSPEC} fit of this observation.} implies $L(2-10\,\kev) = 9.9\times10^{42}\,\ergs$, which is higher than the contemporary X-ray measurement by a factor of $\approx 30$.
%which is comparable to contemporary X-ray measurement.
% We conclude that during the second optical flare, \objname\ was UV-brighter (and/or X-ray weaker) than typical, persistent AGN, similarly to our conclusion regarding the pre-flare state.

%\citep{VandenBerk2001}, \textcolor{red}{we use the UVOT luminostiy as-is.} and assume $\Gamma=2.46$, implies $L(2-10\,\rm keV) = 5.1\times10^{41}\,\ergs$, which is higher than the contemporary X-ray measurement by a factor of $\approx10\times$.}

During both X-ray flares, which occurred well past the optical flare peak, the NUV luminosity stayed rather stable, at $\nu L_\nu (uvm2) \approx 1.9\times 10^{43}\,\ergs$. Following the same conversion to $L(2 \,\kev)$ and extrapolating to $2 - 10\,\kev$ using the corresponding photon indices ($\Gamma = 4.8$ and 4.12, see Table \ref{tab:xspec_models}), we obtained a NUV-based expected X-ray luminosity of $L(2-10\,\kev) = 1.6\times10^{42}\,\ergs$ and $L(2-10\ \kev) = 2\times10^{42}\,\ergs$, for the first and second X-ray flares, respectively. These NUV-based luminosities are higher than the observed $2-10\,\kev$ luminosity of \objname\ by factors of $\approx 80$ and $\approx 20$, for the first and second X-ray flares, respectively. 
% \textcolor{red}{Both $L(2-10\,\kev)$ are at least 20 times smaller than average $L(2-10\,\kev) = 4.8\times10^{43}\ \ergs$ calculated for the sample of 635 nearby AGN ($z<0.1$) \citep{2017ricci}.}

We therefore conclude that the (second) major optical flare associated with \objname\ is also UV-bright, and/or X-ray weak, even during the dramatic X-ray flares observed by Swift, compared with what is typically seen in persistent AGN. This (relative) UV-brightness is consistent with what is seen in other SMBH-related transients showing BF emission lines (i.e. \citealt{2021Malyali, Makrygianni2023}; see Section~\ref{sec:other-bff}).

We note that the relative weakness of the (soft) X-rays could be seen as a surprising feature in this, and other SMBH-related transients, given the strong emission from highly ionized species, which would necessitate a hard continuum with significant emission at $>50$ and indeed $>200$ eV (i.e., within the XRT band; see discussion in Section~\ref{sec:evolution-bff} above).
Comparing the X-ray light-curve (Fig.~\ref{fig:lc_long_zoom})  with that of the high ionization lines (Fig.~\ref{fig:lines_lc}, we notice that the periods of maximum $\heii+\niii$ and \fex\ emission, near the peak and the first bump in the broad-band optical light-curves, showed only modest X-ray brightening. The enhanced $\heii+\niii$ and \fex\ emission (peaking near MJD = 60450) could be associated with the first major X-ray flare, however, there is no enhancement in these lines during the second X-ray flare (which was similarly bright; peaking just prior to MJD = 60550).
Our high cadence spectral monitoring of \objname\ thus shows that the link between the broad-band (soft) X-ray emission and the BF-related and coronal line emission may not be straightforward, in contrast with the conclusion drawn by \cite{2024veres}. 
This is particularly evident by the second bright X-ray flare, which was not covered by the data presented in \cite{2024veres}, which does not seem to correspond to enhanced line emission.

\begin{deluxetable*}{lCCCCCCCC}
\label{tab:xspec_models}
\tablecaption{Best-fit parameters for the X-ray absorbed power-law spectral models}
\tablewidth{\textwidth}
\tablehead{
 \colhead{quantity} & 
 \colhead{first obs} & 
 \colhead{2nd optical peak} & 
 \colhead{all data} & 
 \colhead{pre-X-flares} & 
 \colhead{1st X-flare} & 
 \colhead{2nd X-flare} & 
 \colhead{both X-flares} 
 }
% \decimals
\startdata
$\Gamma$                                        & 2.46\pm0.51 & 3.32\pm0.33   & 4.05\pm0.05   & 3.98\pm0.08   & 4.80\pm0.14     & 4.12\pm0.07   & 4.11\pm0.07     \\
Norm ($10^{-4}\,{\rm counts}\,{\rm s}^{-1}$)    & 1.59\pm0.4  & 2.28\pm0.46   & 2.12\pm0.08   & 1.53\pm0.08   & 2.36\pm0.24     & 4.40\pm0.22     & 3.32\pm0.17     \\
$F_{ 0.3 - 2\,\kev}$ ($10^{-12}\,\ergcms$)      & 0.43\pm0.43 & 0.87\pm0.87   & 1.18\pm1.18 & 0.78\pm0.78 & 1.97\pm1.97   & 2.43\pm2.43   & 1.85\pm1.85   \\
$F_{\rm 2 - 10\,\kev}$ ($10^{-12}\,\ergcms$)    & 0.21\pm0.21 & 0.098\pm0.098 & 0.03\pm0.03 & 0.01\pm0.01 & 0.007\pm0.007 & 0.038\pm0.038 & 0.033\pm0.033 \\
$L_{0.3-2\,\kev}$ ($10^{41}\,\ergs$)            & 12.6            & 25.4              & 34.6            & 22.7            & 57.5              & 70.9              & 54.3 \\
$L_{2-10\,\kev}$ ($10^{41}\,\ergs$)   & 6.1             & 2.86              & 0.89            & 0.40            & 0.20              & 1.12              & 0.95\\  
\enddata
% \tablecomments{The models are of the form \todo}
\end{deluxetable*}

\section{The nature of the flare}
\label{sec:nature}

\begin{figure}
\centering
\includegraphics[scale=0.35]{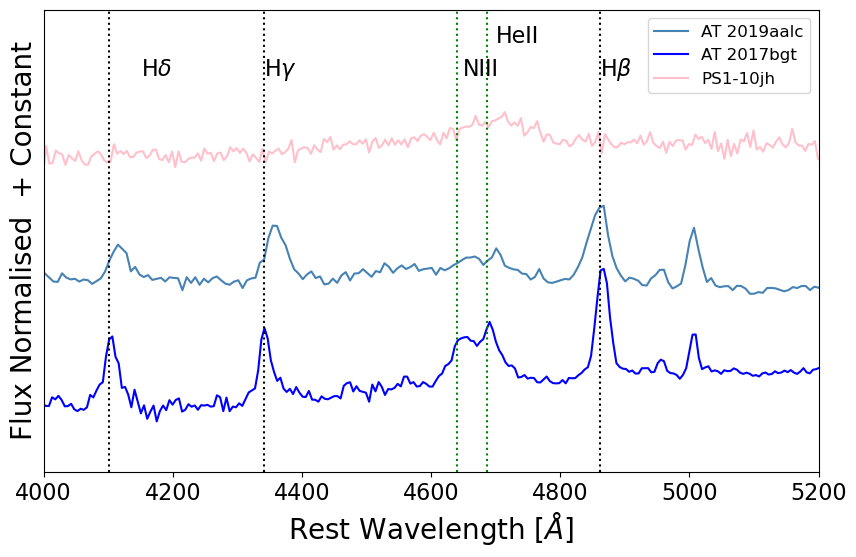}
\caption{Comparison of the optical spectrum of \objname\ (light blue) with two other SMBH-powered transients: the BFF AT\,2017bgt \cite[dark blue][]{2019Trakhtenbrot} and the optically-selected TDE PS1-10jh \cite[pink][]{2012gezari,2022Charalampopoulos}. 
All plotted spectra were taken about a month after the optical peak of each flare, and are normalized, continuum subtracted \& shifted (for clarity). The spectral features of \objname\ resemble those of the BFF AT\,2017bgt, particularly in terms of the $\heii+\niii$ emission feature.}
\label{fig:spec_comp_bff_tde}
\end{figure}

In this Section we discuss the nature of the recurring brightening events observed in the nucleus of \hname, comparing it to other main observed types of SMBH-related transient phenomena. To contextualize this discussion, in Figure \ref{fig:spec_comp_bff_tde} we show an optical spectrum of \objname\ along with representative spectra of other relevant SMBH-related transients, all acquired about a month following the (brightest) optical peak: the canonical TDE PS1-10jh \citep{2012gezari, 2022Charalampopoulos}\footnote{The spectrum of PS1-10jh was downloaded from the Weizmann Interactive Supernova Data Repository (WISeREP; \citealt{2012yaron}).} and the archetypical BFF AT\,2017bgt \citep{2019Trakhtenbrot}.
Note that \objname\ cannot be considered a CL-AGN, as the broad emission lines seen in the spectra obtained during the past few years of dramatic variability were also seen in the archival SDSS spectrum taken back in 2008, and there is no data to support a drastic spectral transition in the X-ray regime.
We also recall that the recent \cite{2024veres} study concluded that \objname\ presents several characteristics of a BFF while showing no clear evidence that would decisively link it to a (single, UV-optical) TDE.
As we discuss below, we broadly agree with the findings of \cite{2024veres}.

\subsection{Comparison to other BFFs}
\label{sec:other-bff}

%\objname\ is a SMBH-related transient presenting Bowen fluorescence features and (at least) two major UV-optical flares.  
\objname\ is a transient presenting Bowen fluorescence features and (at least) two major UV-optical flares. 
We compare it to other Bowen fluorescence flares (BFFs), specifically, the event used to identify this class---AT\, 2017bgt \citep{2019Trakhtenbrot}, and the BFF AT\,2021loi \citep{Makrygianni2023}, as well as the other cases studied by \citet[][F01007–2237]{2017tadhunter}, \citet[][OGLE17aaj]{Gromadzki2019}, and \citet[][AT\,2019avd]{2021Malyali}. 
%\toref.

\begin{figure*}
\centering
\includegraphics[scale=0.7]{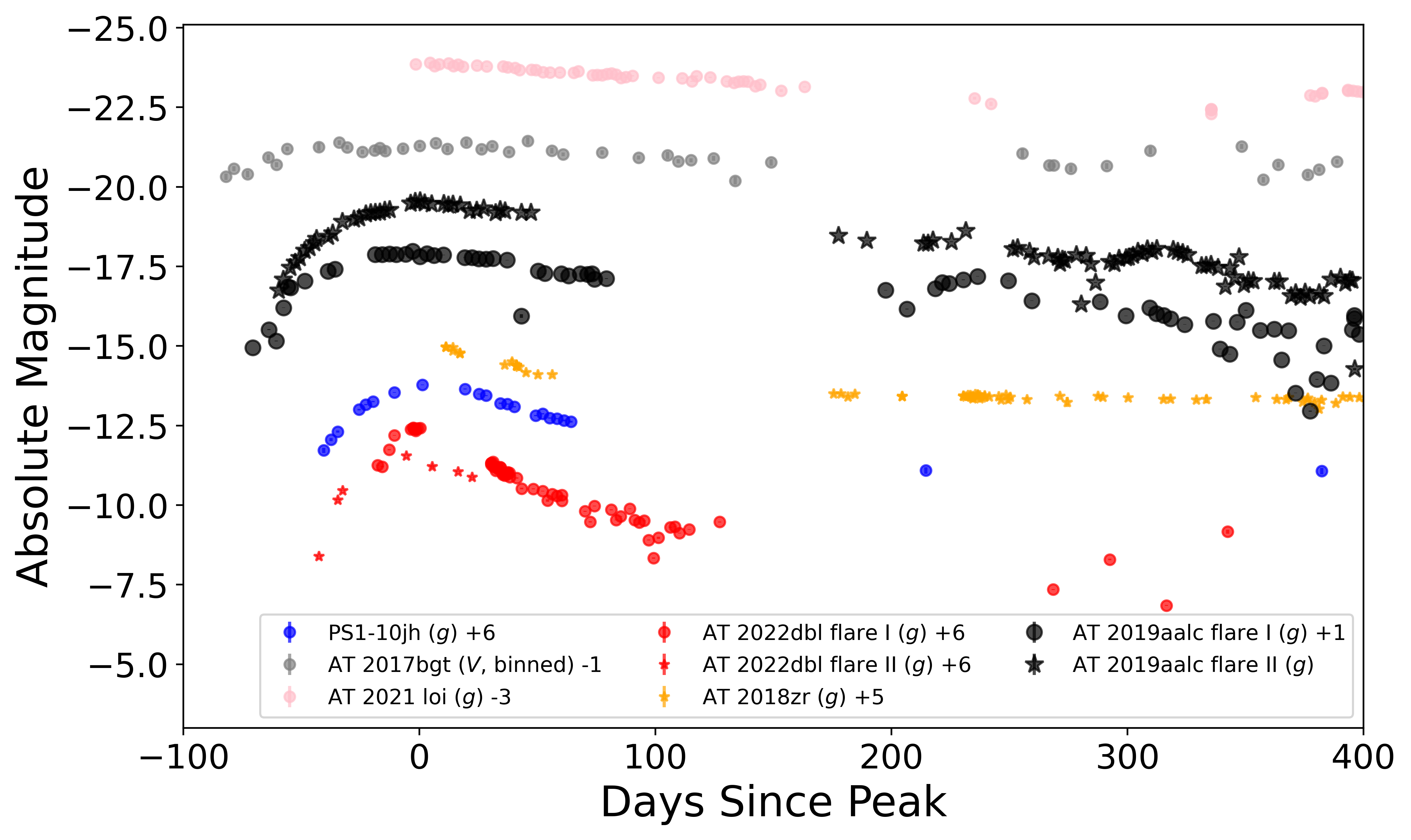}
\caption{Comparison of the optical light-curves of some of the BFFs known so far, and other selected nuclear transients. For \objname\ we show the light-curves of both main flares. In addition we show the BFFs AT\,2017bgt \citep{2019Trakhtenbrot} and AT\,2021loi \citep{Makrygianni2023}; the optically-selected TDE PS1-10jh \citep{2012gezari}; the repeating, partial TDE candidate AT\,2022dbl \citep{Lin2024_pTDE}; and the TDE candidate AT\,2018zr \citep{2019Holoien}, which showed a relatively slow dimming for TDEs. 
\objname, like other BFFs, shows a slower dimming compared with TDEs, accompanied by significant rebrightening events, $\gg100$ days post-peak.}
\label{fig:slopes_comp}
\end{figure*}

Figure~\ref{fig:slopes_comp}\ compares the optical light-curve of \objname\ with some of these other BFFs (extending Fig. 11 in \citealt{Makrygianni2023}).
We first note that while several published BFFs show evidence for recurring brightening over timescales of about a year\footnote{See also the more recent study of the transient brightening event(s) observed in F01007–2237, by \cite{2021tadhunter}.}, the dual flares seen in \objname\ are much more pronounced. In particular, none of the previously published BFFs had a second flare that is brighter than the first (``primary'') one. 
Comparing the strength of the (brighter) UV flares, defined by the ratios of the near-peak UVOT/$uvm2$ to the archival GALEX/NUV measurements, \objname\ showed a relatively mild UV flux increase, by a factor of $\approx 10$, compared with $\approx 75$ for AT\,2017bgt and $\approx 20$ for AT\,2021loi. However, the optical flare strengths exhibit a conflicting trend, with \objname\ increasing by a factor of $\approx4$, AT\, 2017bgt by a factor of $\approx1.5$, and AT\,2021loi by a factor of $\approx2$.
As for the post-peak decline of the UV-optical light-curves, Figure~\ref{fig:slopes_comp} demonstrates that BFFs tend to decline rather slowly. \objname\ is the only one with additional major re-flaring during the decay.

In terms of spectroscopy, in both \objname\ and AT\,2021loi the archival, pre-flare optical spectra show signatures of enhanced pre-flare emission of \HeIIop\ and \NIII, compared to normal AGN. 
For AT\,2021loi, the strength of these features relative to \hb\ is  $\Fbf/\Fhb\approx0.6$, while for \objname\ it is $\Fbf/\Fhb\approx0.3-0.4$ (for the SDSS and Keck spectra). 
In F01007–2237, the \heii\ and \niii\ emission was also enhanced prior to the optical flare(s), however, that system lacked clear signs of broad Balmer line emission in the archival, pre-flare spectroscopy.
In post-flare spectroscopy for the BFFs AT\,2021loi and AT 2017bgt, and for \objname, we find several similar features, including 
broad Balmer lines with $\fwhb\ \approx 2000-3000\,\kms$, enhanced $\heii+\niii$ emission, and the presence of coronal emission lines, such as \fex. 
All these lines have BLR-like line widths, which means they originate from high-density gas located relatively close to the SMBH. 
For \objname, we demonstrated that the line-emitting region(s) are indeed close to the SMBH by the light-curves of the lines, which broadly resemble the broad-band optical light-cuves.
The relative strength of the BF-related $\heii+\niii$ emission in \objname\ is lower than what was found for some of the previously studied BFFs, with $\Fbf/\Fhb\gtrsim1.4$ near the peak of the (second) optical flare, compared with $\approx1.1$, 0.9 and 0.5 (for AT\,2021loi, OGLE17aaj, and AT\,2017bgt, respectively).
%
% However, as mentioned previously, we highlight the lack of Balmer broad components in F01007–2237 \citep{2017tadhunter} and in OGLE17aaj \citep{Gromadzki2019}. 
The presence of relatively broad \fex\ in \objname\ is similar to other SMBH-related transients with BF emission, including the BFFs AT\,2017bgt and AT\,2021loi, as well as AT\,2019avd \citep{2021Malyali}. F01007–2237 exhibited post-flare [Fe\,\textsc{vii}]\,$\lambda$6087 broad emission \citep{2021tadhunter}.

In terms of its relatively weak X-ray emission, \objname\ is again similar to the BFFs AT\,2017bgt (which had $L(2-10\,\kev)\approx 10^{42}\,\ergs$) and AT\,2021loi (which had no X-ray emission detected). 

In summary, \objname\ provides further key evidence in support of several properties that seem to be common for (well observed) BFFs, including extreme UV flares, rebrightening after the first, main flare, shallow decrease of the optical and UV luminosity, appearance of broad coronal lines, and rather weak X-ray activity in comparison to normal AGN.

\subsection{Relevance of TDEs}
\label{sec:tde}

The first optical flare identified in \hname\ was classified as TDE candidate \cite[see, e.g.,][]{vanvelzen2024}.
This was justified by the coincidence of the transient source with the center of the host galaxy, its blue color \citep{2024veres} and the broad emission feature that spectrally coincided with \HeIIop---all of which are typical of TDE candidates identified in the UV-optical regime \cite[e.g.,][``canonical TDEs'' hereafter]{2012gezari,  2014arcavi, Gezari21_rev, 2021SSRvvanVelzen}.

In terms of the optical light-curve behavior, we first note that the post-peak(s) decline phase(s) observed in \objname\ are much slower than most canonical TDEs, and are only comparable to the shallowest (slowest) declines seen in such systems (e.g., Table~1 in \citealp{2020vanvelzen}).
This is demonstrated by the light-curve of the optically-selected TDE PS1-10jh \citep{2012gezari} that is shown in Fig.~\ref{fig:slopes_comp}. 
% again, consistent with the behavior found for most previously reported BFFs.

Moreover, the presence of two bright peaks and additional, milder bumps during the decline phase is also not typical for canonical TDEs.
While such recurring flaring activity is generally expected for (some configurations of) a pTDE \cite[e.g.,][]{Coughlin2019,Zhong2022}, we stress that pTDE light-curves are expected to decay even faster than those of canonical (full) TDEs, making the observed light-curve of \objname\ yet less compatible with a pTDE scenario.
The nuclear transient AT\,2022dbl, which was interpreted as a repeated, partial TDE \citep{Lin2024_pTDE}, indeed presented two distinct (optical) flares, both of which are presented as part of Fig.~\ref{fig:slopes_comp}. 
However, in that case, the first flare was brighter than the second one, while in \objname\ the opposite is true: the second flare is brighter than the first one. 
In the case of AT\,2020vdq \citep{2023somalwar}, the second flare was brighter, as we see for \objname\, but the duration of the second event is significantly shorter (see Figure 1 in \citealt{2023somalwar}).
Another source in which the optical light-curve showed a rich structure is AT\,2021aeuk, which (similarly to \objname) occurred in a previously-known AGN \citep{2025Sun_rpTDE}. This transient showed three flares (the precursor flare and two main ones) within four years with visible bumps during the decay phase and was interpreted to be a repeating pTDE. Unlike \objname, however, in this source, the first major flare is again brighter than the second one, and there are no signs of enhanced BF or coronal line emission.

Another scenario that was proposed to produce dual flares near SMBHs involves the tidal disruption of a binary star system  \citep{Mandel2015}. Such a scenario is, however, unlikely to explain the recurring flares in \objname, as the expected timescale between two peaks is only $\lesssim$150 day (assuming main-sequence binary stars), compared with the $\approx 4$ years separating the two main optical flares observed for \objname.

In terms of spectroscopy, we first note that the broad emission feature coincident with the \HeIIop\ spectral region is not as broad in \objname\ as it is in canonical TDEs ($\sim3000$ vs. $\gtrsim10,000\,\kms$, respectively). Instead, it shows a clear double-peaked (or bimodal) nature, which can be accounted for by the blending of two lines, each with a width similar to that of the broad Balmer lines. 
This is, again, one of the features that sets apart BFFs and canonical TDEs, although several TDEs are known to also show BF emission lines \citep{Holoien2016, Leloudas2019, 2022Charalampopoulos}. 
Coronal lines may also be present in some other SMBH-related transients, including TDEs (e.g., AT\,2019qiz; \citealt{2023Short}) and more ambiguous events (e.g., AT\,2019avd; \citealt{2021Malyali}). In these cases, however, they are significantly narrower than what is observed in BFFs, suggesting the coronal lines are emitted further from the central (ionizing) engine, perhaps similarly to normal, persistent AGN with such emission lines. 
Finally, we recall that \objname\ occurred in a previously-known AGN with broad Balmer lines, which again differentiates it from canonical TDEs identified in inactive galaxies.
While TDEs may occur in, and/or interact with, pre-existing accretion disks \cite[e.g.,][]{Chan2019,Ricci2020_1ES}, 
we note that the number of such ``TDE-in-AGN'' candidate systems identified so far is a factor of a few-to-ten lower than what is expected from theory \citep{2015Merloni,2024kaur}, suggesting that our (theoretical) understanding of such complex systems is still limited.

Another feature seen in some, but definitely not all canonical TDEs, is enhanced radio emission that may be detected with a significant delay after the optical peak \cite[e.g.,][]{2020Alexander, Horesh2021,Sfaradi2022,2024Cendes}. Thus, late-time radio emission from this source would give us more clues about the nature of this source.
After the first optical flare of \objname, its radio flux increased by a factor of 2 between two VLASS observations separated by $\approx$2 years \citep{vanvelzen2024}. The radio emission in the former post-flare VLASS observation was already somewhat higher than the archival, pre-flare FIRST measurement, but only by a factor of $\approx1.5$ or so. Moreover, \objname\ showed a relatively constant radio flux during the second flare, with the appearance of 18 GHz emission \citep{2024veres}. 
Therefore, while \objname\ shows some brightening in its radio emission, which may be interpreted as a newly launched outflow or jet \cite[see detailed analysis in][]{2024veres}, its radio behavior differs from that of canonical TDEs.

\subsection{Recurring flares in AGN disks}
\label{sec:recurring-flares-model}

The recurring flares and bumps seen in \objname, associated with the previously-known AGN \hname, along with the quasi-simultaneous BF emission features that are driven by EUV continuum emission, motivate us to consider scenarios for recurring inabilities and/or disruptions in the inner parts of AGN accretion flows.  

While the timescales of outbursts driven by radiation pressure instabilities in standard thin AGN disks are expected to be of the order of thousands of years, recent studies have shown that they may be shortened significantly, to even tens of years, if one modifies the disk models to include various viscosity laws and/or magnetic fields \citep{grzedzielski2017, sniegowska2023}.
Moreover, setting the outer boundary radius of the unstable part of the accretion disk also allows for further shrinking of the typical outburst timescale.

\begin{figure}
\center
\includegraphics[width=0.475\textwidth]{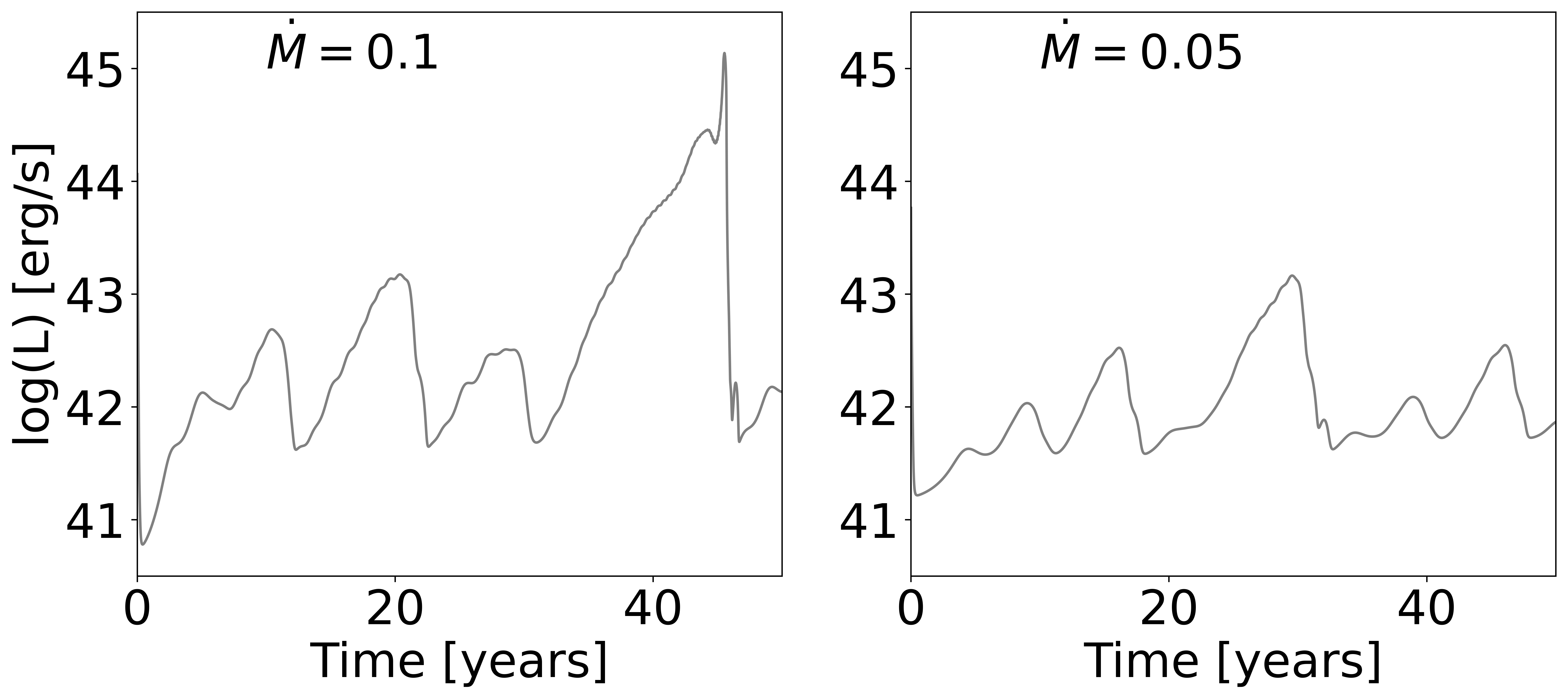}
\caption{Model light-curves of radiation pressure driven instabilities in a thin accretion disk, showing recurring flares and ``bumps'', reminiscent of what is seen in \objname. The two models shown assume a SMBH mass of $\mbh=1.35\times10^{7}\,\Msol$, as estimated for \objname, the outer radii of the unstable regions within the disks $R_{\rm out} = 50\,R_{\rm schw}$, but differ in their accretion rates which are assumed to be either $\dot{M}_{\rm BH} = 0.1\,\mpyr$ (left) or  $0.05\,\mpyr$ (right). 
The structure of the model light-curve seems to become richer when the accretion rate is lower.}
    \label{fig:ad-model}
\end{figure}

To explore the relevance of radiation pressure instabilities in AGN disks to the events seen in \objname, we used the GLADIS code\footnote{https://github.com/agnieszkajaniuk/GLADIS} \citep{2020mbhe.confE..48J}, which was used for calculating radiation pressure instabilities for X-ray binaries and changing-look AGN \citep{sniegowska2023}.
This one-dimensional model assumes a viscous torque proportional to the total pressure, $\alpha P_{\rm tot}$, a vertically averaged accretion disk, a viscosity parameter of $\alpha=0.01$, and a decrease of the outer radius within which instabilities may develop ($R_{\rm out}$ hereafter). For simplicity, for the purposes of the current discussion, we did not modify the disk's vertical structure by a magnetic field. We assumed $\mbh=1.35\times10^7\,\Msol$ and $\dot{M}_{\rm BH} \approx 0.1\,\mpyr$ based on the archival SDSS spectrum (the latter assumes a radiative efficiency of $\eta=0.1$). To be able to explore a range of model outcomes, we calculated a small grid of models with  $\dot{M}_{\rm BH} = 0.05$, 0.1, 0.15, and 0.2, and with two boundary radii of $R_{\rm out} = 30$ and 50 $R_{\rm Schw}$.\footnote{$R_{\rm Schw}$ is the Schwarzschild radius of the SMBH, which implicitly assumes the estimated \mbh.} This inner part of the disk would emit the highest-energy (disk-related) radiation, and therefore is likely the most relevant for driving the BF (and coronal) line emission seen in \objname. 

Among the various calculated models, we find those with $\dot{M}_{\rm BH} = 0.1$  $\dot{M}_{\rm BH} = 0.05$ to be the most relevant in terms of timescales, and we present the output of those models in Figure \ref{fig:ad-model}. For smaller values of $\dot{M}$, we obtain timescales comparable to the observations, in terms of major flares separated by years, combined with a complex, ``bumpy'' structure reminiscent of the observed broad-band optical light-curve of \objname. 
% Moreover, from the model, we obtain complex, bumpy structures, also present in the observations.

We stress that while this sort of AGN disk radiation instabilities model offers a possible avenue to explain recurring flares in AGNs, it also suffers from several important caveats. 
First, the lower luminosity depicted by the model output (Fig.~\ref{fig:ad-model}) stems from the fact that the luminosities represent only the unstable part of the disk, i.e. the integrated radiation out to $R_{\rm out}$ (30 or 50 $R_{\rm Schw}$). Integrating over the entire disk would increase the luminosity but would act to dampen the apparent flares and bumps.
Second, the addition of magnetic fields could further reduce the typical timescales of the recurring flares, but on the other hand, an increasingly strong magnetic field would act to stabilize the disk and dampen the disk variability \citep{sniegowska2023}.
Third, the current version of the model cannot reproduce multi-wavelength properties, such as (soft) X-ray or radio emission.
Perhaps most importantly, radiation instabilities generally produce flares in which the rise time is drastically longer than the decline time (as indeed seen in Fig.~\ref{fig:ad-model}). This is the opposite of what is seen in \objname\ and in most other AGN-or SMBH-related transients reported to date. It is possible that ongoing transient searches are biased against slowly brightening AGNs (with sudden post-peak drops), for which such models may be indeed more relevant.

Other possible types of instabilities, like those related to hydrogen ionization \citep{noda2018}, are generally expected to occur further out in the accretion flow \cite[see, e.g., Fig.~1 in][]{janiuk2011}, and thus the associated timescales are expected to be significantly longer (perhaps by orders of magnitudes; see, e.g., \citealt{2009Hameury}). 
A more detailed discussion of various models for accretion flow instabilities is beyond the scope of the present work.

\section{Conclusions}
\label{sec:conc}

We presented multi-epoch observations of the transient and dramatic brightening event(s) known as \objname\, which occurred in the previously-known AGN \hostname\, with two distinct main flares observed in 2019 and 2023. We focused on a spectroscopic time series showing enhanced Bowen fluorescence and coronal line emission, in addition to (partially public) UV and optical photometric data, and the most up-to-date X-ray and IR data. Our main findings are as follows:

\begin{enumerate}
   
      \item The transient and extreme variability seen in \objname\ occurred in a previously-known, broad-line (unobscured) AGN, as determined from archival SDSS spectroscopy. That archival pre-flare spectrum already showed some mildly enhanced \HeIIop\ and \NIII\ line emission (see Section~\ref{sec:evolution-bff} and Fig.~\ref{fig:key_spec_comp}).
      
      \item During the first optical flare, the optical flux from the nucleus increased by a factor of $\approx2\times$. 
      During the second optical flare, the optical flux increased by $\approx6\times$, while the near-UV flux increased by a factor of $\approx10\times$.
      % , as determined by comparing our Swift data to archival GALEX data. 
      The optical and UV light-curves show milder rebrightening events occurring $\approx$230 days after both the first and second main flares (in 2019 and 2023; see Fig.~\ref{fig:lc_long}).
      
      \item The overall pace of dimming following the two main flares was relatively slow, specifically slower than what is expected for canonical (partial) TDEs. The decline is marginally consistent with some of the slowest-evolving TDE candidates observed to date (see Figure \ref{fig:slopes_comp}). 

      \item The WISE MIR light-curve also shows two clear flares following the optical ones, with delays of 210-230 days. 
      This suggests that the MIR emission rises from reprocession of UV/optical radiation by dusty circumnuclear gas, the presence of which is also supported by AGN-like MIR color of \objname\ (see Figure~\ref{fig:lc_long} and Section \ref{sec:photometry-all-wv}).

      \item After the second flaring event, we observe enhanced, relatively broad ($\approx2000-3000\,\kms$) line emission from the \HeIIop\ and \NIII\ transitions, characteristic of Bowen fluorescence (BF). The BF emission generally follows the same trends as seen in the optical-UV photometric light-curve, but in fact slightly precedes it (Fig.~\ref{fig:lines_lc}). This is consistent with what is expected if the BF features are driven by the EUV emission originating from the innermost parts of an accretion flow, while the NUV-optical continuum emerges from larger scales in the disk. This is the first SMBH-related BF flare (BFF) where such detailed temporal information is available (see Section \ref{sec:balmer-lines-evolution}).
      
      \item We report the presence of relatively broad coronal lines, including \fexiv, \fex\ and [Fe\,\textsc{vii}]\,$\lambda$6087.
      The \fex\ line is significantly broader than the narrow forbidden lines, reaching FWHM $\approx$ 2800\,\kms. 
      This provides further evidence for significantly enhanced EUV emission from the innermost parts of the accretion flow being reprocessed by BLR-scale, high-density gas (see Section \ref{sec:coronal-lines}).

      \item The transient showed two dramatic X-ray flares during the prolonged dimming phase of the second optical flare (MJD = 60443 and MJD = 60551). The accumulated Swift/XRT data suggests a SED that is much softer than what is seen in most AGNs, with $\Gamma\gtrsim2.5$. Overall, \objname\ shows a lower X-ray emission than what is expected from UV--to--X-ray relation of normal AGN, even during its dual X-ray flares. Indeed, during the X-ray flares, the UV luminosity was nearly constant (see Section~\ref{sec:x-ray-evolution}.)
      
      \item The strong and broad ($\approx$2000\,\kms) \HeIIop\ and \NIII\ lines, the slowly declining UV/optical light-curve, the recurring brightening events, and the UV brightness (and/or X-ray weakness) are all in general agreement with what is seen in other BFFs reported so far (see Section~\ref{sec:other-bff}). In contrast, the light-curve and the spectral features of \objname\ are markedly inconsistent with those of canonical, optically selected TDEs, and even with the expected class of repeating TDEs.
      
      \item We considered the possibility that the recurring flares of \objname\ could be explained by radiation pressure instabilities in the inner part of the accretion disk. While we were able to obtain some properties of the event (the required timescales and bumpy structure), we did not cover all of them (for example we did not obtain fast rise/slow-decay shape of the modeled flares) (see Figure~\ref{fig:ad-model} and Section~\ref{sec:recurring-flares-model}). 
   \end{enumerate}

% COMMENT
%
The basic association of \objname\ with a BFF driven by a flare from an enhanced accretion event in a previously known AGN is in line with the findings of \cite{2024veres}, who studied a rich collection of multi-epoch and multi-wavelength data. Our study further highlights the temporal behavior of the Bowen and coronal emission lines and follows the declining phase of the second major flare, including the minor rebrightening events.

\objname\ joins the growing sample of known BFFs, and indeed shares many of their familiar properties. Moreover, together with AT\,2021loi \citep{Makrygianni2023} it adds to the evidence that BFFs with such properties occur in already-active SMBHs (i.e., in AGNs). Thanks to our intensive spectral monitoring, we were able to track the variable BF and coronal line intensity, providing the best opportunity to date to link these high-ionization features to the EUV emission which likely originates at the heart of the highly variable accretion flow, but which is otherwise inaccessible.
\objname\ also showed unique peculiarities, such as the dual optical/UV flare, the recurring milder bumps in the UV/optical light-curve during the dimming phase of the second flare, and the dual X-ray flares (which have no corresponding UV/optical variability). 
% With Swift X-ray monitoring we were able to track reflaring events.

The underlying cause for the dramatic events seen in \objname\ remains unclear, as it is still challenging to associate it with a particular type of physical mechanism. 
% From archival SDSS optical spectrum and multi-wavelength data \citep{2023TNSAN.195....1G}, we know that the host is an unobscured AGN.
Given the rich data collected for this source \citep{2023TNSAN.194....1V, 2024veres}, and its unique multi-wavelength properties (including the associated radio brightening and neutrino event \citep{Reusch2023, vanvelzen2024}), it may very well turn out to be a testbed for some of the models put forward to explain SMBH-related transients, particularly in already-active SMBHs. 

The relatively short-term spectral and multi-wavelength changes seen in \objname, particularly once the initial optical flare has dimmed, underscore the need for continued, high-cadence, and multi-wavelength monitoring of SMBH-related transients. We are indeed continuing our monitoring campaign of \objname, for whatever additional surprises and insights it may provide.

\begin{acknowledgements}
We would like to thank Matthew Graham for providing us a reduced version of the Keck/LRIS spectrum of \objname; Giovanni Miniutti for his help and insightful comments regarding the X-ray observations; and Hagai Netzer for valuable comments.
We thank the Swift team for approving and conducting the target of opportunity observations.
M.S., B.T., S.F., and I.A. acknowledge support from the European Research Council (ERC) under the European Union's Horizon 2020 research and innovation program (grant agreements 852097 and 950533) and from the Israel Science Foundation (grant numbers 1849/19 and 2752/19).
L.M. acknowledges support through a UK Research and Innovation Future Leaders Fellowship (grant number MR/T044136/1).
I.A. is a CIFAR Azrieli Global Scholar in the Gravity and the Extreme Universe Program and acknowledges support from that program, the United States—Israel Binational Science Foundation (BSF), and the Israeli Council for Higher Education Alon Fellowship. B.P. acknowledges the support from the Polish National Science Center grant No. 2021/41/B/ST9/04110. 
This research was supported by the Excellence Cluster ORIGINS and by the Munich Institute for Astro-, Particle and BioPhysics (MIAPbP), which are funded by the Deutsche Forschungsgemeinschaft (DFG, German Research Foundation) under Germany's Excellence Strategy - EXC 2094 - 390783311.
%
% This research was supported by the Munich Institute for Astro-, Particle and BioPhysics (MIAPbP) which is funded by the Deutsche Forschungsgemeinschaft (DFG, German Research Foundation) under Germany´s Excellence Strategy – EXC-2094 – 390783311
%
 B.T. acknowledges the hospitality of the Instituto de Estudios Astrof\'isicos at Universidad Diego Portales, and of the Instituto de Astrof\'isica at Pontificia Universidad Cat\'olica de Chile.

This work made use of the ZTF forced-photometry service, which was funded under the Heising-Simons Foundation grant No. 12540303 (PI: Graham). 
This work makes use of data from the Las Cumbres Observatory global telescope network. The Las Cumbres group is supported by NSF grants AST-1911151 and AST-1911225 and NASA Swift grant 80NSSC19k1639. 

This work also made use of the NASA/IPAC Extragalactic Database (NED), which is funded by the National Aeronautics and Space Administration and operated by the California Institute of Technology, and of data, software, and web tools obtained from the High Energy Astrophysics Science Archive Research Center (HEASARC), a service of the Astrophysics Science Division at NASA/GSFC and of the Smithsonian Astrophysical Observatory’s High Energy Astrophysics Division.

\end{acknowledgements}

% \vspace{3in}
\software{{\tt AstroPy} \citep{astropy:2013, astropy:2018, astropy:2022}, {\tt Matplotlib} \citep{Hunter:2007}, {\tt NumPy} \citep{harris2020array}, {\tt SciPy} \citep{2020SciPy-NMeth}, {\tt lcogtsnpipe} \citep{Valenti2016}, {\tt PyQSOFit} \citep{pyqsofit}, {\tt Lmfit} \citep{2016lmfit} }
\facilities{Las Cumbres (FLOYDS), Keck (LRIS), WISE, Swift (UVOT and XRT), ZTF, ATLAS, VLA}

%%%%%%%%%%%%%%%%%%%%%%%%%%%%%%%%%%%%%%%%%%%%%%%%%%%%%

\clearpage
\appendix

%%%%%%%%%%%%%%%%%%%%%%%%%%%%%%%%%%%%%%%%%%%%%%%%%%%%%
%\newpage
\section{Power-law fits of the declining photometric light-curves}
\label{app:lc_fits}

\begin{figure*}[h!]
\centering
\includegraphics[scale=0.8]{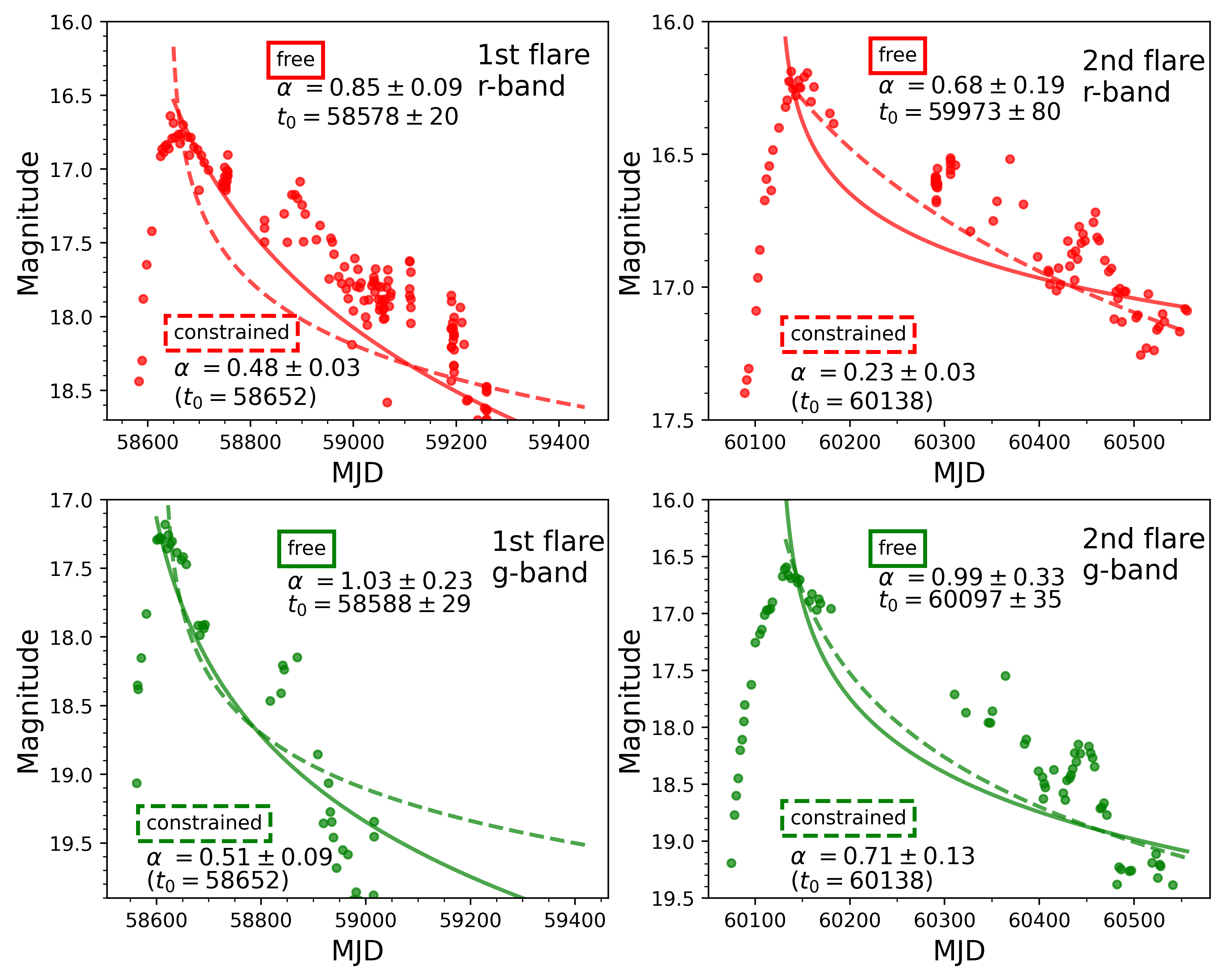}
\caption{Power-law fits to the dimming optical light-curves of the two optical flares. The {\it top} and {\it bottom} panels show the light-curves of \objname\ obtained with the ZTF ($r$ and $g$ bands) for the first flare ({\it left} panels) and second flare ({\it right} panels). For each flare and band, we provide two power-law fits: one where $t_0$ is fixed at the epoch of optical flare detection (marked by dashed vertical lines), and one where $t_0$ is free to vary (solid lines).
Note the different ranges of the vertical axes in the various panels. 
The power-law models do not provide a good fit for the observations. 
% best-fit power laws provide a good 
}
\label{fig:lc_pl_fits}
\end{figure*}

Here we provide details regarding the fitting of the optical light-curves of the two optical flares, using data from ZTF ($g$ and $r$ bands). 
The data, best-fitting models and their parameters are shown in Figure \ref{fig:lc_pl_fits}, and the best-fitting parameters are summarized in Table~\ref{tab:lc_pl_fits}.

For each flare and band, we obtained two kinds of power-law fits: one where $t_0$ is fixed at the epoch of optical flare detection, and the other where $t_0$ is free to vary.

For the first flare, if we fix $t_0$ to the value of the peak of the optical flare (MJD = 58652), we obtain $\alpha =  0.51\pm 0.09$ for the $g$-band and $0.48\pm0.03$ for the $r$-band.
If we allow $t_0$ to vary, we obtain $\alpha = 1.03 \pm 0.23$ and $t_0$ that corresponds to MJD = 58588 for the $g$-band, or $\alpha = 0.85 \pm 0.09$ and $t_0$ corresponding to MJD = 58578 for the $r$-band). 
For both bands, the best-fitting ``free'' $t_0$ is consistent with the first optical peak, but these models do {\it not} provide a good fit for the data.

For the second flare, if we fix $t_0$ to the epoch the peak of the optical flare (MJD = 60138), we obtain $\alpha =  0.71 \pm 0.13$ for the $g$-band and  $\alpha = 0.23\pm 0.03$ for the $r$-band.
For models where $t_0$ is a free parameter, we obtain $\alpha = 0.99 \pm  0.33$ and $t_0$ corresponding to MJD = 60097 for the $g$-band or $\alpha = 0.68 \pm 0.19$ and $t_0$ corresponding to  MJD = 59973 for the $r$-band. 
Although both these $t_0$ are generally consistent with the observed peaks, the best-fitting models again provide an overall poor fit to the data. 

Taken at face value, our best-fitting power-law models indicate  that the emission from \objname\ dimmed at a rate that is slower than what is expected for canonic TDE. 
However we stress again that these power-law models fail to explain the declining phases of the light-curves, both in terms of overall shape and the structure (i.e., the bumps), for both flares and the two bands.  
The only way to significantly improve the agreement between the power-law models and the data is if the disruption is allowed to occur $>$200 days prior to the observed peaks, and with much brighter peak brightness, in sharp contrast with observations.

\begin{deluxetable*}{cc|cc|cc}
\tablewidth{\textwidth}
\tablecaption{Best-fit parameters for power-law fits of optical light-curves}
\label{tab:lc_pl_fits}
\tablehead{
\colhead{Flare} &  \colhead{Band}  & \multicolumn{2}{c}{$t_0$ fixed \tablenotemark{\small a}} & \multicolumn{2}{c}{$t_0$ free \tablenotemark{\small b}} \\ 
 & & $t_0$ & $\alpha$ & $t_0$  & $\alpha$
}
\startdata
1 & $g$ & $(0)$ & $0.51\pm0.09$ & $  83\pm29$ & $1.03\pm0.23$ \\
~ & $r$ & $(0)$ & $0.48\pm0.03$ & $  73\pm20$ & $0.85\pm0.09$ \\ 
\hline
2 & $g$ & $(0)$ & $0.71\pm0.13$ & $  22\pm35$ & $0.99\pm0.33$ \\
~ & $r$ & $(0)$ & $0.23\pm0.03$ & $-102\pm80$ & $0.68\pm0.19$  \\ 
\enddata
\tablecomments{In all fits, $t_0$ is given in days, relative to the date of respective flare detection, i.e. MJD = 58652 and 60138 for the first and second flare.}
\tablenotetext{a}{Fits where $t_0$ is fixed to the date of respective flare detection.}
\tablenotetext{b}{Fits where $t_0$ is free to vary.}
\end{deluxetable*}

% \begin{table}[h!]
% \centering
% \caption{Best-fit parameters for power-law fits of optical light-curves}
% \label{tab:lc_pl_fits_old}
% \begin{tabular}{lccccc}
% \hline
% \hline
% Flare &  Band  & \multicolumn{2}{c}{$t_0$ fixed} & \multicolumn{2}{c}{$t_0$ free} \\ 
%  & & $t_0^a$ & $\alpha$ & $t_0^b$  & $\alpha$ \\ \hline
% 1 & $g$ & $(0)$ & $0.54\pm0.44$ & $  83\pm29$ & $1.03\pm0.23$ \\
% ~ & $r$ & $(0)$ & $0.71\pm0.12$ & $  73\pm20$ & $0.85\pm0.09$ \\ 
% \hline
% 2 & $g$ & $(0)$ & $0.44\pm0.10$ & $  22\pm35$ & $0.99\pm0.33$ \\
% ~ & $r$ & $(0)$ & $0.19\pm0.06$ & $-102\pm80$ & $0.68\pm0.19$   \\ 
% \hline
% \end{tabular}
% \tablecomments{$a$ - $t_0$ fixed to be relative to the date of respective flare detection, i.e. MJD = 58505 and 60075, for the first and second flare; $b$ - $t_0$ given relative to the date of respective flare detection, i.e. MJD = 58505 and 60075.}
% \end{table}

%%%%%%%%%%%%%%%%%%%%%%%%%%%%%%%%%%%%%%%%%%%%%%%%%%%%%

\clearpage
\section{Time evolution of Balmer lines}
\label{app:evolution-balmer}

\begin{figure*}[h!]
    \centering
    \includegraphics[scale=0.5]{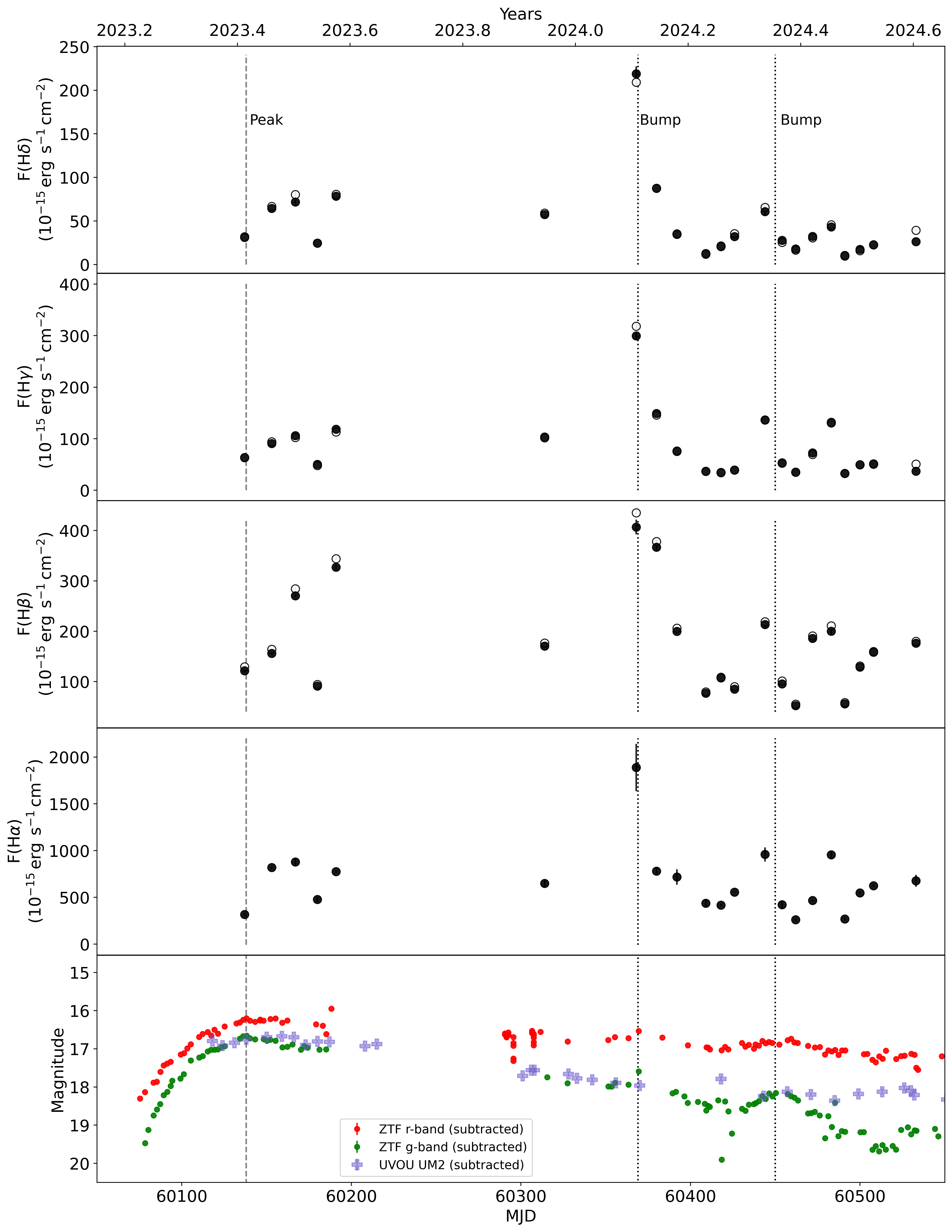}
    \caption{Time evolution of the flux of the broad Balmer emission lines following the second main optical flare of \objname. 
    As in Fig.~\ref{fig:lines_lc}, we show both measurements based on direct integration of the line flux density (except for \ha; empty symbols), and measurements derived from the spectral modeling (full symbols).
    The bottom panel shows the same photometric light-curve as in Fig.~\ref{fig:lines_lc}.
    Vertical lines mark the peak of the second optical flare (left) and the two ``bumps''.
    All broad Balmer lines show significant variations that can be generally associated with the variations seen in the broad-band light-curve, particularly during the two bumps.}
    \label{fig:balmer-evolution}
\end{figure*}

Figure \ref{fig:balmer-evolution} shows the time series of the fluxes of the broad Balmer lines: \Halpha, \Hbeta, \Hgamma, and \Hdelta. 

Their flux varies in a consistent way, including a (delayed) brightening during the optical flare itself,  but with an apparent lag of several weeks, and short periods of enhanced emission that coincide with the bumps seen in the broad-band optical light-curves.

%%%%%%%%%%%%%%%%%%%%%%%%%%%%%%%%%%%%%%%%%%%%%%%%%%%%%

\bibliography{AT19aalc}{}

\begin{thebibliography}{}
\expandafter\ifx\csname natexlab\endcsname\relax\def\natexlab#1{#1}\fi
\providecommand{\url}[1]{\href{#1}{#1}}
\providecommand{\dodoi}[1]{doi:~\href{http://doi.org/#1}{\nolinkurl{#1}}}
\providecommand{\doeprint}[1]{\href{http://ascl.net/#1}{\nolinkurl{http://ascl.net/#1}}}
\providecommand{\doarXiv}[1]{\href{https://arxiv.org/abs/#1}{\nolinkurl{https://arxiv.org/abs/#1}}}

\bibitem[{{Abazajian} {et~al.}(2009){Abazajian}, {Adelman-McCarthy}, {Ag{\"u}eros}, {Allam}, {Allende Prieto}, {An}, {Anderson}, {Anderson}, {Annis}, {Bahcall}, {Bailer-Jones}, {Barentine}, {Bassett}, {Becker}, {Beers}, {Bell}, {Belokurov}, {Berlind}, {Berman}, {Bernardi}, {Bickerton}, {Bizyaev}, {Blakeslee}, {Blanton}, {Bochanski}, {Boroski}, {Brewington}, {Brinchmann}, {Brinkmann}, {Brunner}, {Budav{\'a}ri}, {Carey}, {Carliles}, {Carr}, {Castander}, {Cinabro}, {Connolly}, {Csabai}, {Cunha}, {Czarapata}, {Davenport}, {de Haas}, {Dilday}, {Doi}, {Eisenstein}, {Evans}, {Evans}, {Fan}, {Friedman}, {Frieman}, {Fukugita}, {G{\"a}nsicke}, {Gates}, {Gillespie}, {Gilmore}, {Gonzalez}, {Gonzalez}, {Grebel}, {Gunn}, {Gy{\"o}ry}, {Hall}, {Harding}, {Harris}, {Harvanek}, {Hawley}, {Hayes}, {Heckman}, {Hendry}, {Hennessy}, {Hindsley}, {Hoblitt}, {Hogan}, {Hogg}, {Holtzman}, {Hyde}, {Ichikawa}, {Ichikawa}, {Im}, {Ivezi{\'c}}, {Jester}, {Jiang}, {Johnson}, {Jorgensen}, {Juri{\'c}}, {Kent}, {Kessler}, {Kleinman}, {Knapp},
  {Konishi}, {Kron}, {Krzesinski}, {Kuropatkin}, {Lampeitl}, {Lebedeva}, {Lee}, {Lee}, {French Leger}, {L{\'e}pine}, {Li}, {Lima}, {Lin}, {Long}, {Loomis}, {Loveday}, {Lupton}, {Magnier}, {Malanushenko}, {Malanushenko}, {Mandelbaum}, {Margon}, {Marriner}, {Mart{\'\i}nez-Delgado}, {Matsubara}, {McGehee}, {McKay}, {Meiksin}, {Morrison}, {Mullally}, {Munn}, {Murphy}, {Nash}, {Nebot}, {Neilsen}, {Newberg}, {Newman}, {Nichol}, {Nicinski}, {Nieto-Santisteban}, {Nitta}, {Okamura}, {Oravetz}, {Ostriker}, {Owen}, {Padmanabhan}, {Pan}, {Park}, {Pauls}, {Peoples}, {Percival}, {Pier}, {Pope}, {Pourbaix}, {Price}, {Purger}, {Quinn}, {Raddick}, {Re Fiorentin}, {Richards}, {Richmond}, {Riess}, {Rix}, {Rockosi}, {Sako}, {Schlegel}, {Schneider}, {Scholz}, {Schreiber}, {Schwope}, {Seljak}, {Sesar}, {Sheldon}, {Shimasaku}, {Sibley}, {Simmons}, {Sivarani}, {Allyn Smith}, {Smith}, {Smol{\v{c}}i{\'c}}, {Snedden}, {Stebbins}, {Steinmetz}, {Stoughton}, {Strauss}, {SubbaRao}, {Suto}, {Szalay}, {Szapudi}, {Szkody}, {Tanaka},
  {Tegmark}, {Teodoro}, {Thakar}, {Tremonti}, {Tucker}, {Uomoto}, {Vanden Berk}, {Vandenberg}, {Vidrih}, {Vogeley}, {Voges}, {Vogt}, {Wadadekar}, {Watters}, {Weinberg}, {West}, {White}, {Wilhite}, {Wonders}, {Yanny}, \& {Yocum}}]{2009abazajian}
{Abazajian}, K.~N., {Adelman-McCarthy}, J.~K., {Ag{\"u}eros}, M.~A., {et~al.} 2009, \apjs, 182, 543, \dodoi{10.1088/0067-0049/182/2/543}

\bibitem[{{Aird} {et~al.}(2015){Aird}, {Coil}, {Georgakakis}, {Nandra}, {Barro}, \& {P{\'e}rez-Gonz{\'a}lez}}]{2015Aird}
{Aird}, J., {Coil}, A.~L., {Georgakakis}, A., {et~al.} 2015, \mnras, 451, 1892, \dodoi{10.1093/mnras/stv1062}

\bibitem[{{Alexander} {et~al.}(2020){Alexander}, {van Velzen}, {Horesh}, \& {Zauderer}}]{2020Alexander}
{Alexander}, K.~D., {van Velzen}, S., {Horesh}, A., \& {Zauderer}, B.~A. 2020, \ssr, 216, 81, \dodoi{10.1007/s11214-020-00702-w}

\bibitem[{{Ananna} {et~al.}(2022){Ananna}, {Weigel}, {Trakhtenbrot}, {Koss}, {Urry}, {Ricci}, {Hickox}, {Treister}, {Bauer}, {Ueda}, {Mushotzky}, {Ricci}, {Oh}, {Mej{\'\i}a-Restrepo}, {Brok}, {Stern}, {Powell}, {Caglar}, {Ichikawa}, {Wong}, {Harrison}, \& {Schawinski}}]{Ananna2022}
{Ananna}, T.~T., {Weigel}, A.~K., {Trakhtenbrot}, B., {et~al.} 2022, \apjs, 261, 9, \dodoi{10.3847/1538-4365/ac5b64}

\bibitem[{{Arcavi} {et~al.}(2014){Arcavi}, {Gal-Yam}, {Sullivan}, {Pan}, {Cenko}, {Horesh}, {Ofek}, {De Cia}, {Yan}, {Yang}, {Howell}, {Tal}, {Kulkarni}, {Tendulkar}, {Tang}, {Xu}, {Sternberg}, {Cohen}, {Bloom}, {Nugent}, {Kasliwal}, {Perley}, {Quimby}, {Miller}, {Theissen}, \& {Laher}}]{2014arcavi}
{Arcavi}, I., {Gal-Yam}, A., {Sullivan}, M., {et~al.} 2014, \apj, 793, 38, \dodoi{10.1088/0004-637X/793/1/38}

\bibitem[{{Arcodia} {et~al.}(2021){Arcodia}, {Merloni}, {Nandra}, {Buchner}, {Salvato}, {Pasham}, {Remillard}, {Comparat}, {Lamer}, {Ponti}, {Malyali}, {Wolf}, {Arzoumanian}, {Bogensberger}, {Buckley}, {Gendreau}, {Gromadzki}, {Kara}, {Krumpe}, {Markwardt}, {Ramos-Ceja}, {Rau}, {Schramm}, \& {Schwope}}]{Arcodia21}
{Arcodia}, R., {Merloni}, A., {Nandra}, K., {et~al.} 2021, \nat, 592, 704, \dodoi{10.1038/s41586-021-03394-6}

\bibitem[{{Arcodia} {et~al.}(2024){Arcodia}, {Liu}, {Merloni}, {Malyali}, {Rau}, {Chakraborty}, {Goodwin}, {Buckley}, {Brink}, {Gromadzki}, {Arzoumanian}, {Buchner}, {Kara}, {Nandra}, {Ponti}, {Salvato}, {Anderson}, {Baldini}, {Grotova}, {Krumpe}, {Maitra}, {Miller-Jones}, \& {Ramos-Ceja}}]{Arcodia24}
{Arcodia}, R., {Liu}, Z., {Merloni}, A., {et~al.} 2024, \aap, 684, A64, \dodoi{10.1051/0004-6361/202348881}

\bibitem[{{Arnaud}(1996)}]{1996ASPC..101...17A}
{Arnaud}, K.~A. 1996, in Astronomical Society of the Pacific Conference Series, Vol. 101, Astronomical Data Analysis Software and Systems V, ed. G.~H. {Jacoby} \& J.~{Barnes}, 17

\bibitem[{{Astropy Collaboration} {et~al.}(2013){Astropy Collaboration}, {Robitaille}, {Tollerud}, {Greenfield}, {Droettboom}, {Bray}, {Aldcroft}, {Davis}, {Ginsburg}, {Price-Whelan}, {Kerzendorf}, {Conley}, {Crighton}, {Barbary}, {Muna}, {Ferguson}, {Grollier}, {Parikh}, {Nair}, {Unther}, {Deil}, {Woillez}, {Conseil}, {Kramer}, {Turner}, {Singer}, {Fox}, {Weaver}, {Zabalza}, {Edwards}, {Azalee Bostroem}, {Burke}, {Casey}, {Crawford}, {Dencheva}, {Ely}, {Jenness}, {Labrie}, {Lim}, {Pierfederici}, {Pontzen}, {Ptak}, {Refsdal}, {Servillat}, \& {Streicher}}]{astropy:2013}
{Astropy Collaboration}, {Robitaille}, T.~P., {Tollerud}, E.~J., {et~al.} 2013, \aap, 558, A33, \dodoi{10.1051/0004-6361/201322068}

\bibitem[{{Astropy Collaboration} {et~al.}(2018){Astropy Collaboration}, {Price-Whelan}, {Sip{\H{o}}cz}, {G{\"u}nther}, {Lim}, {Crawford}, {Conseil}, {Shupe}, {Craig}, {Dencheva}, {Ginsburg}, {Vand erPlas}, {Bradley}, {P{\'e}rez-Su{\'a}rez}, {de Val-Borro}, {Aldcroft}, {Cruz}, {Robitaille}, {Tollerud}, {Ardelean}, {Babej}, {Bach}, {Bachetti}, {Bakanov}, {Bamford}, {Barentsen}, {Barmby}, {Baumbach}, {Berry}, {Biscani}, {Boquien}, {Bostroem}, {Bouma}, {Brammer}, {Bray}, {Breytenbach}, {Buddelmeijer}, {Burke}, {Calderone}, {Cano Rodr{\'\i}guez}, {Cara}, {Cardoso}, {Cheedella}, {Copin}, {Corrales}, {Crichton}, {D'Avella}, {Deil}, {Depagne}, {Dietrich}, {Donath}, {Droettboom}, {Earl}, {Erben}, {Fabbro}, {Ferreira}, {Finethy}, {Fox}, {Garrison}, {Gibbons}, {Goldstein}, {Gommers}, {Greco}, {Greenfield}, {Groener}, {Grollier}, {Hagen}, {Hirst}, {Homeier}, {Horton}, {Hosseinzadeh}, {Hu}, {Hunkeler}, {Ivezi{\'c}}, {Jain}, {Jenness}, {Kanarek}, {Kendrew}, {Kern}, {Kerzendorf}, {Khvalko}, {King}, {Kirkby}, {Kulkarni},
  {Kumar}, {Lee}, {Lenz}, {Littlefair}, {Ma}, {Macleod}, {Mastropietro}, {McCully}, {Montagnac}, {Morris}, {Mueller}, {Mumford}, {Muna}, {Murphy}, {Nelson}, {Nguyen}, {Ninan}, {N{\"o}the}, {Ogaz}, {Oh}, {Parejko}, {Parley}, {Pascual}, {Patil}, {Patil}, {Plunkett}, {Prochaska}, {Rastogi}, {Reddy Janga}, {Sabater}, {Sakurikar}, {Seifert}, {Sherbert}, {Sherwood-Taylor}, {Shih}, {Sick}, {Silbiger}, {Singanamalla}, {Singer}, {Sladen}, {Sooley}, {Sornarajah}, {Streicher}, {Teuben}, {Thomas}, {Tremblay}, {Turner}, {Terr{\'o}n}, {van Kerkwijk}, {de la Vega}, {Watkins}, {Weaver}, {Whitmore}, {Woillez}, {Zabalza}, \& {Astropy Contributors}}]{astropy:2018}
{Astropy Collaboration}, {Price-Whelan}, A.~M., {Sip{\H{o}}cz}, B.~M., {et~al.} 2018, \aj, 156, 123, \dodoi{10.3847/1538-3881/aabc4f}

\bibitem[{{Astropy Collaboration} {et~al.}(2022){Astropy Collaboration}, {Price-Whelan}, {Lim}, {Earl}, {Starkman}, {Bradley}, {Shupe}, {Patil}, {Corrales}, {Brasseur}, {N{"o}the}, {Donath}, {Tollerud}, {Morris}, {Ginsburg}, {Vaher}, {Weaver}, {Tocknell}, {Jamieson}, {van Kerkwijk}, {Robitaille}, {Merry}, {Bachetti}, {G{"u}nther}, {Aldcroft}, {Alvarado-Montes}, {Archibald}, {B{'o}di}, {Bapat}, {Barentsen}, {Baz{'a}n}, {Biswas}, {Boquien}, {Burke}, {Cara}, {Cara}, {Conroy}, {Conseil}, {Craig}, {Cross}, {Cruz}, {D'Eugenio}, {Dencheva}, {Devillepoix}, {Dietrich}, {Eigenbrot}, {Erben}, {Ferreira}, {Foreman-Mackey}, {Fox}, {Freij}, {Garg}, {Geda}, {Glattly}, {Gondhalekar}, {Gordon}, {Grant}, {Greenfield}, {Groener}, {Guest}, {Gurovich}, {Handberg}, {Hart}, {Hatfield-Dodds}, {Homeier}, {Hosseinzadeh}, {Jenness}, {Jones}, {Joseph}, {Kalmbach}, {Karamehmetoglu}, {Ka{l}uszy{'n}ski}, {Kelley}, {Kern}, {Kerzendorf}, {Koch}, {Kulumani}, {Lee}, {Ly}, {Ma}, {MacBride}, {Maljaars}, {Muna}, {Murphy}, {Norman}, {O'Steen},
  {Oman}, {Pacifici}, {Pascual}, {Pascual-Granado}, {Patil}, {Perren}, {Pickering}, {Rastogi}, {Roulston}, {Ryan}, {Rykoff}, {Sabater}, {Sakurikar}, {Salgado}, {Sanghi}, {Saunders}, {Savchenko}, {Schwardt}, {Seifert-Eckert}, {Shih}, {Jain}, {Shukla}, {Sick}, {Simpson}, {Singanamalla}, {Singer}, {Singhal}, {Sinha}, {Sip{H{o}}cz}, {Spitler}, {Stansby}, {Streicher}, {{{S}}umak}, {Swinbank}, {Taranu}, {Tewary}, {Tremblay}, {Val-Borro}, {Van Kooten}, {Vasovi{'c}}, {Verma}, {de Miranda Cardoso}, {Williams}, {Wilson}, {Winkel}, {Wood-Vasey}, {Xue}, {Yoachim}, {Zhang}, {Zonca}, \& {Astropy Project Contributors}}]{astropy:2022}
{Astropy Collaboration}, {Price-Whelan}, A.~M., {Lim}, P.~L., {et~al.} 2022, apj, 935, 167, \dodoi{10.3847/1538-4357/ac7c74}

\bibitem[{{Becker} {et~al.}(1995){Becker}, {White}, \& {Helfand}}]{1995FIRST}
{Becker}, R.~H., {White}, R.~L., \& {Helfand}, D.~J. 1995, \apj, 450, 559, \dodoi{10.1086/176166}

\bibitem[{{Bierschenk} {et~al.}(2024){Bierschenk}, {Ricci}, {Temple}, {Satyapal}, {Cann}, {Xie}, {Diaz}, {Ichikawa}, {Koss}, {Bauer}, {Rojas}, {Kakkad}, {Tortosa}, {Ricci}, {Mushotzky}, {Kawamuro}, {Gupta}, {Trakhtenbrot}, {Chang}, {Riffel}, {Oh}, {Harrison}, {Powell}, {Stern}, \& {Urry}}]{2024Bierschenk}
{Bierschenk}, M., {Ricci}, C., {Temple}, M.~J., {et~al.} 2024, \apj, 976, 257, \dodoi{10.3847/1538-4357/ad844a}

\bibitem[{{Blanchard} {et~al.}(2017){Blanchard}, {Nicholl}, {Berger}, {Guillochon}, {Margutti}, {Chornock}, {Alexander}, {Leja}, \& {Drout}}]{blanchard2017}
{Blanchard}, P.~K., {Nicholl}, M., {Berger}, E., {et~al.} 2017, \apj, 843, 106, \dodoi{10.3847/1538-4357/aa77f7}

\bibitem[{{Boller} {et~al.}(2016){Boller}, {Freyberg}, {Tr{\"u}mper}, {Haberl}, {Voges}, \& {Nandra}}]{Boller2016}
{Boller}, T., {Freyberg}, M.~J., {Tr{\"u}mper}, J., {et~al.} 2016, \aap, 588, A103, \dodoi{10.1051/0004-6361/201525648}

\bibitem[{{Boroson} \& {Green}(1992)}]{1992borosongreen}
{Boroson}, T.~A., \& {Green}, R.~F. 1992, \apjs, 80, 109, \dodoi{10.1086/191661}

\bibitem[{{Bowen}(1928)}]{Bowen1928}
{Bowen}, I.~S. 1928, \apj, 67, 1, \dodoi{10.1086/143091}

\bibitem[{{Brown} {et~al.}(2013){Brown}, {Baliber}, {Bianco}, {Bowman}, {Burleson}, {Conway}, {Crellin}, {Depagne}, {De Vera}, {Dilday}, {Dragomir}, {Dubberley}, {Eastman}, {Elphick}, {Falarski}, {Foale}, {Ford}, {Fulton}, {Garza}, {Gomez}, {Graham}, {Greene}, {Haldeman}, {Hawkins}, {Haworth}, {Haynes}, {Hidas}, {Hjelstrom}, {Howell}, {Hygelund}, {Lister}, {Lobdill}, {Martinez}, {Mullins}, {Norbury}, {Parrent}, {Paulson}, {Petry}, {Pickles}, {Posner}, {Rosing}, {Ross}, {Sand}, {Saunders}, {Shobbrook}, {Shporer}, {Street}, {Thomas}, {Tsapras}, {Tufts}, {Valenti}, {Vander Horst}, {Walker}, {White}, \& {Willis}}]{2013PASP..125.1031B}
{Brown}, T.~M., {Baliber}, N., {Bianco}, F.~B., {et~al.} 2013, \pasp, 125, 1031, \dodoi{10.1086/673168}

\bibitem[{{Burrows} {et~al.}(2005){Burrows}, {Hill}, {Nousek}, {Kennea}, {Wells}, {Osborne}, {Abbey}, {Beardmore}, {Mukerjee}, {Short}, {Chincarini}, {Campana}, {Citterio}, {Moretti}, {Pagani}, {Tagliaferri}, {Giommi}, {Capalbi}, {Tamburelli}, {Angelini}, {Cusumano}, {Br{\"a}uninger}, {Burkert}, \& {Hartner}}]{Burrows2005}
{Burrows}, D.~N., {Hill}, J.~E., {Nousek}, J.~A., {et~al.} 2005, \ssr, 120, 165, \dodoi{10.1007/s11214-005-5097-2}

\bibitem[{{Caplar} {et~al.}(2017){Caplar}, {Lilly}, \& {Trakhtenbrot}}]{Caplar17}
{Caplar}, N., {Lilly}, S.~J., \& {Trakhtenbrot}, B. 2017, \apj, 834, 111, \dodoi{10.3847/1538-4357/834/2/111}

\bibitem[{{Cardelli} {et~al.}(1989){Cardelli}, {Clayton}, \& {Mathis}}]{Cardelli1989}
{Cardelli}, J.~A., {Clayton}, G.~C., \& {Mathis}, J.~S. 1989, \apj, 345, 245, \dodoi{10.1086/167900}

\bibitem[{{Cendes} {et~al.}(2024){Cendes}, {Berger}, {Alexander}, {Chornock}, {Margutti}, {Metzger}, {Wieringa}, {Bietenholz}, {Hajela}, {Laskar}, {Stroh}, \& {Terreran}}]{2024Cendes}
{Cendes}, Y., {Berger}, E., {Alexander}, K.~D., {et~al.} 2024, \apj, 971, 185, \dodoi{10.3847/1538-4357/ad5541}

\bibitem[{{Chan} {et~al.}(2019){Chan}, {Piran}, {Krolik}, \& {Saban}}]{Chan2019}
{Chan}, C.-H., {Piran}, T., {Krolik}, J.~H., \& {Saban}, D. 2019, \apj, 881, 113, \dodoi{10.3847/1538-4357/ab2b40}

\bibitem[{{Charalampopoulos} {et~al.}(2022){Charalampopoulos}, {Leloudas}, {Malesani}, {Wevers}, {Arcavi}, {Nicholl}, {Pursiainen}, {Lawrence}, {Anderson}, {Benetti}, {Cannizzaro}, {Chen}, {Galbany}, {Gromadzki}, {Guti{\'e}rrez}, {Inserra}, {Jonker}, {M{\"u}ller-Bravo}, {Onori}, {Short}, {Sollerman}, \& {Young}}]{2022Charalampopoulos}
{Charalampopoulos}, P., {Leloudas}, G., {Malesani}, D.~B., {et~al.} 2022, \aap, 659, A34, \dodoi{10.1051/0004-6361/202142122}

\bibitem[{{Coughlin} \& {Nixon}(2019)}]{Coughlin2019}
{Coughlin}, E.~R., \& {Nixon}, C.~J. 2019, \apjl, 883, L17, \dodoi{10.3847/2041-8213/ab412d}

\bibitem[{{Dai} {et~al.}(2018){Dai}, {McKinney}, {Roth}, {Ramirez-Ruiz}, \& {Miller}}]{2018Dai}
{Dai}, L., {McKinney}, J.~C., {Roth}, N., {Ramirez-Ruiz}, E., \& {Miller}, M.~C. 2018, \apjl, 859, L20, \dodoi{10.3847/2041-8213/aab429}

\bibitem[{{den Brok} {et~al.}(2022){den Brok}, {Koss}, {Trakhtenbrot}, {Stern}, {Cantalupo}, {Lamperti}, {Ricci}, {Ricci}, {Oh}, {Bauer}, {Riffel}, {Rodr{\'\i}guez-Ardila}, {B{\"a}r}, {Harrison}, {Ichikawa}, {Mej{\'\i}a-Restrepo}, {Mushotzky}, {Powell}, {Boissay-Malaquin}, {Stalevski}, {Treister}, {Urry}, \& {Veilleux}}]{denBrok2022}
{den Brok}, J.~S., {Koss}, M.~J., {Trakhtenbrot}, B., {et~al.} 2022, \apjs, 261, 7, \dodoi{10.3847/1538-4365/ac5b66}

\bibitem[{{Dimitrijevi{\'c}} {et~al.}(2007){Dimitrijevi{\'c}}, {Popovi{\'c}}, {Kova{\v{c}}evi{\'c}}, {Da{\v{c}}i{\'c}}, \& {Ili{\'c}}}]{2007MNRAS.374.1181D}
{Dimitrijevi{\'c}}, M.~S., {Popovi{\'c}}, L.~{\v{C}}., {Kova{\v{c}}evi{\'c}}, J., {Da{\v{c}}i{\'c}}, M., \& {Ili{\'c}}, D. 2007, \mnras, 374, 1181, \dodoi{10.1111/j.1365-2966.2006.11238.x}

\bibitem[{{Drake} {et~al.}(2009){Drake}, {Djorgovski}, {Mahabal}, {Beshore}, {Larson}, {Graham}, {Williams}, {Christensen}, {Catelan}, {Boattini}, {Gibbs}, {Hill}, \& {Kowalski}}]{2009drake}
{Drake}, A.~J., {Djorgovski}, S.~G., {Mahabal}, A., {et~al.} 2009, \apj, 696, 870, \dodoi{10.1088/0004-637X/696/1/870}

\bibitem[{{Gehrels} {et~al.}(2004){Gehrels}, {Chincarini}, {Giommi}, {Mason}, {Nousek}, {Wells}, {White}, {Barthelmy}, {Burrows}, {Cominsky}, {Hurley}, {Marshall}, {M{\'e}sz{\'a}ros}, {Roming}, {Angelini}, {Barbier}, {Belloni}, {Campana}, {Caraveo}, {Chester}, {Citterio}, {Cline}, {Cropper}, {Cummings}, {Dean}, {Feigelson}, {Fenimore}, {Frail}, {Fruchter}, {Garmire}, {Gendreau}, {Ghisellini}, {Greiner}, {Hill}, {Hunsberger}, {Krimm}, {Kulkarni}, {Kumar}, {Lebrun}, {Lloyd-Ronning}, {Markwardt}, {Mattson}, {Mushotzky}, {Norris}, {Osborne}, {Paczynski}, {Palmer}, {Park}, {Parsons}, {Paul}, {Rees}, {Reynolds}, {Rhoads}, {Sasseen}, {Schaefer}, {Short}, {Smale}, {Smith}, {Stella}, {Tagliaferri}, {Takahashi}, {Tashiro}, {Townsley}, {Tueller}, {Turner}, {Vietri}, {Voges}, {Ward}, {Willingale}, {Zerbi}, \& {Zhang}}]{Gehrels2004}
{Gehrels}, N., {Chincarini}, G., {Giommi}, P., {et~al.} 2004, \apj, 611, 1005, \dodoi{10.1086/422091}

\bibitem[{{Gezari}(2021)}]{Gezari21_rev}
{Gezari}, S. 2021, \araa, 59, 21, \dodoi{10.1146/annurev-astro-111720-030029}

\bibitem[{{Gezari} {et~al.}(2012){Gezari}, {Chornock}, {Rest}, {Huber}, {Forster}, {Berger}, {Challis}, {Neill}, {Martin}, {Heckman}, {Lawrence}, {Norman}, {Narayan}, {Foley}, {Marion}, {Scolnic}, {Chomiuk}, {Soderberg}, {Smith}, {Kirshner}, {Riess}, {Smartt}, {Stubbs}, {Tonry}, {Wood-Vasey}, {Burgett}, {Chambers}, {Grav}, {Heasley}, {Kaiser}, {Kudritzki}, {Magnier}, {Morgan}, \& {Price}}]{2012gezari}
{Gezari}, S., {Chornock}, R., {Rest}, A., {et~al.} 2012, \nat, 485, 217, \dodoi{10.1038/nature10990}

\bibitem[{{Giustini} {et~al.}(2020){Giustini}, {Miniutti}, \& {Saxton}}]{Giustini20}
{Giustini}, M., {Miniutti}, G., \& {Saxton}, R.~D. 2020, \aap, 636, L2, \dodoi{10.1051/0004-6361/202037610}

\bibitem[{{Graham} {et~al.}(2020){Graham}, {Ross}, {Stern}, {Drake}, {McKernan}, {Ford}, {Djorgovski}, {Mahabal}, {Glikman}, {Larson}, \& {Christensen}}]{graham2020}
{Graham}, M.~J., {Ross}, N.~P., {Stern}, D., {et~al.} 2020, \mnras, 491, 4925, \dodoi{10.1093/mnras/stz3244}

\bibitem[{{Green} {et~al.}(2022){Green}, {Pulgarin-Duque}, {Anderson}, {MacLeod}, {Eracleous}, {Ruan}, {Runnoe}, {Graham}, {Roulston}, {Schneider}, {Ahlf}, {Bizyaev}, {Brownstein}, {del Casal}, {Dodd}, {Hoover}, {Matt}, {Merloni}, {Pan}, {Ramirez}, {Ridder}, \& {Moseley}}]{Green_etal_2022}
{Green}, P.~J., {Pulgarin-Duque}, L., {Anderson}, S.~F., {et~al.} 2022, \apj, 933, 180, \dodoi{10.3847/1538-4357/ac743f}

\bibitem[{{Gromadzki} {et~al.}(2019){Gromadzki}, {Hamanowicz}, {Wyrzykowski}, {Sokolovsky}, {Fraser}, {Kozlowski}, {Guillochon}, {Arcavi}, {Trakhtenbrot}, {Jonker}, {Mattila}, {Udalski}, {Szymanski}, {Soszynski}, {Poleski}, {Pietrukowicz}, {Skowron}, {Mroz}, {Ulaczyk}, {Pawlak}, {Rybicki}, {Sollerman}, {Taddia}, {Kostrzewa Rutkowska}, {Onori}, {Young}, {Maguire}, {Smartt}, {Inserra}, {Gal Yam}, {Rau}, {Chen}, {Angus}, \& {Buckley}}]{Gromadzki2019}
{Gromadzki}, M., {Hamanowicz}, A., {Wyrzykowski}, L., {et~al.} 2019, \aap, 622, L2, \dodoi{10.1051/0004-6361/201833682}

\bibitem[{{Groot} {et~al.}(2024){Groot}, {Bloemen}, {Vreeswijk}, {van Roestel}, {Jonker}, {Nelemans}, {Klein-Wolt}, {Lepoole}, {Pieterse}, {Rodenhuis}, {Boland}, {Haverkorn}, {Aerts}, {Bakker}, {Balster}, {Bekema}, {Dijkstra}, {Dolron}, {Elswijk}, {van Elteren}, {Engels}, {Fokker}, {de Haan}, {Hahn}, {ter Horst}, {Lesman}, {Kragt}, {Morren}, {Nillissen}, {Pessemier}, {Raskin}, {de Rijke}, {Scheers}, {Schuil}, {Timmer}, {Antunes Amaral}, {Arancibia-Rojas}, {Arcavi}, {Blagorodnova}, {Biswas}, {Breton}, {Dawson}, {Dayal}, {De Wet}, {Duffy}, {Faris}, {Fausnaugh}, {Gal-Yam}, {Geier}, {Horesh}, {Johnston}, {Katusiime}, {Kelley}, {Kosakowski}, {Kupfer}, {Leloudas}, {Levan}, {Modiano}, {Mogawana}, {Munday}, {Paice}, {Patat}, {Pelisoli}, {Ramsay}, {Ranaivomanana}, {Ruiz-Carmona}, {Schaffenroth}, {Scaringi}, {Stoppa}, {Street}, {Tranin}, {Uzundag}, {Valenti}, {Veresvarska}, {Vuc̆kovi{\'c}}, {Wichern}, {Wijers}, {Wijnands}, \& {Zimmerman}}]{2024PASP..136k5003G}
{Groot}, P.~J., {Bloemen}, S., {Vreeswijk}, P.~M., {et~al.} 2024, \pasp, 136, 115003, \dodoi{10.1088/1538-3873/ad8b6a}

\bibitem[{{Grzedzielski} {et~al.}(2017){Grzedzielski}, {Janiuk}, {Czerny}, \& {Wu}}]{grzedzielski2017}
{Grzedzielski}, M., {Janiuk}, A., {Czerny}, B., \& {Wu}, Q. 2017, \aap, 603, A110, \dodoi{10.1051/0004-6361/201629672}

\bibitem[{Guillochon {et~al.}(2014)Guillochon, Manukian, \& Ramirez-Ruiz}]{Guillochon2014}
Guillochon, J., Manukian, H., \& Ramirez-Ruiz, E. 2014, The Astrophysical Journal, 783, 23, \dodoi{10.1088/0004-637x/783/1/23}

\bibitem[{{Guo} {et~al.}(2018){Guo}, {Shen}, \& {Wang}}]{pyqsofit}
{Guo}, H., {Shen}, Y., \& {Wang}, S. 2018, {PyQSOFit: Python code to fit the spectrum of quasars}, Astrophysics Source Code Library, record ascl:1809.008.
\newblock \doeprint{1809.008}

\bibitem[{{Guo} {et~al.}(2024){Guo}, {Zou}, {Fawcett}, {Canning}, {Juneau}, {Davis}, {Alexander}, {Jiang}, {Aguilar}, {Ahlen}, {Brooks}, {Claybaugh}, {de la Macorra}, {Doel}, {Fanning}, {Forero-Romero}, {Gontcho A Gontcho}, {Honscheid}, {Kisner}, {Kremin}, {Landriau}, {Meisner}, {Miquel}, {Moustakas}, {Nie}, {Pan}, {Poppett}, {Prada}, {Rezaie}, {Rossi}, {Siudek}, {Sanchez}, {Schubnell}, {Seo}, {Sui}, {Tarl{\'e}}, \& {Zhou}}]{Guo24}
{Guo}, W.-J., {Zou}, H., {Fawcett}, V.~A., {et~al.} 2024, \apjs, 270, 26, \dodoi{10.3847/1538-4365/ad118a}

\bibitem[{{Guolo} \& {Gezari}(2023)}]{2023TNSAN.195....1G}
{Guolo}, M., \& {Gezari}, S. 2023, Transient Name Server AstroNote, 195, 1

\bibitem[{{Hameury} {et~al.}(2009){Hameury}, {Viallet}, \& {Lasota}}]{2009Hameury}
{Hameury}, J.~M., {Viallet}, M., \& {Lasota}, J.~P. 2009, \aap, 496, 413, \dodoi{10.1051/0004-6361/200810928}

\bibitem[{Harris {et~al.}(2020)Harris, Millman, van~der Walt, Gommers, Virtanen, Cournapeau, Wieser, Taylor, Berg, Smith, Kern, Picus, Hoyer, van Kerkwijk, Brett, Haldane, del R{\'{i}}o, Wiebe, Peterson, G{\'{e}}rard-Marchant, Sheppard, Reddy, Weckesser, Abbasi, Gohlke, \& Oliphant}]{harris2020array}
Harris, C.~R., Millman, K.~J., van~der Walt, S.~J., {et~al.} 2020, Nature, 585, 357, \dodoi{10.1038/s41586-020-2649-2}

\bibitem[{{Helfand} {et~al.}(2015){Helfand}, {White}, \& {Becker}}]{First_catalog}
{Helfand}, D.~J., {White}, R.~L., \& {Becker}, R.~H. 2015, \apj, 801, 26, \dodoi{10.1088/0004-637X/801/1/26}

\bibitem[{{Henden} {et~al.}(2009){Henden}, {Welch}, {Terrell}, \& {Levine}}]{2009AAS...21440702H}
{Henden}, A.~A., {Welch}, D.~L., {Terrell}, D., \& {Levine}, S.~E. 2009, in American Astronomical Society Meeting Abstracts, Vol. 214, American Astronomical Society Meeting Abstracts \#214, 407.02

\bibitem[{{HI4PI Collaboration} {et~al.}(2016){HI4PI Collaboration}, {Ben Bekhti}, {Fl{\"o}er}, {Keller}, {Kerp}, {Lenz}, {Winkel}, {Bailin}, {Calabretta}, {Dedes}, {Ford}, {Gibson}, {Haud}, {Janowiecki}, {Kalberla}, {Lockman}, {McClure-Griffiths}, {Murphy}, {Nakanishi}, {Pisano}, \& {Staveley-Smith}}]{2016A&A...594A.116H}
{HI4PI Collaboration}, {Ben Bekhti}, N., {Fl{\"o}er}, L., {et~al.} 2016, \aap, 594, A116, \dodoi{10.1051/0004-6361/201629178}

\bibitem[{{Holoien} {et~al.}(2016){Holoien}, {Kochanek}, {Prieto}, {Stanek}, {Dong}, {Shappee}, {Grupe}, {Brown}, {Basu}, {Beacom}, {Bersier}, {Brimacombe}, {Danilet}, {Falco}, {Guo}, {Jose}, {Herczeg}, {Long}, {Pojmanski}, {Simonian}, {Szczygiel}, {Thompson}, {Thorstensen}, {Wagner}, \& {Wozniak}}]{Holoien2016}
{Holoien}, T.~W.~S., {Kochanek}, C.~S., {Prieto}, J.~L., {et~al.} 2016, \mnras, 455, 2918, \dodoi{10.1093/mnras/stv2486}

\bibitem[{{Holoien} {et~al.}(2019){Holoien}, {Huber}, {Shappee}, {Eracleous}, {Auchettl}, {Brown}, {Tucker}, {Chambers}, {Kochanek}, {Stanek}, {Rest}, {Bersier}, {Post}, {Aldering}, {Ponder}, {Simon}, {Kankare}, {Dong}, {Hallinan}, {Reddy}, {Sanders}, {Topping}, {Pan-STARRS}, {Bulger}, {Lowe}, {Magnier}, {Schultz}, {Waters}, {Willman}, {Wright}, {Young}, {ASAS-SN}, {Dong}, {Prieto}, {Thompson}, {ATLAS}, {Denneau}, {Flewelling}, {Heinze}, {Smartt}, {Smith}, {Stalder}, {Tonry}, \& {Weiland}}]{2019Holoien}
{Holoien}, T.~W.~S., {Huber}, M.~E., {Shappee}, B.~J., {et~al.} 2019, \apj, 880, 120, \dodoi{10.3847/1538-4357/ab2ae1}

\bibitem[{{Hopkins} {et~al.}(2003){Hopkins}, {Miller}, {Nichol}, {Connolly}, {Bernardi}, {G{\'o}mez}, {Goto}, {Tremonti}, {Brinkmann}, {Ivezi{\'c}}, \& {Lamb}}]{2003ApJ...599..971H}
{Hopkins}, A.~M., {Miller}, C.~J., {Nichol}, R.~C., {et~al.} 2003, \apj, 599, 971, \dodoi{10.1086/379608}

\bibitem[{{Horesh} {et~al.}(2021){Horesh}, {Cenko}, \& {Arcavi}}]{Horesh2021}
{Horesh}, A., {Cenko}, S.~B., \& {Arcavi}, I. 2021, Nature Astronomy, 5, 491, \dodoi{10.1038/s41550-021-01300-8}

\bibitem[{Hunter(2007)}]{Hunter:2007}
Hunter, J.~D. 2007, Computing In Science \& Engineering, 9, 90, \dodoi{10.1109/MCSE.2007.55}

\bibitem[{{IceCube Collaboration}(2019)}]{icecube19}
{IceCube Collaboration}. 2019, GRB Coordinates Network, 26258, 1

\bibitem[{{Ivezi{\'c}} {et~al.}(2019){Ivezi{\'c}}, {Kahn}, {Tyson}, {Abel}, {Acosta}, {Allsman}, {Alonso}, {AlSayyad}, {Anderson}, {Andrew}, {Angel}, {Angeli}, {Ansari}, {Antilogus}, {Araujo}, {Armstrong}, {Arndt}, {Astier}, {Aubourg}, {Auza}, {Axelrod}, {Bard}, {Barr}, {Barrau}, {Bartlett}, {Bauer}, {Bauman}, {Baumont}, {Bechtol}, {Bechtol}, {Becker}, {Becla}, {Beldica}, {Bellavia}, {Bianco}, {Biswas}, {Blanc}, {Blazek}, {Blandford}, {Bloom}, {Bogart}, {Bond}, {Booth}, {Borgland}, {Borne}, {Bosch}, {Boutigny}, {Brackett}, {Bradshaw}, {Brandt}, {Brown}, {Bullock}, {Burchat}, {Burke}, {Cagnoli}, {Calabrese}, {Callahan}, {Callen}, {Carlin}, {Carlson}, {Chandrasekharan}, {Charles-Emerson}, {Chesley}, {Cheu}, {Chiang}, {Chiang}, {Chirino}, {Chow}, {Ciardi}, {Claver}, {Cohen-Tanugi}, {Cockrum}, {Coles}, {Connolly}, {Cook}, {Cooray}, {Covey}, {Cribbs}, {Cui}, {Cutri}, {Daly}, {Daniel}, {Daruich}, {Daubard}, {Daues}, {Dawson}, {Delgado}, {Dellapenna}, {de Peyster}, {de Val-Borro}, {Digel}, {Doherty}, {Dubois},
  {Dubois-Felsmann}, {Durech}, {Economou}, {Eifler}, {Eracleous}, {Emmons}, {Fausti Neto}, {Ferguson}, {Figueroa}, {Fisher-Levine}, {Focke}, {Foss}, {Frank}, {Freemon}, {Gangler}, {Gawiser}, {Geary}, {Gee}, {Geha}, {Gessner}, {Gibson}, {Gilmore}, {Glanzman}, {Glick}, {Goldina}, {Goldstein}, {Goodenow}, {Graham}, {Gressler}, {Gris}, {Guy}, {Guyonnet}, {Haller}, {Harris}, {Hascall}, {Haupt}, {Hernandez}, {Herrmann}, {Hileman}, {Hoblitt}, {Hodgson}, {Hogan}, {Howard}, {Huang}, {Huffer}, {Ingraham}, {Innes}, {Jacoby}, {Jain}, {Jammes}, {Jee}, {Jenness}, {Jernigan}, {Jevremovi{\'c}}, {Johns}, {Johnson}, {Johnson}, {Jones}, {Juramy-Gilles}, {Juri{\'c}}, {Kalirai}, {Kallivayalil}, {Kalmbach}, {Kantor}, {Karst}, {Kasliwal}, {Kelly}, {Kessler}, {Kinnison}, {Kirkby}, {Knox}, {Kotov}, {Krabbendam}, {Krughoff}, {Kub{\'a}nek}, {Kuczewski}, {Kulkarni}, {Ku}, {Kurita}, {Lage}, {Lambert}, {Lange}, {Langton}, {Le Guillou}, {Levine}, {Liang}, {Lim}, {Lintott}, {Long}, {Lopez}, {Lotz}, {Lupton}, {Lust}, {MacArthur}, {Mahabal},
  {Mandelbaum}, {Markiewicz}, {Marsh}, {Marshall}, {Marshall}, {May}, {McKercher}, {McQueen}, {Meyers}, {Migliore}, {Miller}, \& {Mills}}]{2019ApJ...873..111I}
{Ivezi{\'c}}, {\v{Z}}., {Kahn}, S.~M., {Tyson}, J.~A., {et~al.} 2019, \apj, 873, 111, \dodoi{10.3847/1538-4357/ab042c}

\bibitem[{{Jana} {et~al.}(2024){Jana}, {Ricci}, {Temple}, {Chang}, {Shablovinskaya}, {Trakhtenbrot}, {Diaz}, {Ilic}, {Nandi}, \& {Koss}}]{Jana24}
{Jana}, A., {Ricci}, C., {Temple}, M.~J., {et~al.} 2024, arXiv e-prints, arXiv:2411.08676, \dodoi{10.48550/arXiv.2411.08676}

\bibitem[{{Janiuk}(2020)}]{2020mbhe.confE..48J}
{Janiuk}, A. 2020, in Multifrequency Behaviour of High Energy Cosmic Sources - XIII. 3-8 June 2019. Palermo, 48, \dodoi{10.22323/1.362.004810.48550/arXiv.1911.05357}

\bibitem[{{Janiuk} \& {Czerny}(2011)}]{janiuk2011}
{Janiuk}, A., \& {Czerny}, B. 2011, \mnras, 414, 2186, \dodoi{10.1111/j.1365-2966.2011.18544.x}

\bibitem[{{Kaur} \& {Stone}(2024)}]{2024kaur}
{Kaur}, K., \& {Stone}, N.~C. 2024, arXiv e-prints, arXiv:2405.18500, \dodoi{10.48550/arXiv.2405.18500}

\bibitem[{{Kellermann} {et~al.}(1989){Kellermann}, {Sramek}, {Schmidt}, {Shaffer}, \& {Green}}]{1989AJ.....98.1195K}
{Kellermann}, K.~I., {Sramek}, R., {Schmidt}, M., {Shaffer}, D.~B., \& {Green}, R. 1989, \aj, 98, 1195, \dodoi{10.1086/115207}

\bibitem[{{Komossa} {et~al.}(2008){Komossa}, {Zhou}, {Wang}, {Ajello}, {Ge}, {Greiner}, {Lu}, {Salvato}, {Saxton}, {Shan}, {Xu}, \& {Yuan}}]{Komossa2008}
{Komossa}, S., {Zhou}, H., {Wang}, T., {et~al.} 2008, \apjl, 678, L13, \dodoi{10.1086/588281}

\bibitem[{{Kova{\v{c}}evi{\'c}} {et~al.}(2018){Kova{\v{c}}evi{\'c}}, {P{\'e}rez-Hern{\'a}ndez}, {Popovi{\'c}}, {Shapovalova}, {Kollatschny}, \& {Ili{\'c}}}]{2018MNRAS.475.2051K}
{Kova{\v{c}}evi{\'c}}, A.~B., {P{\'e}rez-Hern{\'a}ndez}, E., {Popovi{\'c}}, L.~{\v{C}}., {et~al.} 2018, \mnras, 475, 2051, \dodoi{10.1093/mnras/stx3137}

\bibitem[{{Landt} {et~al.}(2015){Landt}, {Ward}, {Steenbrugge}, \& {Ferland}}]{Landt2015}
{Landt}, H., {Ward}, M.~J., {Steenbrugge}, K.~C., \& {Ferland}, G.~J. 2015, \mnras, 449, 3795, \dodoi{10.1093/mnras/stv062}

\bibitem[{{Leloudas} {et~al.}(2019){Leloudas}, {Dai}, {Arcavi}, {Vreeswijk}, {Mockler}, {Roy}, {Malesani}, {Schulze}, {Wevers}, {Fraser}, {Ramirez-Ruiz}, {Auchettl}, {Burke}, {Cannizzaro}, {Charalampopoulos}, {Chen}, {Cikota}, {Della Valle}, {Galbany}, {Gromadzki}, {Heintz}, {Hiramatsu}, {Jonker}, {Kostrzewa-Rutkowska}, {Maguire}, {Mandel}, {Nicholl}, {Onori}, {Roth}, {Smartt}, {Wyrzykowski}, \& {Young}}]{Leloudas2019}
{Leloudas}, G., {Dai}, L., {Arcavi}, I., {et~al.} 2019, \apj, 887, 218, \dodoi{10.3847/1538-4357/ab5792}

\bibitem[{{Li} {et~al.}(2024){Li}, {Ho}, {Ricci}, \& {Trakhtenbrot}}]{Li24}
{Li}, R., {Ho}, L.~C., {Ricci}, C., \& {Trakhtenbrot}, B. 2024, \apj, 975, 50, \dodoi{10.3847/1538-4357/ad77a5}

\bibitem[{{Lin} {et~al.}(2024){Lin}, {Jiang}, {Wang}, {Kong}, {Li}, {He}, {Wang}, {Zhu}, {Li}, {Jiang}, {Singh}, {Teja}, {Sahu}, {Jin}, {Maeda}, \& {Huang}}]{Lin2024_pTDE}
{Lin}, Z., {Jiang}, N., {Wang}, T., {et~al.} 2024, \apjl, 971, L26, \dodoi{10.3847/2041-8213/ad638e}

\bibitem[{{Lu} \& {Quataert}(2023)}]{2023LuQuataert}
{Lu}, W., \& {Quataert}, E. 2023, \mnras, 524, 6247, \dodoi{10.1093/mnras/stad2203}

\bibitem[{{Lusso} \& {Risaliti}(2016)}]{Lusso2016}
{Lusso}, E., \& {Risaliti}, G. 2016, \apj, 819, 154, \dodoi{10.3847/0004-637X/819/2/154}

\bibitem[{{MacLeod} {et~al.}(2012){MacLeod}, {Ivezi{\'c}}, {Sesar}, {de Vries}, {Kochanek}, {Kelly}, {Becker}, {Lupton}, {Hall}, {Richards}, {Anderson}, \& {Schneider}}]{2012macleod}
{MacLeod}, C.~L., {Ivezi{\'c}}, {\v{Z}}., {Sesar}, B., {et~al.} 2012, \apj, 753, 106, \dodoi{10.1088/0004-637X/753/2/106}

\bibitem[{{MacLeod} {et~al.}(2019){MacLeod}, {Green}, {Anderson}, {Bruce}, {Eracleous}, {Graham}, {Homan}, {Lawrence}, {LeBleu}, {Ross}, {Ruan}, {Runnoe}, {Stern}, {Burgett}, {Chambers}, {Kaiser}, {Magnier}, \& {Metcalfe}}]{macleod2019}
{MacLeod}, C.~L., {Green}, P.~J., {Anderson}, S.~F., {et~al.} 2019, \apj, 874, 8, \dodoi{10.3847/1538-4357/ab05e2}

\bibitem[{{Mainzer} {et~al.}(2011){Mainzer}, {Bauer}, {Grav}, {Masiero}, {Cutri}, {Dailey}, {Eisenhardt}, {McMillan}, {Wright}, {Walker}, {Jedicke}, {Spahr}, {Tholen}, {Alles}, {Beck}, {Brandenburg}, {Conrow}, {Evans}, {Fowler}, {Jarrett}, {Marsh}, {Masci}, {McCallon}, {Wheelock}, {Wittman}, {Wyatt}, {DeBaun}, {Elliott}, {Elsbury}, {Gautier}, {Gomillion}, {Leisawitz}, {Maleszewski}, {Micheli}, \& {Wilkins}}]{Mainzer2011}
{Mainzer}, A., {Bauer}, J., {Grav}, T., {et~al.} 2011, \apj, 731, 53, \dodoi{10.1088/0004-637X/731/1/53}

\bibitem[{{Mainzer} {et~al.}(2014){Mainzer}, {Bauer}, {Cutri}, {Grav}, {Masiero}, {Beck}, {Clarkson}, {Conrow}, {Dailey}, {Eisenhardt}, {Fabinsky}, {Fajardo-Acosta}, {Fowler}, {Gelino}, {Grillmair}, {Heinrichsen}, {Kendall}, {Kirkpatrick}, {Liu}, {Masci}, {McCallon}, {Nugent}, {Papin}, {Rice}, {Royer}, {Ryan}, {Sevilla}, {Sonnett}, {Stevenson}, {Thompson}, {Wheelock}, {Wiemer}, {Wittman}, {Wright}, \& {Yan}}]{Mainzer2014}
{Mainzer}, A., {Bauer}, J., {Cutri}, R.~M., {et~al.} 2014, \apj, 792, 30, \dodoi{10.1088/0004-637X/792/1/30}

\bibitem[{{Makrygianni} {et~al.}(2023){Makrygianni}, {Trakhtenbrot}, {Arcavi}, {Ricci}, {Lam}, {Horesh}, {Sfaradi}, {Bostroem}, {Hosseinzadeh}, {Howell}, {Pellegrino}, {Fender}, {Green}, {Williams}, \& {Bright}}]{Makrygianni2023}
{Makrygianni}, L., {Trakhtenbrot}, B., {Arcavi}, I., {et~al.} 2023, \apj, 953, 32, \dodoi{10.3847/1538-4357/ace1ee}

\bibitem[{{Malyali} {et~al.}(2021){Malyali}, {Rau}, {Merloni}, {Nandra}, {Buchner}, {Liu}, {Gezari}, {Sollerman}, {Shappee}, {Trakhtenbrot}, {Arcavi}, {Ricci}, {van Velzen}, {Goobar}, {Frederick}, {Kawka}, {Tartaglia}, {Burke}, {Hiramatsu}, {Schramm}, {van der Boom}, {Anderson}, {Miller-Jones}, {Bellm}, {Drake}, {Duev}, {Fremling}, {Graham}, {Masci}, {Rusholme}, {Soumagnac}, \& {Walters}}]{2021Malyali}
{Malyali}, A., {Rau}, A., {Merloni}, A., {et~al.} 2021, \aap, 647, A9, \dodoi{10.1051/0004-6361/202039681}

\bibitem[{{Mandel} \& {Levin}(2015)}]{Mandel2015}
{Mandel}, I., \& {Levin}, Y. 2015, \apjl, 805, L4, \dodoi{10.1088/2041-8205/805/1/L4}

\bibitem[{{Marconi} {et~al.}(2004){Marconi}, {Risaliti}, {Gilli}, {Hunt}, {Maiolino}, \& {Salvati}}]{marconi2004}
{Marconi}, A., {Risaliti}, G., {Gilli}, R., {et~al.} 2004, \mnras, 351, 169, \dodoi{10.1111/j.1365-2966.2004.07765.x}

\bibitem[{{Martin} {et~al.}(2005){Martin}, {Fanson}, {Schiminovich}, {Morrissey}, {Friedman}, {Barlow}, {Conrow}, {Grange}, {Jelinsky}, {Milliard}, {Siegmund}, {Bianchi}, {Byun}, {Donas}, {Forster}, {Heckman}, {Lee}, {Madore}, {Malina}, {Neff}, {Rich}, {Small}, {Surber}, {Szalay}, {Welsh}, \& {Wyder}}]{2005martin}
{Martin}, D.~C., {Fanson}, J., {Schiminovich}, D., {et~al.} 2005, \apjl, 619, L1, \dodoi{10.1086/426387}

\bibitem[{{Masci} {et~al.}(2019){Masci}, {Laher}, {Rusholme}, {Shupe}, {Groom}, {Surace}, {Jackson}, {Monkewitz}, {Beck}, {Flynn}, {Terek}, {Landry}, {Hacopians}, {Desai}, {Howell}, {Brooke}, {Imel}, {Wachter}, {Ye}, {Lin}, {Cenko}, {Cunningham}, {Rebbapragada}, {Bue}, {Miller}, {Mahabal}, {Bellm}, {Patterson}, {Juri{\'c}}, {Golkhou}, {Ofek}, {Walters}, {Graham}, {Kasliwal}, {Dekany}, {Kupfer}, {Burdge}, {Cannella}, {Barlow}, {Van Sistine}, {Giomi}, {Fremling}, {Blagorodnova}, {Levitan}, {Riddle}, {Smith}, {Helou}, {Prince}, \& {Kulkarni}}]{2019PASP..131a8003M}
{Masci}, F.~J., {Laher}, R.~R., {Rusholme}, B., {et~al.} 2019, \pasp, 131, 018003, \dodoi{10.1088/1538-3873/aae8ac}

\bibitem[{{Mazzalay} {et~al.}(2010){Mazzalay}, {Rodr{\'\i}guez-Ardila}, \& {Komossa}}]{2010Mazzalay}
{Mazzalay}, X., {Rodr{\'\i}guez-Ardila}, A., \& {Komossa}, S. 2010, \mnras, 405, 1315, \dodoi{10.1111/j.1365-2966.2010.16533.x}

\bibitem[{{Mej{\'\i}a-Restrepo} {et~al.}(2022){Mej{\'\i}a-Restrepo}, {Trakhtenbrot}, {Koss}, {Oh}, {den Brok}, {Stern}, {Powell}, {Ricci}, {Caglar}, {Ricci}, {Bauer}, {Treister}, {Harrison}, {Urry}, {Ananna}, {Asmus}, {Assef}, {B{\"a}r}, {Bessiere}, {Burtscher}, {Ichikawa}, {Kakkad}, {Kamraj}, {Mushotzky}, {Privon}, {Rojas}, {Sani}, {Schawinski}, \& {Veilleux}}]{MR22}
{Mej{\'\i}a-Restrepo}, J.~E., {Trakhtenbrot}, B., {Koss}, M.~J., {et~al.} 2022, \apjs, 261, 5, \dodoi{10.3847/1538-4365/ac6602}

\bibitem[{{Merloni} {et~al.}(2015){Merloni}, {Dwelly}, {Salvato}, {Georgakakis}, {Greiner}, {Krumpe}, {Nandra}, {Ponti}, \& {Rau}}]{2015Merloni}
{Merloni}, A., {Dwelly}, T., {Salvato}, M., {et~al.} 2015, \mnras, 452, 69, \dodoi{10.1093/mnras/stv1095}

\bibitem[{{Metzger} {et~al.}(2022){Metzger}, {Stone}, \& {Gilbaum}}]{2022metzger}
{Metzger}, B.~D., {Stone}, N.~C., \& {Gilbaum}, S. 2022, \apj, 926, 101, \dodoi{10.3847/1538-4357/ac3ee1}

\bibitem[{{Mil{\'a}n Veres} {et~al.}(2024){Mil{\'a}n Veres}, {Franckowiak}, {van Velzen}, {Adebahr}, {Taziaux}, {Necker}, {Stein}, {Kier}, {Mueller}, {Bomans}, {Jordana-Mitjans}, {Kowalski}, {Hammerstein}, {Marci-Boehncke}, {Reusch}, {Garrappa}, {Rose}, \& {Kashyap Das}}]{2024veres}
{Mil{\'a}n Veres}, P., {Franckowiak}, A., {van Velzen}, S., {et~al.} 2024, arXiv e-prints, arXiv:2408.17419, \dodoi{10.48550/arXiv.2408.17419}

\bibitem[{{Miller} {et~al.}(2025){Miller}, {Abrams}, {Aldering}, {Anand}, {Angus}, {Arcavi}, {Baltay}, {Bauer}, {Brethauer}, {Bloom}, {Bommireddy}, {Catelan}, {Chornock}, {Clark}, {Collett}, {Dimitriadis}, {Faris}, {Forster}, {Franckowiak}, {Frohmaier}, {Galbany}, {Galleguillos}, {Goobar}, {Gutierrez}, {Hall}, {Hammerstein}, {Herner}, {Hook}, {Huston}, {Johansson}, {Kilpatrick}, {Kim}, {Knop}, {Kowalski}, {Kwok}, {LeBaron}, {Lin}, {Liu}, {Lu}, {Lu}, {Lunnan}, {Maguire}, {Makrygianni}, {Margutti}, {Maoz}, {Milan Veres}, {Moore}, {Nayana}, {Nicholl}, {Nordin}, {Pignata}, {Polin}, {Poznanski}, {Prieto}, {Rabinowitz}, {Rehemtulla}, {Rigault}, {Ryczanowski}, {Sarin}, {Schulze}, {Shah}, {Sheng}, {Shilling}, {Simmons}, {Singh}, {Smith}, {Smith}, {Sollerman}, {Soumagnac}, {Stubbs}, {Sullivan}, {Suresh}, {Trakhtenbrot}, {Ward}, {Wiston}, {Xiong}, {Yao}, \& {Nugent}}]{2025ls4}
{Miller}, A.~A., {Abrams}, N.~S., {Aldering}, G., {et~al.} 2025, arXiv e-prints, arXiv:2503.14579, \dodoi{10.48550/arXiv.2503.14579}

\bibitem[{{Minezaki} {et~al.}(2019){Minezaki}, {Yoshii}, {Kobayashi}, {Sugawara}, {Sakata}, {Enya}, {Koshida}, {Tomita}, {Suganuma}, {Aoki}, \& {Peterson}}]{2019minezaki}
{Minezaki}, T., {Yoshii}, Y., {Kobayashi}, Y., {et~al.} 2019, \apj, 886, 150, \dodoi{10.3847/1538-4357/ab4f7b}

\bibitem[{{Miniutti} {et~al.}(2019){Miniutti}, {Saxton}, {Giustini}, {Alexander}, {Fender}, {Heywood}, {Monageng}, {Coriat}, {Tzioumis}, {Read}, {Knigge}, {Gandhi}, {Pretorius}, \& {Ag{\'\i}s-Gonz{\'a}lez}}]{Miniutti19}
{Miniutti}, G., {Saxton}, R.~D., {Giustini}, M., {et~al.} 2019, \nat, 573, 381, \dodoi{10.1038/s41586-019-1556-x}

\bibitem[{{Netzer} {et~al.}(1985){Netzer}, {Elitzur}, \& {Ferland}}]{Netzer1985}
{Netzer}, H., {Elitzur}, M., \& {Ferland}, G.~J. 1985, \apj, 299, 752, \dodoi{10.1086/163741}

\bibitem[{{Newsome} {et~al.}(2024){Newsome}, {Arcavi}, {Howell}, {McCully}, {Terreran}, {Hosseinzadeh}, {Bostroem}, {Dgany}, {Farah}, {Faris}, {Padilla-Gonzalez}, {Pellegrino}, \& {Andrews}}]{2024newsome}
{Newsome}, M., {Arcavi}, I., {Howell}, D.~A., {et~al.} 2024, \apj, 977, 258, \dodoi{10.3847/1538-4357/ad8a69}

\bibitem[{{Newville} {et~al.}(2016){Newville}, {Stensitzki}, {Allen}, {Rawlik}, {Ingargiola}, \& {Nelson}}]{2016lmfit}
{Newville}, M., {Stensitzki}, T., {Allen}, D.~B., {et~al.} 2016, {Lmfit: Non-Linear Least-Square Minimization and Curve-Fitting for Python}, Astrophysics Source Code Library, record ascl:1606.014

\bibitem[{{Noda} \& {Done}(2018)}]{noda2018}
{Noda}, H., \& {Done}, C. 2018, \mnras, 480, 3898, \dodoi{10.1093/mnras/sty2032}

\bibitem[{{Oke} {et~al.}(1995){Oke}, {Cohen}, {Carr}, {Cromer}, {Dingizian}, {Harris}, {Labrecque}, {Lucinio}, {Schaal}, {Epps}, \& {Miller}}]{Oke1995}
{Oke}, J.~B., {Cohen}, J.~G., {Carr}, M., {et~al.} 1995, \pasp, 107, 375, \dodoi{10.1086/133562}

\bibitem[{{Oknyansky} {et~al.}(2019){Oknyansky}, {Winkler}, {Tsygankov}, {Lipunov}, {Gorbovskoy}, {van Wyk}, {Buckley}, \& {Tyurina}}]{oknyansky2019}
{Oknyansky}, V.~L., {Winkler}, H., {Tsygankov}, S.~S., {et~al.} 2019, \mnras, 483, 558, \dodoi{10.1093/mnras/sty3133}

\bibitem[{{Pan} {et~al.}(2021){Pan}, {Li}, \& {Cao}}]{Pan21}
{Pan}, X., {Li}, S.-L., \& {Cao}, X. 2021, \apj, 910, 97, \dodoi{10.3847/1538-4357/abe766}

\bibitem[{{Pasham}(2023)}]{2023ATel16118....1P}
{Pasham}, D. 2023, The Astronomer's Telegram, 16118, 1

\bibitem[{{Peterson} \& {Wandel}(2000)}]{Peterson2000}
{Peterson}, B.~M., \& {Wandel}, A. 2000, \apjl, 540, L13, \dodoi{10.1086/312862}

\bibitem[{{Petrushevska} {et~al.}(2023){Petrushevska}, {Leloudas}, {Ili{\'c}}, {Bronikowski}, {Charalampopoulos}, {Jaisawal}, {Paraskeva}, {Pursiainen}, {Raki{\'c}}, {Schulze}, {Taggart}, {Wedderkopp}, {Anderson}, {de Boer}, {Chambers}, {Chen}, {Damljanovi{\'c}}, {Fraser}, {Gao}, {Gomboc}, {Gromadzki}, {Ihanec}, {Maguire}, {Mar{\v{c}}un}, {M{\"u}ller-Bravo}, {Nicholl}, {Onori}, {Reynolds}, {Smartt}, {Sollerman}, {Smith}, {Wevers}, \& {Wyrzykowski}}]{petrushevska2023}
{Petrushevska}, T., {Leloudas}, G., {Ili{\'c}}, D., {et~al.} 2023, \aap, 669, A140, \dodoi{10.1051/0004-6361/202244623}

\bibitem[{{Piran} {et~al.}(2015){Piran}, {Svirski}, {Krolik}, {Cheng}, \& {Shiokawa}}]{2015piran}
{Piran}, T., {Svirski}, G., {Krolik}, J., {Cheng}, R.~M., \& {Shiokawa}, H. 2015, \apj, 806, 164, \dodoi{10.1088/0004-637X/806/2/164}

\bibitem[{{Raj} \& {Nixon}(2021)}]{2021rajnixon}
{Raj}, A., \& {Nixon}, C.~J. 2021, \apj, 909, 82, \dodoi{10.3847/1538-4357/abdc25}

\bibitem[{{Rees}(1988)}]{Rees1988}
{Rees}, M.~J. 1988, \nat, 333, 523, \dodoi{10.1038/333523a0}

\bibitem[{{Reusch}(2023)}]{Reusch2023}
{Reusch}, S. 2023, arXiv e-prints, arXiv:2307.00902, \dodoi{10.48550/arXiv.2307.00902}

\bibitem[{{Ricci} \& {Trakhtenbrot}(2023)}]{2023riccitrakhtenbrot}
{Ricci}, C., \& {Trakhtenbrot}, B. 2023, Nature Astronomy, 7, 1282, \dodoi{10.1038/s41550-023-02108-4}

\bibitem[{{Ricci} {et~al.}(2017){Ricci}, {Trakhtenbrot}, {Koss}, {Ueda}, {Del Vecchio}, {Treister}, {Schawinski}, {Paltani}, {Oh}, {Lamperti}, {Berney}, {Gandhi}, {Ichikawa}, {Bauer}, {Ho}, {Asmus}, {Beckmann}, {Soldi}, {Balokovi{\'c}}, {Gehrels}, \& {Markwardt}}]{2017ricci}
{Ricci}, C., {Trakhtenbrot}, B., {Koss}, M.~J., {et~al.} 2017, \apjs, 233, 17, \dodoi{10.3847/1538-4365/aa96ad}

\bibitem[{{Ricci} {et~al.}(2020){Ricci}, {Kara}, {Loewenstein}, {Trakhtenbrot}, {Arcavi}, {Remillard}, {Fabian}, {Gendreau}, {Arzoumanian}, {Li}, {Ho}, {MacLeod}, {Cackett}, {Altamirano}, {Gandhi}, {Kosec}, {Pasham}, {Steiner}, \& {Chan}}]{Ricci2020_1ES}
{Ricci}, C., {Kara}, E., {Loewenstein}, M., {et~al.} 2020, \apjl, 898, L1, \dodoi{10.3847/2041-8213/ab91a1}

\bibitem[{{Rodr{\'\i}guez-Ardila} \& {Cerqueira-Campos}(2025)}]{Rodriguez2025_CLR_rev}
{Rodr{\'\i}guez-Ardila}, A., \& {Cerqueira-Campos}, F. 2025, Frontiers in Astronomy and Space Sciences, 12, 1548632, \dodoi{10.3389/fspas.2025.1548632}

\bibitem[{{Roth} {et~al.}(2016){Roth}, {Kasen}, {Guillochon}, \& {Ramirez-Ruiz}}]{Roth2016}
{Roth}, N., {Kasen}, D., {Guillochon}, J., \& {Ramirez-Ruiz}, E. 2016, \apj, 827, 3, \dodoi{10.3847/0004-637X/827/1/3}

\bibitem[{{Rumbaugh} {et~al.}(2018){Rumbaugh}, {Shen}, {Morganson}, {Liu}, {Banerji}, {McMahon}, {Abdalla}, {Benoit-L{\'e}vy}, {Bertin}, {Brooks}, {Buckley-Geer}, {Capozzi}, {Carnero Rosell}, {Carrasco Kind}, {Carretero}, {Cunha}, {D'Andrea}, {da Costa}, {DePoy}, {Desai}, {Doel}, {Frieman}, {Garc{\'\i}a-Bellido}, {Gruen}, {Gruendl}, {Gschwend}, {Gutierrez}, {Honscheid}, {James}, {Kuehn}, {Kuhlmann}, {Kuropatkin}, {Lima}, {Maia}, {Marshall}, {Martini}, {Menanteau}, {Plazas}, {Reil}, {Roodman}, {Sanchez}, {Scarpine}, {Schindler}, {Schubnell}, {Sheldon}, {Smith}, {Soares-Santos}, {Sobreira}, {Suchyta}, {Swanson}, {Walker}, {Wester}, \& {DES Collaboration}}]{2018rumbaugh}
{Rumbaugh}, N., {Shen}, Y., {Morganson}, E., {et~al.} 2018, \apj, 854, 160, \dodoi{10.3847/1538-4357/aaa9b6}

\bibitem[{{Sand} {et~al.}(2011){Sand}, {Brown}, {Haynes}, \& {Dubberley}}]{Sand2011}
{Sand}, D.~J., {Brown}, T., {Haynes}, R., \& {Dubberley}, M. 2011, in American Astronomical Society Meeting Abstracts, Vol. 218, American Astronomical Society Meeting Abstracts \#218, 132.03

\bibitem[{{Scepi} {et~al.}(2021){Scepi}, {Begelman}, \& {Dexter}}]{2021scepi}
{Scepi}, N., {Begelman}, M.~C., \& {Dexter}, J. 2021, \mnras, 502, L50, \dodoi{10.1093/mnrasl/slab002}

\bibitem[{{Schlafly} \& {Finkbeiner}(2011)}]{Schlafly2011}
{Schlafly}, E.~F., \& {Finkbeiner}, D.~P. 2011, \apj, 737, 103, \dodoi{10.1088/0004-637X/737/2/103}

\bibitem[{{Sfaradi} {et~al.}(2022){Sfaradi}, {Horesh}, {Fender}, {Green}, {Williams}, {Bright}, \& {Schulze}}]{Sfaradi2022}
{Sfaradi}, I., {Horesh}, A., {Fender}, R., {et~al.} 2022, \apj, 933, 176, \dodoi{10.3847/1538-4357/ac74bc}

\bibitem[{{Shappee} {et~al.}(2014){Shappee}, {Prieto}, {Grupe}, {Kochanek}, {Stanek}, {De Rosa}, {Mathur}, {Zu}, {Peterson}, {Pogge}, {Komossa}, {Im}, {Jencson}, {Holoien}, {Basu}, {Beacom}, {Szczygiel}, {Brimacombe}, {Adams}, {Campillay}, {Choi}, {Contreras}, {Dietrich}, {Dubberley}, {Elphick}, {Foale}, {Giustini}, {Gonzalez}, {Hawkins}, {Howell}, {Hsiao}, {Koss}, {Leighly}, {Morrell}, {Mudd}, {Mullins}, {Nugent}, {Parrent}, {Phillips}, {Pojmanski}, {Rosing}, {Ross}, {Sand}, {Terndrup}, {Valenti}, {Walker}, \& {Yoon}}]{Shappee14}
{Shappee}, B.~J., {Prieto}, J.~L., {Grupe}, D., {et~al.} 2014, \apj, 788, 48, \dodoi{10.1088/0004-637X/788/1/48}

\bibitem[{Shen(2013)}]{Shen2013_rev}
Shen, Y. 2013, Bulletin of the Astronomical Society of India, 41, 61

\bibitem[{{Short} {et~al.}(2023){Short}, {Lawrence}, {Nicholl}, {Ward}, {Reynolds}, {Mattila}, {Yin}, {Arcavi}, {Carnall}, {Charalampopoulos}, {Gromadzki}, {Jonker}, {Kim}, {Leloudas}, {Mandel}, {Onori}, {Pursiainen}, {Schulze}, {Villforth}, \& {Wevers}}]{2023Short}
{Short}, P., {Lawrence}, A., {Nicholl}, M., {et~al.} 2023, \mnras, 525, 1568, \dodoi{10.1093/mnras/stad2270}

\bibitem[{{Shvartzvald} {et~al.}(2024){Shvartzvald}, {Waxman}, {Gal-Yam}, {Ofek}, {Ben-Ami}, {Berge}, {Kowalski}, {B{\"u}hler}, {Worm}, {Rhoads}, {Arcavi}, {Maoz}, {Polishook}, {Stone}, {Trakhtenbrot}, {Ackermann}, {Aharonson}, {Birnholtz}, {Chelouche}, {Guetta}, {Hallakoun}, {Horesh}, {Kushnir}, {Mazeh}, {Nordin}, {Ofir}, {Ohm}, {Parsons}, {Pe'er}, {Perets}, {Perdelwitz}, {Poznanski}, {Sadeh}, {Sagiv}, {Shahaf}, {Soumagnac}, {Tal-Or}, {Santen}, {Zackay}, {Guttman}, {Rekhi}, {Townsend}, {Weinstein}, \& {Wold}}]{2024Shvartzvald}
{Shvartzvald}, Y., {Waxman}, E., {Gal-Yam}, A., {et~al.} 2024, \apj, 964, 74, \dodoi{10.3847/1538-4357/ad2704}

\bibitem[{{{\'S}niegowska} {et~al.}(2020){{\'S}niegowska}, {Czerny}, {Bon}, \& {Bon}}]{Sniegowska20}
{{\'S}niegowska}, M., {Czerny}, B., {Bon}, E., \& {Bon}, N. 2020, \aap, 641, A167, \dodoi{10.1051/0004-6361/202038575}

\bibitem[{{{\'S}niegowska} {et~al.}(2023){{\'S}niegowska}, {Grz{\c{e}}dzielski}, {Czerny}, \& {Janiuk}}]{sniegowska2023}
{{\'S}niegowska}, M., {Grz{\c{e}}dzielski}, M., {Czerny}, B., \& {Janiuk}, A. 2023, \aap, 672, A19, \dodoi{10.1051/0004-6361/202243828}

\bibitem[{{Somalwar} {et~al.}(2023){Somalwar}, {Ravi}, {Yao}, {Guolo}, {Graham}, {Hammerstein}, {Lu}, {Nicholl}, {Sharma}, {Stein}, {van Velzen}, {Bellm}, {Coughlin}, {Groom}, {Masci}, \& {Riddle}}]{2023somalwar}
{Somalwar}, J.~J., {Ravi}, V., {Yao}, Y., {et~al.} 2023, arXiv e-prints, arXiv:2310.03782, \dodoi{10.48550/arXiv.2310.03782}

\bibitem[{{Steinberg} \& {Stone}(2024)}]{Steinberg24}
{Steinberg}, E., \& {Stone}, N.~C. 2024, \nat, 625, 463, \dodoi{10.1038/s41586-023-06875-y}

\bibitem[{{Stern} {et~al.}(2012){Stern}, {Assef}, {Benford}, {Blain}, {Cutri}, {Dey}, {Eisenhardt}, {Griffith}, {Jarrett}, {Lake}, {Masci}, {Petty}, {Stanford}, {Tsai}, {Wright}, {Yan}, {Harrison}, \& {Madsen}}]{Stern2012}
{Stern}, D., {Assef}, R.~J., {Benford}, D.~J., {et~al.} 2012, \apj, 753, 30, \dodoi{10.1088/0004-637X/753/1/30}

\bibitem[{{STScI Development Team}(2020)}]{stsynphot}
{STScI Development Team}. 2020, {stsynphot: synphot for HST and JWST}, Astrophysics Source Code Library, record ascl:2010.003.
\newblock \doeprint{2010.003}

\bibitem[{{Sun} {et~al.}(2025){Sun}, {Guo}, {Gu}, {Li}, {Chen}, {Gonz{\'a}lez-Buitrago}, {Wang}, {Li}, {Feng}, {Xiong}, {Wang}, {Yuan}, {Jin}, {Zhang}, {Deng}, \& {Zhang}}]{2025Sun_rpTDE}
{Sun}, J., {Guo}, H., {Gu}, M., {et~al.} 2025, arXiv e-prints, arXiv:2501.01824, \dodoi{10.48550/arXiv.2501.01824}

\bibitem[{{Tadhunter} {et~al.}(2021){Tadhunter}, {Patel}, \& {Mullaney}}]{2021tadhunter}
{Tadhunter}, C., {Patel}, M., \& {Mullaney}, J. 2021, \mnras, 504, 4377, \dodoi{10.1093/mnras/stab1105}

\bibitem[{{Tadhunter} {et~al.}(2017){Tadhunter}, {Spence}, {Rose}, {Mullaney}, \& {Crowther}}]{2017tadhunter}
{Tadhunter}, C., {Spence}, R., {Rose}, M., {Mullaney}, J., \& {Crowther}, P. 2017, Nature Astronomy, 1, 0061, \dodoi{10.1038/s41550-017-0061}

\bibitem[{{Tonry} {et~al.}(2018){Tonry}, {Denneau}, {Heinze}, {Stalder}, {Smith}, {Smartt}, {Stubbs}, {Weiland}, \& {Rest}}]{2018torny}
{Tonry}, J.~L., {Denneau}, L., {Heinze}, A.~N., {et~al.} 2018, \pasp, 130, 064505, \dodoi{10.1088/1538-3873/aabadf}

\bibitem[{{Trakhtenbrot} \& {Netzer}(2012)}]{TrakhtenbrotNetzer12}
{Trakhtenbrot}, B., \& {Netzer}, H. 2012, \mnras, 427, 3081, \dodoi{10.1111/j.1365-2966.2012.22056.x}

\bibitem[{{Trakhtenbrot} {et~al.}(2019{\natexlab{a}}){Trakhtenbrot}, {Arcavi}, {Ricci}, {Tacchella}, {Stern}, {Netzer}, {Jonker}, {Horesh}, {Mej{\'\i}a-Restrepo}, {Hosseinzadeh}, {Hallefors}, {Howell}, {McCully}, {Balokovi{\'c}}, {Heida}, {Kamraj}, {Lansbury}, {Wyrzykowski}, {Gromadzki}, {Hamanowicz}, {Cenko}, {Sand}, {Hsiao}, {Phillips}, {Diamond}, {Kara}, {Gendreau}, {Arzoumanian}, \& {Remillard}}]{2019Trakhtenbrot}
{Trakhtenbrot}, B., {Arcavi}, I., {Ricci}, C., {et~al.} 2019{\natexlab{a}}, Nature Astronomy, 3, 242, \dodoi{10.1038/s41550-018-0661-3}

\bibitem[{{Trakhtenbrot} {et~al.}(2019{\natexlab{b}}){Trakhtenbrot}, {Arcavi}, {MacLeod}, {Ricci}, {Kara}, {Graham}, {Stern}, {Harrison}, {Burke}, {Hiramatsu}, {Hosseinzadeh}, {Howell}, {Smartt}, {Rest}, {Prieto}, {Shappee}, {Holoien}, {Bersier}, {Filippenko}, {Brink}, {Zheng}, {Li}, {Remillard}, \& {Loewenstein}}]{Trakhtenbrot2019_1ES}
{Trakhtenbrot}, B., {Arcavi}, I., {MacLeod}, C.~L., {et~al.} 2019{\natexlab{b}}, \apj, 883, 94, \dodoi{10.3847/1538-4357/ab39e4}

\bibitem[{{Truemper}(1993)}]{1993truemper}
{Truemper}, J. 1993, Science, 260, 1769, \dodoi{10.1126/science.260.5115.1769}

\bibitem[{{Valenti} {et~al.}(2014){Valenti}, {Sand}, {Pastorello}, {Graham}, {Howell}, {Parrent}, {Tomasella}, {Ochner}, {Fraser}, {Benetti}, {Yuan}, {Smartt}, {Maund}, {Arcavi}, {Gal-Yam}, {Inserra}, \& {Young}}]{2014Valenti}
{Valenti}, S., {Sand}, D., {Pastorello}, A., {et~al.} 2014, \mnras, 438, L101, \dodoi{10.1093/mnrasl/slt171}

\bibitem[{{Valenti} {et~al.}(2016){Valenti}, {Howell}, {Stritzinger}, {Graham}, {Hosseinzadeh}, {Arcavi}, {Bildsten}, {Jerkstrand}, {McCully}, {Pastorello}, {Piro}, {Sand}, {Smartt}, {Terreran}, {Baltay}, {Benetti}, {Brown}, {Filippenko}, {Fraser}, {Rabinowitz}, {Sullivan}, \& {Yuan}}]{Valenti2016}
{Valenti}, S., {Howell}, D.~A., {Stritzinger}, M.~D., {et~al.} 2016, \mnras, 459, 3939, \dodoi{10.1093/mnras/stw870}

\bibitem[{{van Velzen} {et~al.}(2020){van Velzen}, {Holoien}, {Onori}, {Hung}, \& {Arcavi}}]{2020vanvelzen}
{van Velzen}, S., {Holoien}, T. W.~S., {Onori}, F., {Hung}, T., \& {Arcavi}, I. 2020, \ssr, 216, 124, \dodoi{10.1007/s11214-020-00753-z}

\bibitem[{{van Velzen} {et~al.}(2021{\natexlab{a}}){van Velzen}, {Pasham}, {Komossa}, {Yan}, \& {Kara}}]{2021SSRvvanVelzen}
{van Velzen}, S., {Pasham}, D.~R., {Komossa}, S., {Yan}, L., \& {Kara}, E.~A. 2021{\natexlab{a}}, \ssr, 217, 63, \dodoi{10.1007/s11214-021-00835-6}

\bibitem[{{van Velzen} {et~al.}(2021{\natexlab{b}}){van Velzen}, {Gezari}, {Hammerstein}, {Roth}, {Frederick}, {Ward}, {Hung}, {Cenko}, {Stein}, {Perley}, {Taggart}, {Foley}, {Sollerman}, {Blagorodnova}, {Andreoni}, {Bellm}, {Brinnel}, {De}, {Dekany}, {Feeney}, {Fremling}, {Giomi}, {Golkhou}, {Graham}, {Ho}, {Kasliwal}, {Kilpatrick}, {Kulkarni}, {Kupfer}, {Laher}, {Mahabal}, {Masci}, {Miller}, {Nordin}, {Riddle}, {Rusholme}, {van Santen}, {Sharma}, {Shupe}, \& {Soumagnac}}]{vanvelzen2021}
{van Velzen}, S., {Gezari}, S., {Hammerstein}, E., {et~al.} 2021{\natexlab{b}}, \apj, 908, 4, \dodoi{10.3847/1538-4357/abc258}

\bibitem[{{van Velzen} {et~al.}(2024){van Velzen}, {Stein}, {Gilfanov}, {Kowalski}, {Hayasaki}, {Reusch}, {Yao}, {Garrappa}, {Franckowiak}, {Gezari}, {Nordin}, {Fremling}, {Sharma}, {Yan}, {Kool}, {Stern}, {Veres}, {Sollerman}, {Medvedev}, {Sunyaev}, {Bellm}, {Dekany}, {Duev}, {Graham}, {Kasliwal}, {Kulkarni}, {Laher}, {Riddle}, \& {Rusholme}}]{vanvelzen2024}
{van Velzen}, S., {Stein}, R., {Gilfanov}, M., {et~al.} 2024, \mnras, 529, 2559, \dodoi{10.1093/mnras/stae610}

\bibitem[{{Vanden Berk} {et~al.}(2001){Vanden Berk}, {Richards}, {Bauer}, {Strauss}, {Schneider}, {Heckman}, {York}, {Hall}, {Fan}, {Knapp}, {Anderson}, {Annis}, {Bahcall}, {Bernardi}, {Briggs}, {Brinkmann}, {Brunner}, {Burles}, {Carey}, {Castander}, {Connolly}, {Crocker}, {Csabai}, {Doi}, {Finkbeiner}, {Friedman}, {Frieman}, {Fukugita}, {Gunn}, {Hennessy}, {Ivezi{\'c}}, {Kent}, {Kunszt}, {Lamb}, {Leger}, {Long}, {Loveday}, {Lupton}, {Meiksin}, {Merelli}, {Munn}, {Newberg}, {Newcomb}, {Nichol}, {Owen}, {Pier}, {Pope}, {Rockosi}, {Schlegel}, {Siegmund}, {Smee}, {Snir}, {Stoughton}, {Stubbs}, {SubbaRao}, {Szalay}, {Szokoly}, {Tremonti}, {Uomoto}, {Waddell}, {Yanny}, \& {Zheng}}]{VandenBerk2001}
{Vanden Berk}, D.~E., {Richards}, G.~T., {Bauer}, A., {et~al.} 2001, \aj, 122, 549, \dodoi{10.1086/321167}

\bibitem[{{Vanden Berk} {et~al.}(2004){Vanden Berk}, {Wilhite}, {Kron}, {Anderson}, {Brunner}, {Hall}, {Ivezi{\'c}}, {Richards}, {Schneider}, {York}, {Brinkmann}, {Lamb}, {Nichol}, \& {Schlegel}}]{VdB04}
{Vanden Berk}, D.~E., {Wilhite}, B.~C., {Kron}, R.~G., {et~al.} 2004, \apj, 601, 692, \dodoi{10.1086/380563}

\bibitem[{{Velzen}(2021)}]{2021TNSTR3680....1V}
{Velzen}, S.~V. 2021, Transient Name Server Discovery Report, 2021-3680, 1

\bibitem[{{Veres} {et~al.}(2023){Veres}, {Reusch}, {Stein}, {Necker}, {Hammerstein}, {Franckowiak}, {Adebahr}, {M{\"u}ller}, {Taziaux}, {Kowalski}, {Jordana-Mitjans}, {Velzen}, \& {Garrappa}}]{2023TNSAN.194....1V}
{Veres}, P.~M., {Reusch}, S., {Stein}, R., {et~al.} 2023, Transient Name Server AstroNote, 194, 1

\bibitem[{Virtanen {et~al.}(2020)Virtanen, Gommers, Oliphant, Haberland, Reddy, Cournapeau, Burovski, Peterson, Weckesser, Bright, {van der Walt}, Brett, Wilson, Millman, Mayorov, Nelson, Jones, Kern, Larson, Carey, Polat, Feng, Moore, {VanderPlas}, Laxalde, Perktold, Cimrman, Henriksen, Quintero, Harris, Archibald, Ribeiro, Pedregosa, {van Mulbregt}, \& {SciPy 1.0 Contributors}}]{2020SciPy-NMeth}
Virtanen, P., Gommers, R., Oliphant, T.~E., {et~al.} 2020, Nature Methods, 17, 261, \dodoi{10.1038/s41592-019-0686-2}

\bibitem[{{Wang} {et~al.}(2025){Wang}, {Woo}, {Gallo}, {Son}, {Yang}, {Jin}, {Guo}, \& {Kong}}]{Wang25}
{Wang}, S., {Woo}, J.-H., {Gallo}, E., {et~al.} 2025, \apj, 981, 129, \dodoi{10.3847/1538-4357/adadf3}

\bibitem[{{Wang} {et~al.}(2020){Wang}, {Shen}, {Jiang}, {Grier}, {Horne}, {Homayouni}, {Peterson}, {Trump}, {Brandt}, {Hall}, {Ho}, {Li}, {Hernandez Santisteban}, {Kinemuchi}, {McGreer}, \& {Schneider}}]{Wang2020}
{Wang}, S., {Shen}, Y., {Jiang}, L., {et~al.} 2020, \apj, 903, 51, \dodoi{10.3847/1538-4357/abb36d}

\bibitem[{{Wang} {et~al.}(2012){Wang}, {Zhou}, {Komossa}, {Wang}, {Yuan}, \& {Yang}}]{Wang2012}
{Wang}, T.-G., {Zhou}, H.-Y., {Komossa}, S., {et~al.} 2012, \apj, 749, 115, \dodoi{10.1088/0004-637X/749/2/115}

\bibitem[{{Wevers} {et~al.}(2024){Wevers}, {French}, {Zabludoff}, {Fischer}, {Rowlands}, {Guolo}, {Dalla Barba}, {Arcodia}, {Berton}, {Bian}, {Linial}, {Miniutti}, \& {Pasham}}]{Wevers24}
{Wevers}, T., {French}, K.~D., {Zabludoff}, A.~I., {et~al.} 2024, \apjl, 970, L23, \dodoi{10.3847/2041-8213/ad5f1b}

\bibitem[{{Winter} \& {Lunardini}(2023)}]{Winter23}
{Winter}, W., \& {Lunardini}, C. 2023, \apj, 948, 42, \dodoi{10.3847/1538-4357/acbe9e}

\bibitem[{{Yang} {et~al.}(2013){Yang}, {Wang}, {Ferland}, {Yuan}, {Zhou}, \& {Jiang}}]{Yang2013}
{Yang}, C.-W., {Wang}, T.-G., {Ferland}, G., {et~al.} 2013, \apj, 774, 46, \dodoi{10.1088/0004-637X/774/1/46}

\bibitem[{{Yaron} \& {Gal-Yam}(2012)}]{2012yaron}
{Yaron}, O., \& {Gal-Yam}, A. 2012, \pasp, 124, 668, \dodoi{10.1086/666656}

\bibitem[{{York} {et~al.}(2000){York}, {Adelman}, {Anderson}, {Anderson}, {Annis}, {Bahcall}, {Bakken}, {Barkhouser}, {Bastian}, {Berman}, {Boroski}, {Bracker}, {Briegel}, {Briggs}, {Brinkmann}, {Brunner}, {Burles}, {Carey}, {Carr}, {Castander}, {Chen}, {Colestock}, {Connolly}, {Crocker}, {Csabai}, {Czarapata}, {Davis}, {Doi}, {Dombeck}, {Eisenstein}, {Ellman}, {Elms}, {Evans}, {Fan}, {Federwitz}, {Fiscelli}, {Friedman}, {Frieman}, {Fukugita}, {Gillespie}, {Gunn}, {Gurbani}, {de Haas}, {Haldeman}, {Harris}, {Hayes}, {Heckman}, {Hennessy}, {Hindsley}, {Holm}, {Holmgren}, {Huang}, {Hull}, {Husby}, {Ichikawa}, {Ichikawa}, {Ivezi{\'c}}, {Kent}, {Kim}, {Kinney}, {Klaene}, {Kleinman}, {Kleinman}, {Knapp}, {Korienek}, {Kron}, {Kunszt}, {Lamb}, {Lee}, {Leger}, {Limmongkol}, {Lindenmeyer}, {Long}, {Loomis}, {Loveday}, {Lucinio}, {Lupton}, {MacKinnon}, {Mannery}, {Mantsch}, {Margon}, {McGehee}, {McKay}, {Meiksin}, {Merelli}, {Monet}, {Munn}, {Narayanan}, {Nash}, {Neilsen}, {Neswold}, {Newberg}, {Nichol}, {Nicinski},
  {Nonino}, {Okada}, {Okamura}, {Ostriker}, {Owen}, {Pauls}, {Peoples}, {Peterson}, {Petravick}, {Pier}, {Pope}, {Pordes}, {Prosapio}, {Rechenmacher}, {Quinn}, {Richards}, {Richmond}, {Rivetta}, {Rockosi}, {Ruthmansdorfer}, {Sandford}, {Schlegel}, {Schneider}, {Sekiguchi}, {Sergey}, {Shimasaku}, {Siegmund}, {Smee}, {Smith}, {Snedden}, {Stone}, {Stoughton}, {Strauss}, {Stubbs}, {SubbaRao}, {Szalay}, {Szapudi}, {Szokoly}, {Thakar}, {Tremonti}, {Tucker}, {Uomoto}, {Vanden Berk}, {Vogeley}, {Waddell}, {Wang}, {Watanabe}, {Weinberg}, {Yanny}, {Yasuda}, \& {SDSS Collaboration}}]{2000york}
{York}, D.~G., {Adelman}, J., {Anderson}, John~E., J., {et~al.} 2000, \aj, 120, 1579, \dodoi{10.1086/301513}

\bibitem[{{Zeltyn} {et~al.}(2022){Zeltyn}, {Trakhtenbrot}, {Eracleous}, {Runnoe}, {Trump}, {Stern}, {Shen}, {Hern{\'a}ndez-Garc{\'\i}a}, {Bauer}, {Yang}, {Dwelly}, {Ricci}, {Green}, {Anderson}, {Assef}, {Guolo}, {MacLeod}, {Davis}, {Fries}, {Gezari}, {Grogin}, {Homan}, {Koekemoer}, {Krumpe}, {LaMassa}, {Liu}, {Merloni}, {Mart{\'\i}nez-Aldama}, {Schneider}, {Temple}, {Brownstein}, {Ibarra-Medel}, {Burke}, {Pellegrino}, \& {Kollmeier}}]{Zeltyn2022}
{Zeltyn}, G., {Trakhtenbrot}, B., {Eracleous}, M., {et~al.} 2022, \apjl, 939, L16, \dodoi{10.3847/2041-8213/ac9a47}

\bibitem[{{Zeltyn} {et~al.}(2024){Zeltyn}, {Trakhtenbrot}, {Eracleous}, {Yang}, {Green}, {Anderson}, {LaMassa}, {Runnoe}, {Assef}, {Bauer}, {Brandt}, {Davis}, {Frederick}, {Fries}, {Graham}, {Grogin}, {Guolo}, {Hern{\'a}ndez-Garc{\'\i}a}, {Koekemoer}, {Krumpe}, {Liu}, {Mart{\'\i}nez-Aldama}, {Ricci}, {Schneider}, {Shen}, {{\'S}niegowska}, {Temple}, {Trump}, {Xue}, {Brownstein}, {Dwelly}, {Morrison}, {Bizyaev}, {Pan}, \& {Kollmeier}}]{2024zeltyn}
---. 2024, \apj, 966, 85, \dodoi{10.3847/1538-4357/ad2f30}

\bibitem[{{Zhong} {et~al.}(2022){Zhong}, {Li}, {Berczik}, \& {Spurzem}}]{Zhong2022}
{Zhong}, S., {Li}, S., {Berczik}, P., \& {Spurzem}, R. 2022, \apj, 933, 96, \dodoi{10.3847/1538-4357/ac71ad}

\end{thebibliography}
\bibliographystyle{aasjournal}

\end{document}